\documentclass[usenatbib]{mn2e}

\usepackage{times,graphics,amssymb,epsfig,subfigure,color,cases}
\usepackage[latin1]{inputenc} 
\def\labequn #1{\label{eq:#1}}

\def\mic{$\mu$m}
\def\unit #1{\,{\rm #1}}
\def\ev{\unit{eV}}
\def\kev{\unit{keV}}
\def\etal{et al.\ }

\def\zsol{Z_{\odot}}

\newcommand\ion[2]{#1$\;${\scshape{#2}}}
\def\h1{\ion{H\,}{i}}

\def\Hbeta{\ion{H}{\,$beta$}}

\def\lya{\ion{Ly$\,$}{$\alpha$}}

\def\he1{\ion{He$\,$}{i}}
\def\c2{\ion{C$\,$}{ii}}
\def\c3{\ion{C$\,$}{iii}}
\def\c4{\ion{C$\,$}{iv}}
\def\n5{\ion{N$\,$}{v}}
\def\si4{\ion{Si$\,$}{iv}}
\def\fe2{\ion{Fe$\,$}{ii}}
\def\o3{$[$\ion{O$\,$}{iii}$]$}
\def\warm1{{\sc WA-1}}
\def\wa2{{\sc WA-2}}
\def\w3{{\sc WA-3}}

\def\etal{{et al.\ }}
\def\ga{\stackrel{>}{\sim}}
\def\la{\stackrel{<}{\sim}}

\def\arcsec{$''$}

\def\cm{{\rm\thinspace cm}}

\def\erg{{\rm\thinspace erg}}

\def\ga{{\rm\thinspace gauss}}

\def\keV{{\rm\thinspace keV}}
\def\km{{\rm\thinspace km}}

\def\Lsun{\hbox{$\rm\thinspace L_{\odot}$}}

\def\Msun{\hbox{$\rm\thinspace M_{\odot}$}}

\def\ph{{\rm\thinspace ph}}
\def\s{{\rm\thinspace s}}

\def\ergpcmsqpspA{\hbox{$\erg\cm^{-2}\s^{-1}$\AA$^{-1}\,$}}
\def\ergps{\hbox{$\erg\s^{-1}\,$}}

\def\kmps{\hbox{$\km\s^{-1}\,$}}

\def\phpcmsqps{\hbox{$\ph\cm^{-2}\s^{-1}\,$}}

\def\psqcm{\hbox{$\cm^{-2}\,$}}

\def\powerlawfluxat1kev{\hbox{$\ph\cm^{-2}\s^{-1}\keV^{-1}$}}

\def\lapp{\ifmmode\stackrel{<}{_{\sim}}\else$\stackrel{<}{_{\sim}}$\fi}

\def\gapp{\ifmmode\stackrel{>}{_{\sim}}\else$\stackrel{>}{_{\sim}}$\fi}
\def\spose#1{\hbox to 0pt{#1\hss}}

\def\approxlt{\mathrel{\spose{\lower 3pt\hbox{$\sim$}}
        \raise 2.0pt\hbox{$<$}}}
\def\approxgt{\mathrel{\spose{\lower 3pt\hbox{$\sim$}}
        \raise 2.0pt\hbox{$>$}}}

\def\lapp{\ifmmode\stackrel{<}{_{\sim}}\else$\stackrel{<}{_{\sim}}$\fi}
\def\gapp{\ifmmode\stackrel{>}{_{\sim}}\else$\stackrel{>}{_{\sim}}$\fi}

\def\mcg6{MCG~-6-30-15}
\def\pg1211{PG~1211+143}
\def\1624{4U~1624$-$490}
\def\mrk766{Markarian~766}
\def\mr2251{MRC~2251-178}
\def\ngc2110{NGC~2110}
\def\iras13349{IRAS~13349+2438}
\def\iras18325{IRAS~18325--5926}
\def\grs1915{GRS~1915+105}
\def\cirx1{Cir~X-1}
\def\xtej1748{XTE~J1748-288}
\def\chandra{{\it Chandra }}
\def\hst{{\it HST }}
\def\XMM{{\it XMM-Newton }}
\def\xmm{{\it XMM }}

\def\asca{{\it ASCA }}

\def\rosat{{\it ROSAT }}

\def\xtegammamcg6{$\Gamma=1.9$}

\def\ovii{O~{\sc vii }\,}

\def\fe25{Fe~{\sc xxv}\,}
\def\fe26{Fe~{\sc xxvi}\,}
\def\ne9{Ne~{\sc ix }\,}
\def\ne10{Ne~{\sc x }\,}
\def\mg11{Mg~{\sc xi }\,}
\def\si13{Si~{\sc xiii }\,}

\def\nj{$N_{\rm j}\;$}

\def\micron{$\mu {\rm m}\;$}

\def\ovii{{\sc O~vii}$\,$}

\def\ne9{{\sc Ne~ix}}
\def\ne10{{\sc Ne~x}}

\def\si4{Si~{\sc iv}}
\def\fe25{Fe~{\sc xxv}}
\def\fe26{Fe~{\sc xxvi}}
\def\mg2{Mg~{\sc ii}}
\def\Msun{\ifmmode M_{\odot} \else $M_{\odot}$\fi}
\def\Lsun{\ifmmode L_{\odot} \else $L_{\odot}$\fi}

\def\h1{H {\sc i}}

\def\Hbeta{H$\beta$}

\def\lya{Ly$\alpha$}

\def\he1{He\,{\sc i}}
\def\c2{C\,{\sc ii}}
\def\c3{C\,{\sc iii}}
\def\c4{C\,{\sc iv}}
\def\n5{N\,{\sc v}}
\def\si4{Si\,{\sc iv}}
\def\fe2{Fe\,{\sc ii}}
\def\o3{[{\sc O~iii}]}

\def\ir13349{IRAS~13349}
\def\iras13349{IRAS~13349+2438}

\title[IRAS 13349+2438]
{The Ionized Absorber and Nuclear Environment of IRAS 13349+2438: 
Multi-wavelength insights from coordinated Chandra HETGS, HST STIS, HET, and Spitzer IRS 
}

\author[Lee \etal]
{\parbox[]{6.in} 
{
Julia C. Lee$^{1,2}$\thanks{E-mail: jclee@cfa.harvard.edu},
Gerard A. Kriss$^{3,4}$, 
Susmita Chakravorty$^{1,2}$,\
Farid Rahoui$^{1,2}$,
Andrew~J.~Young$^{5}$, 
William N. Brandt$^{6,7}$, 
Dean~C.~Hines$^{3}$, 
Patrick~M. Ogle$^{8}$, 
Christopher S. Reynolds$^{9}$ \\
\footnotesize \it $^{1}$Harvard University Department of Astronomy; jclee@cfa.harvard.edu \\ 
\it $^{2}$Harvard-Smithsonian Center for Astrophysics, 60 Garden Street MS-6, Cambridge, MA 02138 USA \\
\it $^{3}$Space Telescope Science Institute, Baltimore, MD 21218 USA \\
\it $^{4}$Department of Physics \& Astronomy, The Johns Hopkins University, 3400 North Charles Street, Baltimore, MD 21218 USA \\
\it $^{5}$HH Wills Physics Laboratory, University of Bristol, Tyndall Avenue, Bristol BS8 1TL, U.K. \\
\it $^{6}$Department of Astronomy \& Astrophysics, 525 Davey Lab, The Pennsylvania State University, University Park, PA, 16802 USA \\
\it $^{7}$Institute for Gravitation and the Cosmos, The Pennsylvania State University, University Park, PA 16802, USA \\
\it $^{8}$IPAC, MS 100-22, California Institute of Technology, Pasadena, CA 91125 USA \\
\it $^{9}$Department of Astronomy, University of Maryland, College Park, MD 20742 USA \\
}
}
\begin{document}
\maketitle

%%%%%%%%%%%%%%%%%%%%%%%%%%%%%%%%%%%%%%%%%%%%%%%%%%%%%%%%

\begin{abstract}

We present results from a multiwavelength IR--to--X-ray campaign of the
infrared bright (but highly optical-UV extincted) QSO \iras13349\
obtained with the \chandra High Energy
Transmission Grating Spectrometer (HETGS), the {\it Hubble Space
Telescope} Space Telescope Imaging Spectrograph (STIS), the
Hobby-Eberly Telescope (HET) 8-meter, and the {\it Spitzer} Infrared
Spectrometer (IRS).   Based on HET optical spectra of $[$\ion{O}{iii}$]$,
we refine the redshift of \ir13349 to be $z = 0.10853$.
The weakness of the $[$\ion{O}{iii}$]$ in combination with strong \ion{Fe}{ ii}
in the HET spectra reveal extreme Eigenvector-1 characteristics
in \ir13349, but the 2468 \kmps width of the \Hbeta\ line argues
against a narrow-line Seyfert~1 classification;
on average, IR, optical and UV spectra
show \ir13349 to be a typical QSO.
Independent estimates based on the \Hbeta\ line width and fits to
the \ir13349 SED both give a
black hole mass of $M_{\rm BH} = 10^{9}~\Msun$.
The heavily reddened STIS UV
spectra reveal for the first time blue-shifted
absorption from \ion{Ly}{$\alpha$}, \n5\ and  \c4, with components
at systemic velocities of $-950~\kmps$ and $-75~\kmps$.
The higher velocity UV lines are coincident with the lower-ionisation
($\xi \sim 1.6$) \warm1 warm absorber lines seen in the X-rays with the
HETGS. In addition, a $\xi \sim 3.4$ \wa2 is also required by the data,
while a $\xi \sim 3$ \w3 is predicted by theory, and seen at less significance;
all detected X-ray absorption lines are blueshifted by  $\sim700-900$\kmps.
Theoretical models comparing different ionising SEDs reveal that including
the UV (i.e., the accretion disc) as part of the ionising continuum has
strong implications for  the conclusions one would draw about the
thermodynamic stability of the warm absorber.  Specific to \ir13349,
we find that an X-ray-UV ionising SED favors a continuous distribution
of ionisation states in a smooth flow (this paper), versus discrete clouds
in pressure equilibrium (previous work by other authors).  Direct detections of dust are
seen in both the IR and X-rays. We see weak PAH emission at 7.7\,\micron
and 11.3\,\micron which may also be blended with forsterite,
and 10\,\micron and 18\,\micron silicate emission, as well as an Fe L edge at 700 eV
indicative of iron-base dust with a dust-to-gas ratio $> 90$\%.
We develop a geometrical model in which we view the nuclear
regions of the QSO along a line of sight that passes through the upper
atmosphere of an obscuring torus.  This sight line is largely
transparent in X-rays since the gas is ionised, but it is completely
obscured by dust that blocks a direct view of the UV/optical emission
region. In the context of our model,  20\% of the intrinsic UV/optical 
continuum is scattered into our sight line by the far wall of an obscuring torus. 
An additional 2.4\% of the direct light, which likely dominates the UV emission, 
is Thomson-scattered into our line-of-sight by another off-plane component of highly ionized gas.
\end{abstract}

%%%%%%%%%%%%%%%%%%%%%%%%%%%%%%%%%%%%%%%%%%%%%%%%%%%%%%%%%%%%%%%%%%%


%%%%%%%%%%%%%%%%%%%%%%%%%%%%%%%%%%%%%%%%%%%%%%%%%%%%%%%%%%%%%%%%%%%%

\section{Introduction}

IRAS~13349+2438, hereafter ``\ir13349'', ($z=0.10764$ -- \citealt{ir13349-z};
updated here to $z = 0.108530$ based on the 3$\times$ higher resolution HET
spectra reported in this paper) is a prototype infrared-luminous quasar with a
bolometric luminosity of $\approxgt 2\times 10^{46}$~erg~s$^{-1}$
\citep{ir13349:discovery}.   Images of the host galaxy and nearby environment
show the galaxy to be spiral-like, with a possible companion at $\sim$
5\arcsec\ along the minor axis (e.g., \citealt{hutchings:90}).  Evidence for
tidal structure suggests that the object may have interacted with the
companion, and this could supply gas and dust to the nucleus, fueling quasar
activity and enhancing nuclear obscuration and scattering. Indeed, \ir13349 has
a broad-emission line optical spectrum that becomes heavily reddened at shorter
wavelengths, and exhibits high optical continuum and emission-line polarisation
\citep[hereafter W92]{wills92}. The observed polarisation rises strongly toward
shorter wavelengths, but the optical polarised flux spectrum is
indistinguishable from a typical, unreddened quasar (Hines et al. 2001).  These
polarisation properties indicate that observers see the quasar's nucleus via
both a direct, but attenuated, light path and a scattered light path
\citep[W92;][]{hines01}. In addition to its high luminosity and spectral
variability, \ir13349 also exhibits strong Eigenvector-1 characteristics
(strong optical \ion{Fe}{ii} emission and weak $[$\ion{O}{iii}$]$ relative to $\rm
H\beta$; \citealt{borosongreen92}).  It is typically identified to be
radio-quiet, although weak 4.87~mJy radio emission has been reported at 6~GHz
by \cite{ir13349-radio}.

The observed high X-ray flux and large-amplitude X-ray variability indicate
that a large fraction of the X-ray emission is seen via a direct, rather than a
scattered, path.  As such, \ir13349 has been the subject of a number of X-ray
studies that have helped to clarify the physical processes in the inner regions
of the  quasar nucleus.  Using \XMM EPIC data, \cite{longinotti-ir13349fe03}
made a detailed analysis of the \ir13349 ionised reflection spectrum and
relativistic Fe emission line.  At lower energies, X-ray studies based on the
\rosat\ PSPC \citep{brandt96} and \asca \citep{brandt97}, combined with
optical/near-infrared extinction estimates argue for obscuration by dusty,
ionised gas. Studies with modern day high spectral resolution instruments on
board \XMM \citep{sako13349:01}, and \chandra (\citealt{holczer07,behar09}; also this
paper) reveal additional complexity in the absorbers  (most notable an
unresolved transition array of 2p-3d inner-shell absorption by iron  M-shell
ions, dubbed the UTA by its discoverers, Sako et al.),  and allow direct
measurements of the dust composition (this paper) in the host galaxy.

In this paper, we present a comprehensive analysis of the  absorber properties
of this quasar, based on our high spectral resolution, multi-wavelength
campaign, involving  X-ray (with the \chandra High Energy Transmission Grating
Spectrometer; HETGS), ultraviolet (with HST STIS-MAMA), optical (HET:
Hobby-Eberly Telescope 8-m), and infrared (Spitzer IRS) observations. The
\chandra and HST observations are simultaneous, HET near simultaneous, and
Spitzer IRS taken 1.25~years later as part of the GTO program of G.~Rieke.
These high-quality data allow us to address, in detail for a specific luminous
system, several issues of broader importance in active-galaxy absorption
studies. These include the apparent presence of dust in some ionized absorbers
and its implications for interpreting observations; the relation of ionized
absorbers to other nuclear components, including accretion disks, tori, and
scattering material; the relation of the spectral energy distribution (SED) to
the structure of ionized absorbers; and the potential for feedback into host
galaxies by outflowing nuclear winds.

The paper is organised as follows. In Section~\ref{sec:data}, the
multi-wavelength data are presented, to be followed by a plasma diagnostic
approach to the line analysis in Section~\ref{sec:analysis}.  In
Section~\ref{sec:SED:obsth} we combine our multi-wavelength data with
additional ISO and IRAS archived observations to produce an observed SED that
is used to determine a theoretically motivated ionising spectrum affecting the
warm absorber properties; this then is used to generate ion populations with
{\sc xstar} for our spectral fitting.  This will set the stage for
Section~\ref{sec:discussion} considerations on the warm absorber behaviour as
established by thermodynamic stability arguments
(Section~\ref{subsec:wathermo}), and complex line-of-sight geometry through
dust (Section~\ref{subsec-dwa}).  We adopt $H_0=73$~km~s$^{-1}$ Mpc$^{-1}$,
$\Omega_{\rm M}=0.25$, and $\Omega_{\Lambda}=0.75$ throughout this paper. 

\section{Observations \& Data Reduction} 
\label{sec:data}

As part of a multi-wavelength campaign to better understand the
global physical processes, absorber kinematics, and geometry of \ir13349,
we observed this nearby quasar using spectrographs on \chandra (X-ray),
the {\it Hubble Space Telescope} (ultraviolet), the Hobby-Eberly Telescope 
8-meter (optical), and the {\it Spitzer Space Telescope} (infrared;
\citealt{spitzertelescope}).  Observations performed with \chandra,
HST and HET were nearly simultaneous and of comparable spectral
resolution ($R \sim 1000$), while ``high" (R~$\sim 600$) and 
low-resolution (R~$\sim 70-120$) IR spectra
from Spitzer were obtained $\sim 1.25$~years later as part of the GTO
program of George Rieke.

%%%%%%%%%%%%%%%%%%%%%%%%%%%%%%%%%%%%%%%%
%Table 1
\begin{table}
\begin{center}
        \caption[]{Chandra Observations}
        \label{XObsTbl}
%~~\\
\begin{tabular}{l c c c}
\hline\hline
Date      & Start Time & End Time & Useable Time\\
          &   (GMT)    &    (GMT) & (ks)\\
\hline
2004-02-22 & 06:42:06 & ...      &     \\
2004-02-23 &    ...   & 12:15:15 & 160 \\
2004-02-24 & 13:56:00 & ...      &     \\
2004-02-26 &    ...   & 11:16:41 & 135 \\
\hline
\end{tabular}
\end{center}
\end{table}

%%%%%%%%%%%%%%%%%%%%%%%%%%%%%%%%%%%%%%%%

\subsection{X-ray Observations with the Chandra HETGS}

\ir13349 was observed with the \chandra HETGS (High Energy Transmission Grating
Spectrometer: \citealt{hetgref, weisskopf02chandra}) over several days
in 2004 February for a total of $\sim$~295~ks of usable data as
summarised in Table \ref{XObsTbl}.  Plus and minus first order ($\pm
1$) spectra were extracted using the latest {\sc ciao} release ({\sc
ciao}~4.3 with {\sc caldb}~4.4.3), starting with the L1 (raw
unfiltered event) files, which we reprocess to remove hot pixels and
afterglow events.  In order to maximise signal-to-noise (S/N), we
combine plus and minus 1st orders for both the HEG and MEG.  The
resolving power of the HETGS is $\rm E/\Delta E = 1000$.  We focus on
the  $\sim 1.5-26$~\AA\, (0.45$-$8~keV) spectral region in this paper,
with particular emphasis on the soft (0.5-5~keV; 2.5-26\AA) band of
the warm absorber.  Analysis of the \chandra line spectra was done
within the {\sc isis}\footnote{http://space.mit.edu/CXC/ISIS/
\citep{isisref}} analysis package.

\subsection{Simultaneous UV Observations with the Hubble Space Telescope STIS MAMA}
\label{subsecn:HST}
Simultaneous \hst\ ({\it Hubble Space Telescope}) UV spectra were obtained
using the Space Telescope Imaging Spectrograph (STIS) \citep{HSTSTIS}.  The
faintness of \ir13349\ in the UV made the STIS echelle mode an impractical
choice, so we observed \ir13349\ through the $52 \times 0.5 ~ $\arcsec slit
with the UV MAMA detectors and the low resolution UV gratings (R$\sim 1000$)
G140L and G230L in a 4-orbit visit on 2004 Feb 22 starting at GMT 08:09:48.
Table~\ref{ObsTbl} summarises the STIS observations.

%%%%%%%%%%%%%%%%%%%%%%%%%%%%%%%%%%%%%%%%
%Table 2
\begin{table}
\caption[]{STIS Observations}
\label{ObsTbl}
\begin{tabular}{l c c c c}
\hline\hline       
Data Set Name & Grating  & Date & Start Time & Exposure Time\\
              &               &      &    (GMT)   & (s)\\
\hline
o8wp01010 & G140L & 2004-02-22 & 08:09:48 & 2000 \\
o8wp01020 & G140L & 2004-02-22 & 09:31:42 & 2630 \\
o8wp01030 & G140L & 2004-02-22 & 11:08:22 & 2630 \\
o8wp01040 & G230L & 2004-02-22 & 12:44:03 & 2630 \\
\hline                  
\end{tabular}
\end{table}

%%%%%%%%%%%%%%%%%%%%%%%%%%%%%%%%%%%%%%%%

Three exposures with G140L totaling 7260 s and one exposure with G230L
totaling 2630 s gave a spectrum covering the 1150--3180\AA\, spectral
range.  The G140L spectrum has a resolving power that varies with
wavelength of 960--1440 over its full spectral range; for G230L the
resolving power is slightly lower, 530--1040.

We used the extracted one-dimensional spectra as produced by the STScI
pipeline for our analysis.  Our data required only a slight zero-point
adjustment ($< 1$ pixel, corresponding to $< 100$~\kmps in each exposure) to the wavelength scale
which we determined using the Galactic absorption lines in our
spectra.  Observations of Galactic \h1\ along this sight-line by \cite{murphy_Galnh:96}
show an optically thin column density of $1.07 \times
10^{20}~\rm cm^{-2}$ at a mean heliocentric velocity of  $-19~\rm
km~s^{-1}$.  In the region of wavelength overlap between the G140L and
the G230L spectra, the flux levels agree to $<$0.7\%. We have made no
adjustments to either flux scale, given that an absolute accuracy of
better than 4\%  is expected \citep{bostroem10}.

\subsection{Near Simultaneous Optical Observations with the Hobby-Eberly 8-meter}
Since \ir13349\ shows large-amplitude variability at X-ray
wavelengths, we also obtained a near-simultaneous spectrum of it with
the 8-meter Hobby-Eberly Telescope (HET; \citealt{het}) to constrain
its optical properties during the \chandra\ and \hst\ observations.
(We note that while the HET\footnote{http://www.as.utexas.edu/mcdonald/het/het\_gen\_01.html}
is officially designated a 9.2-meter telescope, 
we conservatively refer to it here as an 8-meter class given that this is the 
average equivalent aperture for a typical observation -- see \citealt{het2}.)
HET observations were obtained on 2004 Feb 26 with the Marcario
Low-Resolution Spectrograph \citep{HET_lrsdesign, HET_lrs}.  The
observations were taken with a 600~line~mm$^{-1}$ grating, a slit
width of $1^{\prime\prime}$, and a GG385 blocking filter, resulting in
a resolving power of 1300 over the observed wavelength range
\hbox{4300--7300~\AA.} 
The total exposure time of 1283.5 s was split
among four sub-exposures.  For flux calibration, we observed the
white-dwarf standard Feige~34 during the middle of the night, two
hours prior to our observation of \ir13349.  Exposures on a Cd-Ne arc
lamp were used to determine the wavelength scale, with zero points
adjusted using the night-sky lines in our spectra.  Final
flux-calibrated spectra were extracted with the standard IRAF
reduction codes for single-slit data.

\subsection{Infrared observations with the Spitzer IRS}
\label{subsec:spitzer}

Low-resolution ($R = 70-120$) and high-resolution ($R\sim600$) infrared spectra
were obtained as part of the GTO program of Rieke (PID: 61) on 2005
June 7 with the Infrared Spectrograph \citep[IRS,][]{spitzerirs} on
the {\it Spitzer Space Telescope} \citep{spitzertelescope} using the
standard staring mode. In each spectroscopic module, the source was
observed at two nodding positions, and exposure times were set to
$2\times5\times14$~s in SL (Short-Low) and LL (Long-Low), as well as $2\times5\times120$~s
and $2\times10\times60$~s in SH (Short-High) and LH (Low-High),
respectively.

The low-resolution spectral data (LRS) were reduced, from the basic
calibration data (BCD) files, following the standard procedure given
in the IRS Data
Handbook\footnote{http://ssc.spitzer.caltech.edu/irs/irsinstrumenthandbook/IRS\_Instrument\_Handbook.pdf}.
The basic steps were bad-pixel correction with {\sc irsclean v2.1},
background subtraction using the off-source slit, as well as optimal
extraction and wavelength/flux calibration with the
\textit{Spectroscopic Modeling Analysis and Reduction Tool } suite
({\sc SMART v8.2.1}; \citealt{spitzer-smartreduce}), for each nodding
position. Spectra were then nod-averaged to improve the
signal-to-noise ratio (S/N).

The flux calibration appeared consistent between the different
modules, with differences less than 5\%, and this allowed us to
combine all the spectra to cover the 5.2$-$35~\mic\ spectral range
(4.7$-$32~\mic\ rest frame).  Nevertheless, the continuum at 25~\mic\
showed a multiplicative offset from the IRAS~13349 photometric flux at
25~\mic\, obtained from the
\footnote{http://tdc-www.harvard.edu/catalogs/iras.html}
IRAS~Point~Source~Catalog (hereafter, IRAS-PSC).  Unfortunately, there
is no associated MIPS 24~\mic\ photometry during the epoch of the
Spitzer IRS observation, so we cannot rule out the possibility that
the object has varied in this energy band.  However, given that
IRAS~13349 is radio quiet, and that the shape of the SED suggests that
the mid-IR is produced by thermal emission from dust heated by the
central engine, we assume that the mid-IR has not varied.  We
therefore scaled the entire spectrum by a factor of 1.077 to match the
%%%%% DCH
%%%%% IRAS-PSC flux measurement of 0.84$\pm$0.06 Jy at 25~\mic.
(un-color-corrected) IRAS-FSC flux density measurement of 0.84$\pm$0.06 Jy at 25~\mic.

%----------------------------------------------

%%%%%%%%%%%%%%%%%%%%%%%%%%%%%%%%%%%%%%%%
%Table 3
\begin{table*}
\begin{center}
\caption{Ionic columns densities based on fits to Chandra observed lines from the full transition series}
\label{tab:xseries}
\begin{tabular}{lcccccccccc}
\hline\hline
{$\;\;$Ion}      &
{{$^a$}$\;\;n_{\rm detected}$ } &
{{$^b$}$\;\;\rm v_{blue}$ } &
{{$^c$}$\;\;\rm b$ } &
{{$^d$}$\;\;\rm \, log \, N_{j} $} &
{{$^e$}$\;\;\rm b$ } &
{{$^f$}$\;\;\rm \, log \, N_{j}  $} &
{{$^g$}$\;\;\rm \, N_{H}$} &
\\
&
&
{($\rm km\, s^{-1}$)} &
{($\rm km\, s^{-1}$)} &
{\small ($\rm cm^{-2}$)} &
{($\rm km\, s^{-1}$)} &
{\small ($\rm cm^{-2}$)} &
{\small ($\rm 10^{21} \, cm^{-2}$)} & \\
\hline
\ion{O}{vii} & 2 - $\infty$ & $-748^{+103}_{-126}$ &  $137^{+39}_{-74}$ & $18.1 \pm 0.5$ & 100 & $17.8 \pm 0.5$ & $ 1.48^{}_{}$ \\ \\
\ion{O}{viii} & 2 - $\infty$ & $-595^{+99}_{-107}$ & $288^{+89}_{-92}$ & $17.8^{+0.2}_{-0.3}$ & 100 & $17.6^{+0.4}_{-0.5}$ & $ 0.94^{}_{}$ \\ \\
\ion{Ne}{ix} & 2 - $\infty$ & $-788^{+131}_{-168}$ & $519^{+117}_{-133}$ & $17.3^{+0.1}_{-0.2}$  & 100 & $17.7^{+0.7}_{-0.5}$ &   $ 8.15^{}_{}$   \\ \\
\ion{Ne}{x} & 2 - 6 & $-833^{+354}_{-160}$ & $284^{+164}_{-139}$ & $17.3^{+0.3}_{-0.4}$  & 100 & $17.4^{+0.4}_{-0.6}$ &  $ 4.08^{}_{}$ \\ \\
\ion{Mg}{xi} & 2 - 4 & $-578^{+212}_{-145}$ & $150^{+79}_{-140}$ & $17.4^{+3.8}_{-0.4}$  & 100 & $17.4^{+0.4}_{-0.5}$ & $ 13.0^{}_{}$  \\ \\
\ion{Mg}{xii} & 2 - 4 & $-675^{+157}_{-123}$ & $133^{+87}_{-78}$ & $17.4^{+0.4}_{-0.7}$ & 100 & $17.4^{+0.4}_{-0.7}$ &$ 13.0^{}_{}$   \\ \\
\ion{Si}{xiii} & 2 - 3 & $-537^{+156}_{-134}$ & $48^{+447}_{-38}$ & $^{+}_{-}$  & 100 & $17.6^{+1.0}_{-1.1}$ & $ 22^{}_{}$  \\ \\
\hline 
\end{tabular}
\end{center}
{ See corresponding curve-of-growth plots in Figure~\ref{fig:cog}.} \\
{$^a$}{ Transitions in series which contribute to the best fit $N_{\rm j}$; $\infty$ correspond to series limit. } \\ 
{$^b$}{ Outflowing velocity measured against the laboratory wavelengths tabulated by \cite{vlineref} } \\
{$^c$}{ Turbulent velocity width ($\rm km \, s^{-1}$)  as determined from the series fit. See \S\ref{subsec:xplasmastudy} for discussion on caveats.} \\
{$^d$}{ Corresponding ionic column density ($\rm cm^{-2}$). } \\
{$^e$}{ Turbulent velocity width frozen at 100~\kmps. See \S\ref{subsec:xplasmastudy} for discussion on reasoning behind this.} \\
{$^f$}{ Corresponding ionic column density ($\rm cm^{-2}$). } \\
{$^g$}{ Corresponding equivalent Hydrogen column assuming the solar abundance values given by \cite{ism_ref:00}.  Note however that the numbers assume an ionization fraction for {\it all} ions to be 50\%;  as such, quoted numbers are only rough guides, and {\it not} to be directly taken at face value. 
}
\end{table*}

%%%%%%%%%%%%%%%%%%%%%%%%%%%%%%%%%%%%%%%%

%----------------------------------------------

We reduce the high-resolution spectral data (HRS) in much the same way
as previously described for the LRS, with the exception that
full-aperture instead of optimal extraction is employed, the
distinction being that pixels along the dispersion axis have equal
(full-aperture) versus varied (optimal) weighting.  Moreover, we could
not remove background by on-source/off-source subtraction, due to the
lack of off-source HRS background files in the IRS archives associated
with the epoch of this observation. Therefore, we extract SL1, LL2,
and LL1 backgrounds from the low-resolution data and fit with
first-order polynomials. We then use these fits to remove the
background contribution from the high-resolution spectra, giving a
good match between the SH and LH continuum flux density
level. Finally, the overall high-resolution spectrum is scaled by
1.077 to match the continuum flux density level at 25~\mic\  
to the IRAS~13349 IRAS-PSC photometric flux density at the same wavelength.

%
%%%%%%%%%%%%%%%%%%%%%%%%%%%%%%%%%%%%%%%%
\begin{figure*}
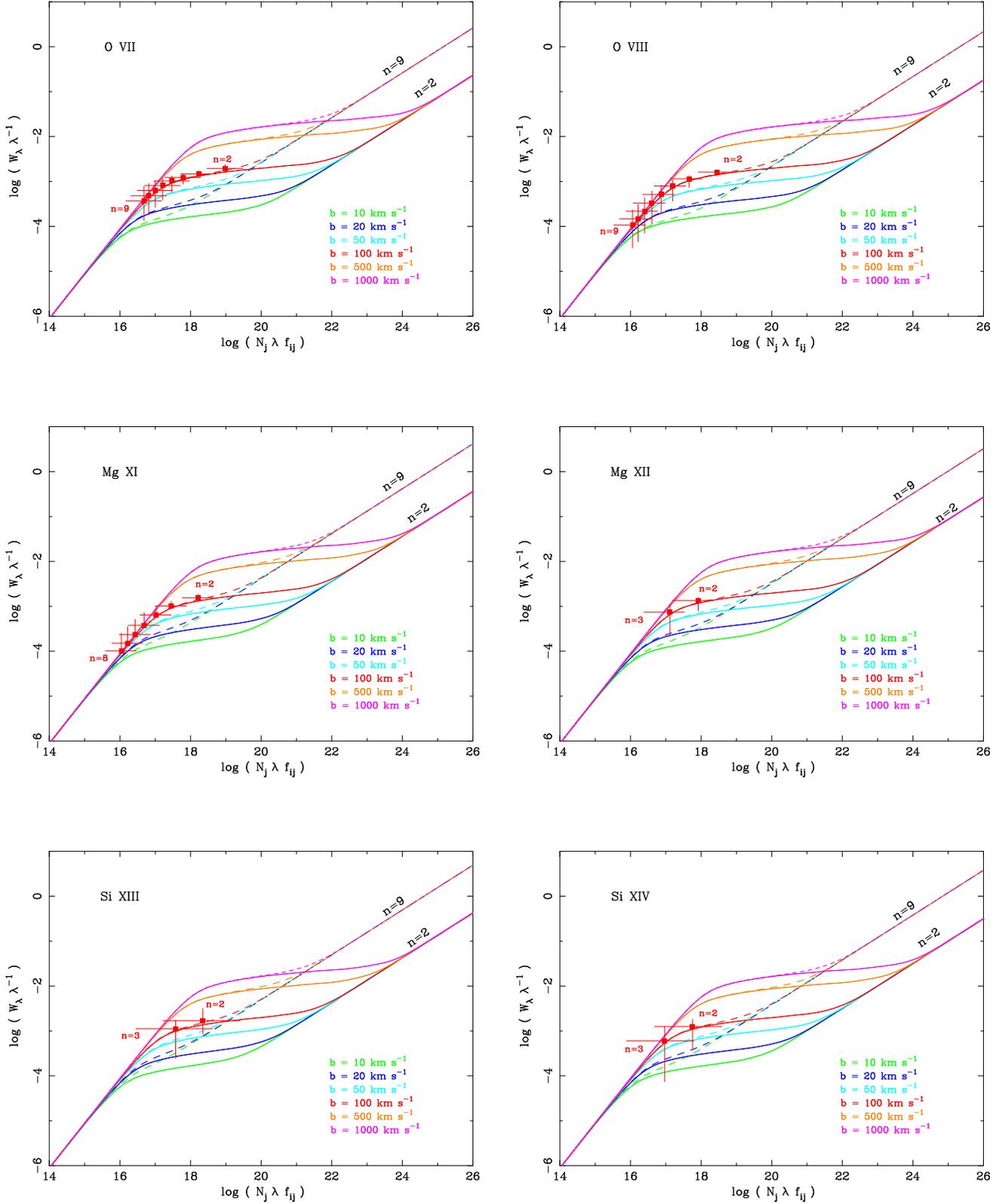

   \hbox{
	\includegraphics[scale=0.35,angle=270]{FIGS/chandra/O7cog_2to9_ir13349.ps} 
     \hspace{0.5cm} 
	\includegraphics[scale=0.35,angle=270]{FIGS/chandra/O8cog_2to9_ir13349.ps} 
     }
   \vspace{0.5in}
%   \hbox{
%        \includegraphics[scale=0.30,angle=270]{FIGS/chandra/Ne9cog_2to8_ir13349.ps}
%     \hspace{0.5cm}
%        \includegraphics[scale=0.30,angle=270]{FIGS/chandra/Ne10cog_2to9_ir13349.ps}
%     }
%   \vspace{0.5in}
   \hbox{
	\includegraphics[scale=0.35,angle=270]{FIGS/chandra/Mg11cog_2to9_ir13349.ps} 
     \hspace{0.5cm} 
	\includegraphics[scale=0.35,angle=270]{FIGS/chandra/Mg12cog_2to9_ir13349.ps} 
     }
   \vspace{0.5in}
   \hbox{
	\includegraphics[scale=0.35,angle=270]{FIGS/chandra/Si13cog_2to9_ir13349.ps} 
     \hspace{0.5cm}
	\includegraphics[scale=0.35,angle=270]{FIGS/chandra/Si14cog_2to9_ir13349.ps} 
     }
\caption[h]{Curve-of-growth calculations representing $n=2$ and $n=9$ transitions
for H-like 1s-1s$n$p and He-like $1s^2$--$1snp$ transitions for different values of the turbulent 
velocity width $b$. Plotted data point represent \nj values measured based on a simultaneous
fit to all the lines detected for a given H- or He-like resonance series, assuming $b=100$\kmps. E.g. plotted points
represent a fit (with oscillator strengths locked in the appropriate ratios) to the detected $n=2$~to~$n=9$ 
lines of He-like $1s^2$--$1snp$ \ion{O}{vii} whereas a fit to the H-like \ion{Mg}{xii} 
only includes a fit to the $n=2$~and~$n=3$ lines. (See Section~\ref{subsec:xplasmastudy} 
and Table~\ref{tab:xseries} for details.) The purpose of the exercise is to demonstrate
the importance of the higher order lines for establishing true \nj derivations, where fits
to single lines (especially the $n=2$ $\alpha$-transition which 
may lie on the saturated part of the COG) may give an \nj with possible values that span
up to two orders of magnitude.
 }
\label{fig:cog}
\end{figure*}
%%%%%%%%%%%%%%%%%%%%%%%%%%%%%%%%%%%%%%%%

\section{Data Analysis and Line Detections} \label{sec:analysis} 
The intent of this paper is to provide the highest spectral resolution
multi-wavelength characterisation of the \ir13349 spectrum to date, as
a  step toward a more detailed understanding of the structure of this
quasar and its host galaxy.    For this paper, we employ both a
plasma-diagnostic approach to the analysis (this section), as well as
photoionization modeling with {\sc xstar}  \citep{xstar_ref} for a
global analysis (\S\ref{sec:SED:obsth}) of the spectra based on the
observed multi-wavelength spectral energy distribution (SED).

\subsection{The X-ray Chandra HETGS spectrum}
%\label{subsecn:xcont}
\label{subsec:xplasmastudy}

%%%%%%%%%%%%%%%%%%%%%%%%%%%%%%%%%%%%%%%%
\begin{figure*}
\centerline{\includegraphics[height=5in,width=5.5in, trim = 125 90 125 75, clip, angle=0]{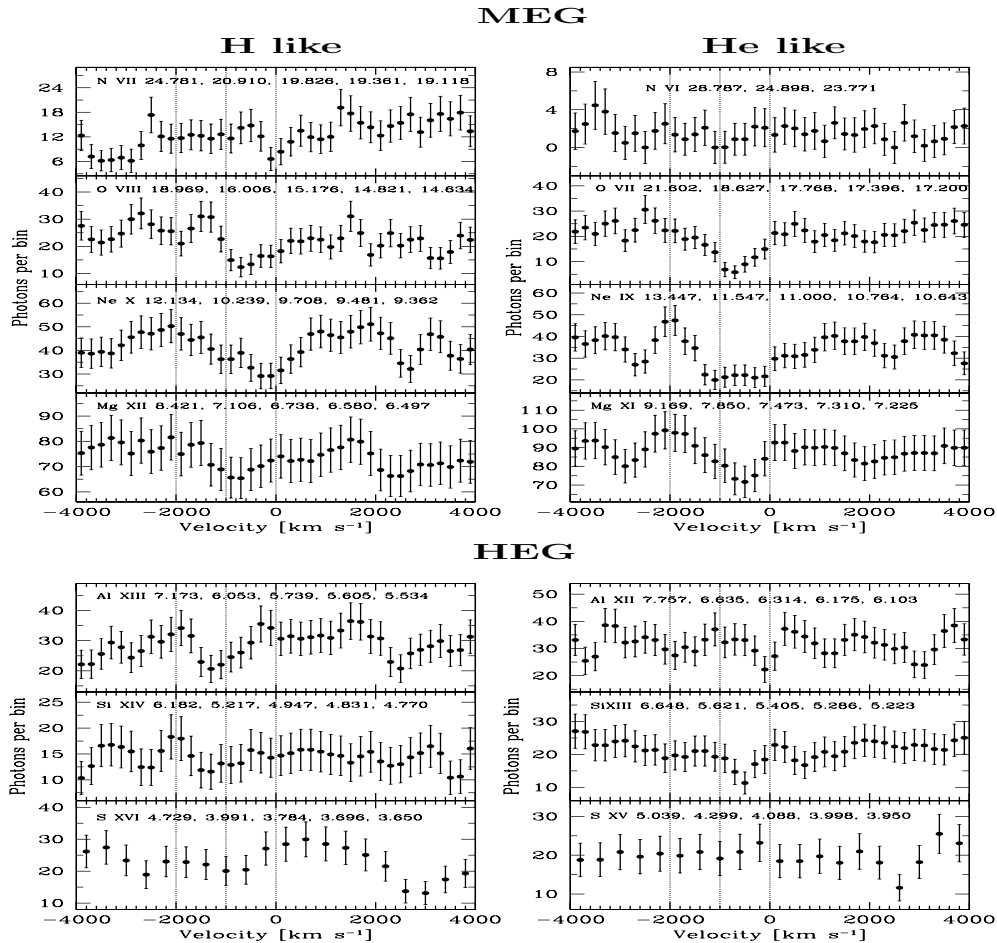}}
%\centerline{\includegraphics[scale=0.7,width=0.5\textwidth, trim = 125 90 125 75, clip, angle=0]{FIGS/chandra/VelocityProfile.ps}}
\caption{ Velocity spectra obtained by co-adding the 5 strongest
resonance lines for H and He like ions of nitrogen, oxygen, neon,
magnesium, aluminium, silicon, and sulfur. For N, O, Ne, and
Mg ions, MEG grating spectra are co-added, whereas HEG data are 
used for Al, Si, and S.  We bin at 200~$\rm{km \, s^{-1}}$ in velocity;
S is binned at 400~$\rm{km \, s^{-1}}$.~--~Velocity bins are chosen
to reflect the approximate FWHM resolution of the HETGS at the energies where these 
ions are found.  To guide the eye, vertical lines denote
velocities corresponding to 0, -1000 and -2000 $\rm{km \, s^{-1}}$.
Multiple absorption components appear strongest for O, Ne, Mg, and Si.
%Absorption is clearly detected for O, Ne, Mg,
%Al and Si. Nitrogen shows hints of absorption, whereas sulphur shows
%no detection. 
%Note the traces of the P Cygni profile for the Al ions
%and Si~{\sc xiii}. 
The errors are Poisson and each bin is independent despite appearances to
the contrary. 
%, although they might appear
%uniform due to the small fluctuations in the signal over the narrow
%range of wavelength considered in each panel.
\label{fig:VelProfile}
}
\end{figure*}
%%%%%%%%%%%%%%%%%%%%%%%%%%%%%%%%%%%%%%%%

We begin initially with a plasma diagnostic approach to the line fits.
To search for individual absorption and emission lines, HETGS spectra
were binned  by two, corresponding to half the spectral resolution,
respectively 0.006\AA\, and 0.012\AA\ FWHM, for the HEG and MEG.  For
these fits, we use a fifth order polynomial to fit local continua,
rather than a phenomenological broad-band continuum.  Such an approach
allows us to remove the broader fluctuations, thereby  maximising our
ability to measure the narrow lines without assumptions about details
of the continuum, which can be complex. Using this analysis procedure,
we find the X-ray spectrum to be dominated by an ionised 
absorber  giving rise to prominent H-
and He-like absorption lines of oxygen, neon, magnesium, and silicon,
as well as absorption lines from a variety of iron ions,  at a
bulk outflow velocity in the range $-900~\kmps \approxlt  v_{\rm
x-ray} \approxlt -500~\kmps$.

Since a comprehensive photoionization treatment of the \ir13349 absorber as
tied to its SED and theory considerations, is explored in detail in
Section~\ref{sec:SED:obsth}, we concentrate here primarily on the key H- and
He-like ions as a model-independent way of accessing the QSO hot plasma
conditions for comparison with later analysis.  For many of these key ions,
high order transitions, corresponding to a high column density warm absorber is
detected (see Table~\ref{tab:xseries}).
% (Table~\ref{tab:xseries}, Figure~\ref{fig:cog}).
For these key ions, we derive the individual ionic column densities (\nj) based
on a fit to the {\it entire detected resonance series} of lines from
$n$=2--$xx$, where $xx$ is the highest transition detected; e.g. H-like $1s-np$
\ion{Mg}{xii} and He-like $1s^2-1snp\,$\ion{Mg}{xi} are detected to $n=4$ so
that our fits to those series are to the Ly/He-($\alpha, \beta, \gamma,
\delta)$ lines with their oscillator strengths locked to their tabulated value
as found e.g.  in \cite{vlineref}. The individual lines in the series are fit
using Voigt profiles, although the derived \nj is based on a simultaneous fit
to all the detected lines in the series.  In this way, we ensure the best
possible determination of \nj  (Table~\ref{tab:xseries}) by reducing the
possibility for false estimates resulting from saturated or contaminated lines.
(Both saturated and contaminated lines can falsely indicate a line width and
strength which is broader and has higher flux than what it truly is for the
ion, thereby resulting  in lower estimates for any given ionic column.)  We
note however, that  despite these precautions, the spectral resolution of the
\chandra HETGS, while the best available to date, is still not sufficient  for
probing at the level of measuring the $10-20$~\kmps thermal widths of these
ions. As such, \nj estimates may only be lower limits.  For thoroughness, we
adopt a value for the turbulent velocity width $b = 100$~\kmps to approximate
the \chandra HETGS spectral resolving capabilities, for deriving the ionic
columns noted in Table~\ref{tab:xseries};  we also detail  the associated
ionic columns based on fits where $b$ is allowed as a free parameter.  For
\ion{Ne}{ix} (and to a lesser extent \ion{Ne}{x}), Fe contaminates many of the
stronger lines in the series, resulting in a larger measured $b$, and hence
smaller \nj.

It is also clear that the measured value of \nj strongly depends on $b$ as best
illustrated by the curve-of-growth (COG) plots in Figure~\ref{fig:cog}.  Here,
we plot COG calculations (which include the damping parameter as presented e.g.
in \citealt{spitzer:cog}) for different values of the turbulent width,  which
range from $b=10-20$~\kmps ($\approx$ the thermal width value of the ion), to
$b=100$~\kmps (maximal resolving power of the HETGS), to $b=1000$~\kmps  (the
approximate value derived by \citealt{sako13349:01} in their analysis of the
\xmm RGS spectrum).  For each ion, two sets of curves corresponding to the $1
\rightarrow 2$ (Ly~/~\ion{He}{$\alpha$}) and $1 \rightarrow 9$ (representing
the series limit) transitions are shown.  Over plotted as individual points on
the curves are the values for the different transitions which are detected in
our data, and which  contribute in the series fitting of the \nj values listed
in Table~\ref{tab:xseries}, for fixed values of $b=100$~\kmps.  It is clear
that for any given ion, a detection of the higher transition series better aids
in the determination of \nj. This is  well illustrated in a comparison of the
H- and He-like oxygen series which are detected  to the series limit  versus
the H- and He-like silicon series, where only the $\alpha$ ($1 \rightarrow 2$)
and $\beta$ ($1 \rightarrow 3$) transitions contribute significantly to the
fitting.  H- and He-like Ne have been excluded from the figure due to
significant contamination from Fe.  For the other ions in the figures,
contamination to the line series comes from other velocity components of the
same ions -- e.g. for He-like \ovii, components 1, 2, 3 of the warm absorbers
discussed in \S\ref{subsubsec:WaModel} all contribute to the line series at
different velocities.

To assess further the presence of multiple velocity components, we generate
velocity spectra based on the 5 strongest resonance transitions for the most
prominent X-ray ions, namely H- and He-like ions of nitrogen,
oxygen, neon, magnesium, aluminium, silicon and sulphur. (note that not all
these are necessarily detected.) For each individual absorption line, a
velocity grid from -4000 to 4000\,\kmps is generated centered around the rest
wavelength of the particular line, i.e.  zero velocity corresponds to the rest
wavelength of the line of choice. The HETGS spectra of counts vs wavelength
(initially binned by 4) is remapped (through interpolation) to convert to the
velocity space. Standard error propagation rules are invoked and the remapping
retains the Gaussian nature of the errors (i.e. $\sqrt(\rm{counts})$).  The
velocity spectra of the five strongest resonance lines are then combined
(errors are added in quadrature) to represent the velocity spectrum of the
concerned ion. Figure \ref{fig:VelProfile} shows the velocity profiles derived
in the aforementioned way.
%(The difference between our methods and that of
%\citet{ngc4051:01,kaspi3783:02} is that (1) they build the velocity spectra on
%a photon by photon basis from raw data, instead of interpolating counts
%spectra, instead of starting with spectra already binned in wavelength; (2)
%they ignore lines which might be contaminated.  
The velocity is binned at 200 $\rm{km \, s^{-1}}$ except for the bottom most
panel (i.e. for sulphur) where the binning is at 400 $\rm{km \, s^{-1}}$. At
wavelengths corresponding to the sulphur lines, which have been considered, the
velocity resolution for HEG  is $> 400 \rm{km \, s^{-1}}$. To achieve the best
results we have used MEG spectra for N, O, Ne and Mg while resorting to HEG for
Al, Si and S having higher energy transitions. The figure clearly shows
detection of absorption for O, Ne, Mg, Al and Si. The N ions show a hint of
absorption, whereas no absorption is detected for S ions. We have also looked
for absorption in carbon, argon and calcium with no significant detection. For
most of the ions the velocity profiles do not have a symmetric Gaussian
distribution indicating the presence of more than one velocity component. We
will return to a discussion of the X-ray lines in \S\ref{subsubsec:WaModel}.

%--------------------------------------
%%%%%%%%%%%%%%%%%%%%%%%%%%%%%%%%%%%%%%%%%%%%%%%%%%%%%%%%%%%%%%%%%%%%%%%%%%%%%%
\begin{figure}
\centerline{\includegraphics[scale=1.0,width=8cm, angle=0]{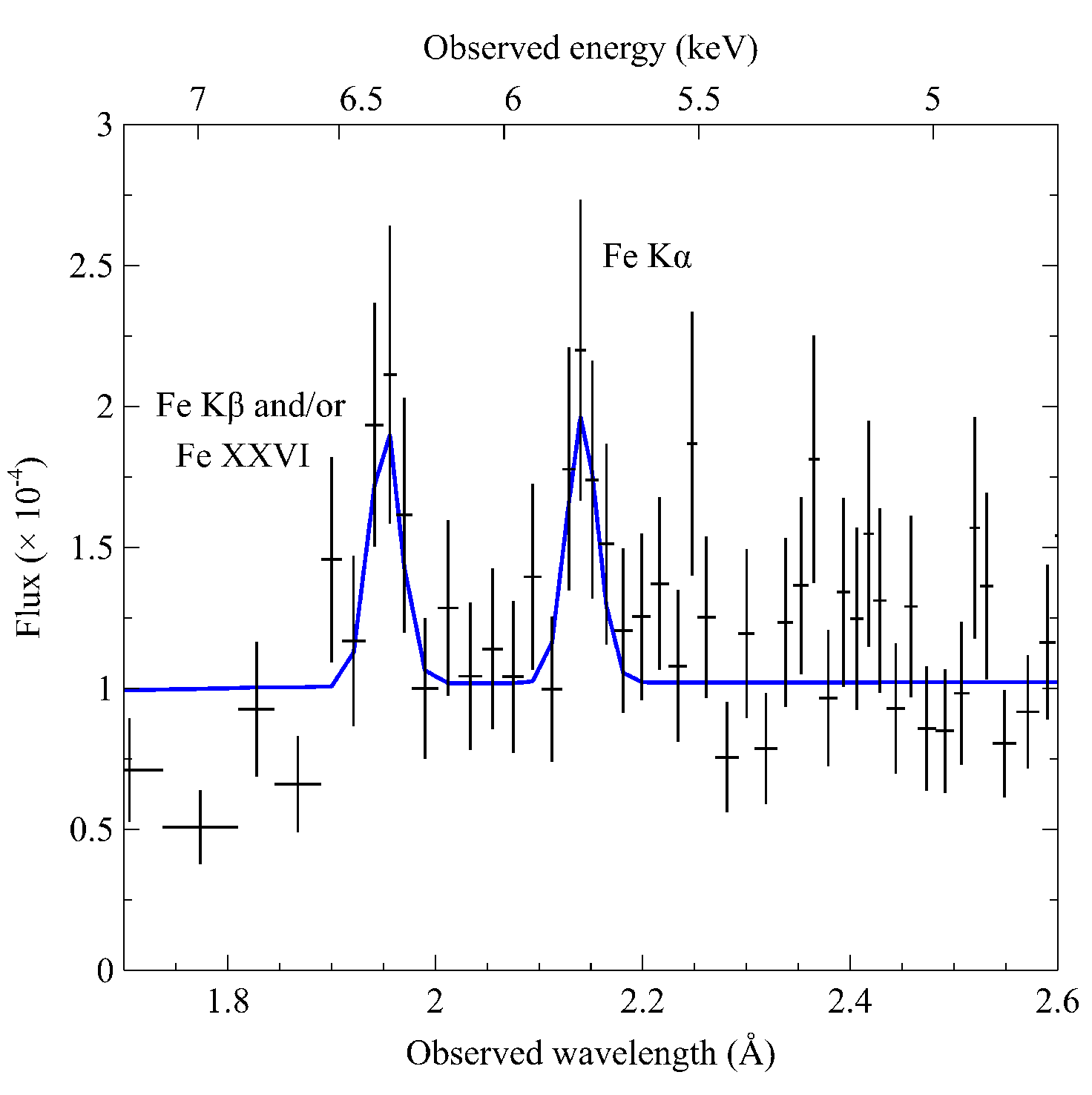}}
\caption{{The Chandra HETGS spectra at $> 6$~keV reveal two prominent emission lines,
one due to \ion{Fe}{i}~K$\alpha$ at rest energy 6.4~keV and another similarly strong
line at $\sim 7$~keV that may be identified with H-like \ion{Fe}{xxvi} or 
\ion{Fe}{i}~K$\beta$ blended with a broadened component of the \ion{Fe}{i}~K$\alpha$ line. }
\label{fig:FeKEmission}
}
\end{figure}
%%%%%%%%%%%%%%%%%%%%%%%%%%%%%%%%%%%%%%%%%%%%%%%%%%%%%%%%%%%%%%%%%%%%%%%%%%%%%%

%%%%%%%%%%%%%%%%%%%%%%%%%%%%%%%%%%%%%%%%%%%%%%%%%%%%%%%%%%%%%%%%%%%%%%%%%%%%%%
\begin{figure}
\centerline{\includegraphics[scale=0.7,width = 7 cm, angle=270]{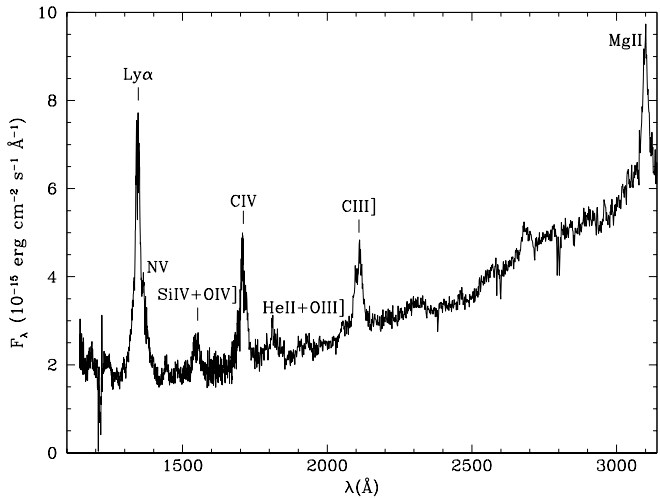}}
\caption{HST/STIS low-resolution spectrum of IRAS13349+2438 in observed wavelengths. Prominent emission lines are marked.
\label{fig:hst}
}
\end{figure}
%%%%%%%%%%%%%%%%%%%%%%%%%%%%%%%%%%%%%%%%%%%%%%%%%%%%%%%%%%%%%%%%%%%%%%%%%%%%%%

\subsubsection{Ionized and neutral emission lines}  

By far the strongest detected {\it narrow X-ray emission} line is the
\ion{Fe}{i}~K$\alpha$ fluorescent line at $\sim 6.4$~keV (rest) with
flux $F_{\rm {K}\alpha} \sim  4 \times 10^{-6}$~\phpcmsqps (Fig.~\ref{fig:FeKEmission};
also Fig.~\ref{fig:hetgxstarfit}).  
Another strong line of approximately equal strength  appears at rest
energy $\sim 7$~keV (Table~\ref{tab:xstar_mod}).  This energy is best
matched to \ion{Fe}{i}~K$\beta$\ but this gives  an unphysical
K$\alpha$:K$\beta\sim 1$, when the ratio should be $\sim 10$.
However, given the claim by \citet{longinotti-ir13349fe03} based on
XMM data for a complex iron line that has a narrow component on top of
a relativistically broadened line, the most likely explanation for the
measured similarities in the K$\alpha$ and K$\beta$ line fluxes here
is that what we have measured in the Chandra data is only a small
portion of the broadened line. Since a discussion of relativistic
effects is not the goal of this paper, we will leave it at a reporting
of the line measurements at high spectral resolution.  An alternative
possibility, especially given the line strength is that the emission is in fact due to
\ion{Fe}{xxvi}~K$\alpha$ at 6.96~keV. We have also checked the
$\sim$~6--8~keV spectral region for emission from other ions (e.g. 6.7~keV
\ion{Fe}{xxv}~K$\alpha$, 7.5~keV \ion{Ni}{i}~K$\alpha$, and 8.26~keV
\ion{Ni}{i}~K$\alpha$), but find no significant detections.

%%%%%%%%%%%%%%%%%%%%%%%%%%%%%%%%%%%%%%%%%%%%%%%%%%%%%%%%%%%%%%%%%%%%%%%%%%%%%%
\begin{figure}
\centerline{\includegraphics[scale=0.5, width = 7 cm, angle=270]{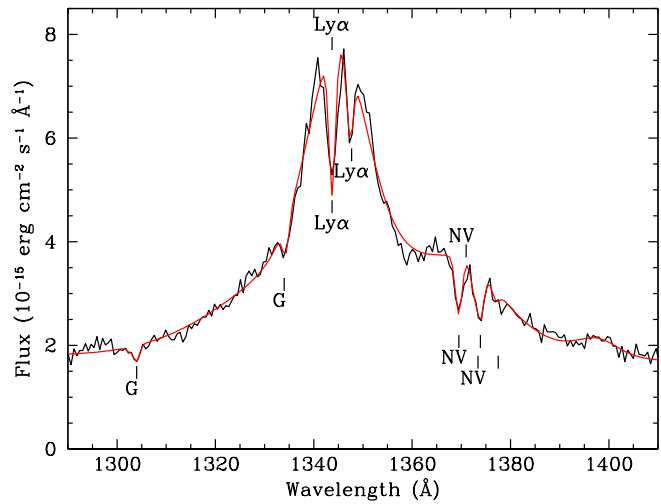}}
\centerline{\includegraphics[scale=0.5, width = 7 cm, angle=270]{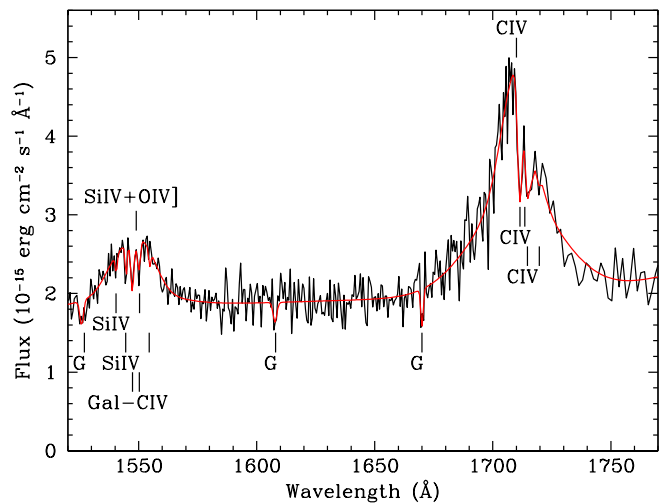}}
\caption{Detail of the STIS G140L spectrum of IRAS13349+2438, in observed wavelengths, in the \lya\
region (top panel) and the \ion{C}{iv} region (bottom panel). The black curves
are the observed data; the red curves are the best fit to the spectrum.
Emission features are labeled above the spectra, and absorption components are
marked below. The ``G'' labels denote foreground Galactic interstellar
absorption lines.
\label{fig:hstzoom}
}
\end{figure}
%%%%%%%%%%%%%%%%%%%%%%%%%%%%%%%%%%%%%%%%%%%%%%%%%%%%%%%%%%%%%%%%%%%%%%%%%%%%%%

%-------------------------------------------
\subsection{The UV HST STIS-MAMA Spectra}
\label{subsecn:HSTspectra}

To measure the UV fluxes, widths, and redshifts of the emission and
absorption lines, we used the IRAF\footnote{IRAF
(http://iraf.noao.edu/) is distributed by the National Optical
Astronomy Observatory, which is operated by the Association of
Universities for Research in Astronomy, Inc., under cooperative
agreement with the National Science Foundation.} task {\sc specfit}
\citep{specfit}.  As discussed in \S\ref{subsec-dwa}, the full
continuum is a complex mix of direct light from the active nucleus
that is heavily reddened combined with light scattered from dust, and
perhaps free electrons also, that also suffers some extinction.  We
were unable to arrive at a model that fully matched the observed shape
of the full continuum, so, to characterise the emission and absorption
lines, we used an empirical approach that simply fit a power law that
was locally optimised around each emission-line blend.  Nearly all
emission lines required both a broad ($\sim 2500 \kmps$ in width) and
a very broad ($\approxgt 7500 \kmps$) component to obtain an adequate
fit.  Table \ref{tbl-elines} gives the fluxes, velocities, and
full-width at half-maximum (FWHM) for the fitted emission
lines. Velocities are relative to the systemic redshift of
$z=0.10853$, based on the observed redshift of the \o3\ $\lambda 5007$
emission line in our HET spectrum (see \S\ref{sec:opticallines} for
details).  The tabulated line widths are corrected for instrumental
resolution by subtracting the resolution in quadrature from the
measured widths.  Wavelengths $< 1720$ \AA\ have the resolution of the
STIS G140L grating as given in the STIS Instrument Handbook, ranging
from 0.93 \AA\ at 1350 \AA\ to 0.81 \AA\ at 1700 \AA.  Longer
wavelengths have the resolution of the G230L grating, which is 3.40
\AA.

%%%%%%%%%%%%%%%%%%%%%%%%%%%%%%%%%%%%%%%%%%%%%%%%%%%%%%%%%%%%%%%%%%%%%%%%%%%%%%
\begin{figure}
\centerline{\includegraphics[scale=0.7,width = 7 cm, angle=270]{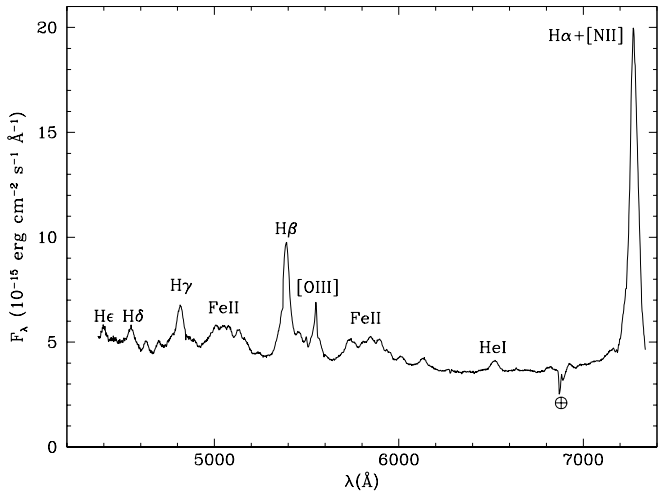}}
\caption{The HET spectrum of \ir13349, in observed wavelengths, shows a typical
quasar spectrum dominated by broad Balmer emission lines and
$[$\ion{O}{iii}$]$, with the exception of particularly strong \ion{Fe}{ii}.
The lone absorption can be attributed to atmospheric Franhofer B-band which we
have not corrected for, since it has no effect on the conclusions derived from
our spectrum.
\label{fig:hetspec}
}
\end{figure}
%%%%%%%%%%%%%%%%%%%%%%%%%%%%%%%%%%%%%%%%%%%%%%%%%%%%%%%%%%%%%%%%%%%%%%%%%%%%%%

%%%%%%%%%%%%%%%%%%%%%%%%%%%%%%%%%%%%%%%%
%Table 4
\begin{table}
\begin{center}
\caption{UV \& Optical Emission Lines in IRAS13349+2438}
\label{tbl-elines}
\begin{tabular}{lcccc}
\hline\hline
{Line} &
{$\rm \lambda_{0}${$^a$}} &
{Flux{$^b$}} &
{Velocity{$^c$}} &
{FWHM} \\
{} &
{(\AA)} &
{} &
{$\rm ( km~s^{-1} )$} &
{$\rm ( km~s^{-1} )$} \\
\hline
\multicolumn{5}{c}{Hubble Space Telescope UV Spectrum (Figure \ref{fig:hstzoom})} \\ \\
\ion{Ly}{$\alpha$} & 1215.67 & $\phantom{0}5.3 \pm 0.1$ & $-710 \pm 368$ & $3026 \pm 150$ \\
\ion{Ly}{$\alpha$} & 1215.67 &           $10.2 \pm 0.1$ & $-710 \pm 368$ & $10018 \pm 140$ \\
\ion{N}{v}        & 1240.15 & $\phantom{0}1.1 \pm 0.1$ & $-1130 \pm 138$ & $3189 \pm 210$ \\
\ion{N}{v}        & 1240.15 & $\phantom{0}2.0 \pm 0.3$ & $-1130 \pm 138$ & $10018 \pm 140$ \\
\ion{Si}{iv}+\ion{O}{iv}]  & 1400    & $\phantom{0}1.4 \pm 0.2$ & $-801 \pm 203$ & $2554 \pm 360$ \\
\ion{Si}{iv}+\ion{O}{iv}]  & 1400    & $\phantom{0}0.8 \pm 0.4$ & $-801 \pm 203$ & $7568 \pm 730$ \\
\ion{C}{iv}       & 1549.05 & $\phantom{0}2.6 \pm 0.7$ & $-1358 \pm 105$ & $2555 \pm 360$ \\
\ion{C}{iv}       & 1549.05 & $\phantom{0}5.7 \pm 0.6$ & $-1358 \pm 105$ & $7568 \pm 730$ \\
\ion{He}{ii}      & 1640.70 & $\phantom{0}0.5 \pm 0.3$ & $-1358 \pm 105$ & $2498 \pm 360$ \\
\ion{He}{ii}      & 1640.70 & $\phantom{0}1.8 \pm 0.6$ & $-1358 \pm 105$ & $7549 \pm 730$ \\
\ion{O}{iii}$]$     & 1663.48 & $\phantom{0}0.3 \pm 0.2$ & $-1358 \pm 105$ & $2500 \pm 360$ \\
\ion{O}{iii}$]$     & 1663.48 & $\phantom{0}0.1 \pm 0.6$ & $-1358 \pm 105$ & $7550 \pm 730$ \\
\ion{Al}{iii}     & 1857.40 & $\phantom{0}0.4 \pm 0.2$ & $-891 \pm 500$ & $2174 \pm 2190$ \\
\ion{Si}{iii}$]$    & 1892.03 & $\phantom{0}0.8 \pm 0.4$ & $-210 \pm 190$ & $1482 \pm 360$ \\
\ion{C}{iii}$]$     & 1908.73 & $\phantom{0}2.4 \pm 1.3$ & $-380 \pm 100$ & $2474 \pm 590$ \\
\ion{Si}{iii}$]$+\ion{C}{iii}$]$ & $\dots$ & $\phantom{0}3.2 \pm 0.6$ & $\ldots$     & $8987 \pm 7500$ \\
\ion{Mg}{ii} & 2798.74 & $\phantom{0}6.5 \pm 0.4$ & $-318 \pm 100$ & $2130 \pm 120$ \\

\hline \\

\multicolumn{5}{c}{Hobby-Eberly Spectrum (Figure \ref{fig:hetspec})} \\ \\

\ion{H}{$\epsilon$} & 3970.07 & $\phantom{0}2.0 \pm 0.8$ & $20 \pm 30$ & $1876 \pm 40$ \\
\ion{H}{$\epsilon$} & 3970.07 & $\phantom{0}2.7 \pm 1.2$ & $-140 \pm 30$ & $5085 \pm 40$ \\
\ion{H}{$\delta$}  & 4101.73 & $\phantom{0}0.5 \pm 0.5$ & $20 \pm  30$ & $1876 \pm 40$ \\
\ion{H}{$\delta$}  & 4101.73 & $\phantom{0}5.4 \pm 0.5$ & $-140 \pm  30$ & $5085 \pm 40$ \\
\ion{H}{$\gamma$}  & 4340.46 & $\phantom{0}0.1 \pm 0.5$ & $20 \pm  30$ & $1876 \pm 40$ \\
\ion{H}{$\gamma$}  & 4340.46 & $\phantom{0}8.0 \pm 0.7$ & $-140 \pm  30$ & $5085 \pm 40$ \\
\ion{Fe}{ii}   & 4434--4684  & $          34.1 \pm 0.7$ & $264 \pm  24$ & $2514 \pm 66$ \\
\ion{He}{ii}      & 4686.74 & $             <     0.8$ & $-140$ (fixed) & $1876$ (fixed) \\
\ion{H}{$\beta$}   & 4861.32 & $          13.4 \pm 0.6$ & $20 \pm  30$ & $1878 \pm 58$ \\
\ion{H}{$\beta$}   & 4861.32 & $          21.6 \pm 0.5$ & $-140 \pm  30$ & $5807 \pm 142$ \\
$[$\ion{O}{iii}$]$    & 4958.9  & $\phantom{0}0.41 \pm 0.05$ & $0 \pm  30$ & $421 \pm 45$ \\
$[$\ion{O}{iii}$]$    & 4958.9  & $\phantom{0}2.2 \pm 0.07$ & $-330 \pm  30$ & $2556 \pm 154$ \\
$[$\ion{O}{iii}$]$    & 5006.8  & $\phantom{0}1.23 \pm 0.15$ & $0 \pm  30$ & $421 \pm 30$ \\
$[$\ion{O}{iii}$]$    & 5006.8  & $\phantom{0}6.4 \pm 0.21$ & $-330 \pm  30$ & $2556 \pm 154$ \\
%\ion{Fe}{ii}   & 5090--5588  & $          43.1 \pm 0.9$ & $264 \pm  24$ & $2514 \pm 66$ \\
%\ion{He}{i}       & 5875.65 & $\phantom{0}0.5 \pm 0.5$ & $30 \pm  40$ & $1725 \pm 330$ \\
%\ion{He}{i}       & 5875.65 & $\phantom{0}0.5 \pm 0.6$ & $30 \pm  40$ & $5015 \pm 930$ \\
$[$\ion{N}{ii}$]$     & 6548.1  & $\phantom{0}0.6 \pm 0.1$ & $0 \pm  30$ & $421 \pm 30$ \\
$[$\ion{N}{ii}$]$     & 6548.1  & $\phantom{0}8.0 \pm 1.7$ & $-330 \pm  30$ & $2650 \pm 70$ \\
%\ion{H}${\alpha$}  & 6562.80 & $          52.8 \pm 3.9$ & $-50 \pm  30$ & $1725 \pm  50$ \\
%\ion{H}${\alpha$}  & 6562.80 & $          47.0 \pm 4.6$ & $-20 \pm  40$ & $5015 \pm 760$ \\
$[$\ion{N}{ii}$]$     & 6583.4  & $\phantom{0}1.8 \pm 0.3$ & $0 \pm  30$ & $421 \pm 30$ \\
$[$\ion{N}{ii}$]$     & 6583.4  & $          24.0 \pm 5.1$ & $-330 \pm  30$ & $2650 \pm 70$ \\
\hline
\end{tabular}
\end{center}
We note that the fits include the absorption lines, so there is no fitting bias that would artificially blueshift the centroids. Furthermore, the absorption lines are too weak to significantly impact fitting results even if unaccounted for. 
{$^a$}{Rest wavelengths are in vacuum for $\lambda < 3200$~\AA\, and in air for $\lambda > 3200$~\AA.} 
{$^b$}{Observed flux in units of $\rm 10^{-14}~erg~cm^{-2}~s^{-1}$.} 
{$^c$}{Velocity is relative to a systemic redshift of $z = 0.10853$, the redshift of the $[$\ion{O}{iii}$]$ $\lambda$5007 emission line.}
\end{table}

%%%%%%%%%%%%%%%%%%%%%%%%%%%%%%%%%%%%%%%%

Figure~\ref{fig:hst} shows the merged UV spectrum of  \ir13349\ from 1150--3180
\AA\ with the most prominent emission lines labeled.  At the scale of this
figure, Galactic and intrinsic absorption lines are not easily visible, aside
from Galactic \ion{Ly}{$\alpha$} absorption at 1216 \AA, which is blended with geocoronal
\ion{Ly}{$\alpha$} emission.  Otherwise, our \hst\ spectrum of \ir13349\ shows two prominent
blue-shifted absorption-line systems in \ion{Ly}{$\alpha$}, \n5, and \c4.  To measure the
equivalent width, position, and FWHM of each line, we use simple Gaussian
absorption lines in our model to fit the spectrum.  Figure~\ref{fig:hstzoom}
shows full-resolution plots of the \ion{Ly}{$\alpha$}, \n5, \si4, and \c4\ regions of the
spectrum overlaid with the best-fit model.  Table \ref{tab:UvALines}
summarises our measurements for each of the detected lines.  Line widths have
been corrected for the instrumental resolution by subtracting the resolution of
231 $\kmps$ in quadrature from the fitted value.  We also quote 2$\sigma$ upper
limits for the \ion{Si}{iv} transitions at the same velocities as the other detected
components since these are useful in constraining the ionisation state of the
absorbing gas.  The lines are slightly broader than the resolution of the
L-mode gratings, with intrinsic widths that are consistent with the Doppler
parameter of $100~\kmps$ found for the X-ray absorbers.  Assuming $b = 100
~\kmps$, we obtain the ionic column densities given in the last column of
Table~\ref{tab:UvALines}.  Since the intrinsic widths of the UV absorption
features are not broader than $b = 100 ~\kmps$, these column densities can be
considered lower limits for \ion{Ly}{$\alpha$}, \n5\ and \c4.  The highest velocity
component, at $-950 \kmps$, has roughly the same velocity as the bulk of the
X-ray absorption.  The lower-velocity component, at $-75 \kmps$, has a lower
outflow velocity than most detected X-ray features.

\subsection{The optical Hobby-Eberly 8-m spectrum}
\label{sec:opticallines}

The flux calibrated HET spectrum of \ir13349\ shown in Figure~\ref{fig:hetspec}
reveals a fairly typical quasar, where broad Balmer and \ion{Fe}{ii} emission
lines and narrow \o3\ emission lines are superposed on a blue continuum.  Given
the luminosity of \ir13349, any starburst component or contribution from the
host galaxy is completely overwhelmed by the QSO emission.

%%%%%%%%%%%%%%%%%%%%%%%%%%%%%%%%%%%%%%%%
%Table 5
\input{TABLES/uv_abs_070511.table}
%%%%%%%%%%%%%%%%%%%%%%%%%%%%%%%%%%%%%%%%

As we did for the \hst\ spectrum, we use {\sc specfit} to measure the
emission lines and continuum in the HET spectrum.  For the continuum
we use a power law in $F_\lambda$.  For the emission lines (excluding
\ion{Fe}{ii} ) two Gaussian components are needed to characterise the
line profile.  For $\rm H\beta$ and the higher-order Balmer lines, we tie
the velocities and widths of all the components together. This is
necessary and helpful in deblending these features from the ubiquitous
\ion{Fe}{ii} emission.  To fit the complex \ion{Fe}{ii} emission itself,
we use the template derived by \cite{veroncetty04} and convolve it
with a Gaussian to match with the broadened width observed in our
spectrum.  The best-fit to the continuum yields $F_\lambda =3.83
\times 10^{-14}\times (\lambda / 5653.5)^{0.69}~\ergpcmsqpspA$.  Table
\ref{tbl-elines} lists the best-fit parameters for the emission lines
in the spectrum.

In addition to these best-fit parameters, we have made some empirical
measures of the observed $\rm H\beta$ profile and the \ion{Fe}{ii} emission
for use in evaluating the mass of the central black hole in \ir13349\
(\S \ref{subsec:ModelSed}) and for interpreting Eigenvector-1
(comprised of the relative equivalent widths of \o3\ $\lambda5007$
and \ion{Fe}{ii} $\lambda 4434-4684$ relative to $\rm H\beta$) as described
by \citet{borosongreen92}.  To measure the dispersion of the $\rm H\beta$
profile directly from the data, we subtracted the fitted continuum,
the fitted \ion{Fe}{ii} emission, and all other fitted emission lines
from the original spectrum.  We then computed the empirical full-width
at half maximum (FWHM) and the dispersion of this net spectrum over
the 5266--5500 \AA\, (4750--4962 \AA, rest) wavelength range of the $\rm H\beta$
profile, to obtain FWHM$=2468 \pm 4.7~\kmps$, velocity dispersion
$\sigma_{\rm H\beta}=1948 \pm 3.7~\kmps$, and
equivalent width $W_{\rm H\beta}=88.1 \pm 4.5$~\AA.  
(Note that the aforementioned $\rm H\beta$
values are derived using a more complex decomposition of the two
components, i.e. a narrow core which dominates the empirical FWHM, and
a broader base needed to describe the  $\rm H\beta$ line profile (see
Table~\ref{tbl-elines} and Figure~\ref{fig:hetspec}).  Similarly, we
subtracted the fitted continuum and all other emission lines from the
original spectrum to obtain the observed \ion{Fe}{ii} spectrum.
Integrated over the 4434--4684 \AA\ rest-wavelength range as defined
by \citet{borosongreen92}, we obtain an \ion{Fe}{ii} flux of
$3.41\times10^{-13}~\rm erg~cm^{-2}~s^{-1}$, and $W_{\textrm{\ion{Fe}{ii}}} = 82.4$
\AA.  For \o3\  $\lambda5007$ we obtain $W_{\textrm{[\ion{O}{iii}]}} = 14.0$\AA.  In
the context of Eigenvector-1 and Eigenvector-2 characteristics as
initially discussed by Boroson \& Green (1992\nocite{borosongreen92};
see also \citealt{boroson02}), \ir13349\ behaves as a quasar at one
extreme of the Eigenvector-1 correlations, e.g., the ratios
\ion{O}{iii}:$\rm H\beta = 0.22$, \ion{He}{ii}:$\rm H\beta < 0.022$, and
\ion{Fe}{ii}:$\rm H\beta = 0.97$.  Using the \o3\ lines in
\ir13349, we also update the redshift to $z=0.10853 \pm 0.0001$. 
We note that the very high S/N HET spectrum enables
the determination of the centroids of discrete spectral features to much better than
a resolution element. For the sharp, bright [O III] 5007~line, we can determine the
centroid to an accuracy of 1.5~pixels corresponding to 30~\kmps.

The strong \fe2\ and weak \o3\ emission that characterise
extreme Eigenvector-1 objects are common to Narrow-Line Seyfert~1
(NLS1) galaxies (e.g., see \citealt{pogge11} and \citealt{boroson11}),
and \citet{vcv12} classify \ir13349 as a NLS1 based on its line width.
However, our HET spectrum shows that \ir13349 does not satisfy two out
of the three criteria primarily used to classify objects as NLS1s. The
width of $\rm H\beta$ is too broad and exceeds the 2000~$\kmps$
boundary, and the width of $\rm H\beta$ is also much broader than
that of \o3\, by nearly a factor of three; NLS1s should have
permitted and broad lines of similar width.  Therefore, despite its
extreme Eigenvector-1 characteristics, we do not consider \ir13349 to
be a NLS1, in agreement with the determination of \citet{grupe04,
grupe10}.

%%%%% DCH
The HET and {\it HST} spectra are very similar to
the prior observations of IRAS13349.~--~The continuum flux density is
nearly identical to that observed by \cite{wills92}, and $\sim15$\%
brighter than that observed by \cite{hines01}. Our emission-line
fluxes are also comparable (but note that the units for the last
column of Table 3 in Hines et al. 2001 should be $10^{-14}~\rm
erg~cm^{-2}~s^{-1}$), and agree to within 15\% for \ion{H}{$\alpha$}. However,
there are differences of 50--100\% for lines such as \ion{H}{$\beta$} and $[$\ion{O}{iii}$]$
, which are badly blended with the optical \ion{Fe}{ii}
multiplets. Although we have taken care to model explicitly the full
\ion{Fe}{ii} emission spectrum, as have Hines et al. (2001), the fits can be
highly model dependent, especially when one includes multiple
components as we have for some lines (such as the Balmer lines,
$[$\ion{O}{iii}$]$, and $[$\ion{N}{ii}$]$). We suspect that the source
of our differences in line fluxes are due to the methods used in deblending these features.

%-------------------------------------------
\subsection{The infrared Spitzer IRS spectra}
\label{subsec:iranalysis}

\begin{figure}
\centerline{\includegraphics[scale=0.5,width = 10 cm, angle=0]{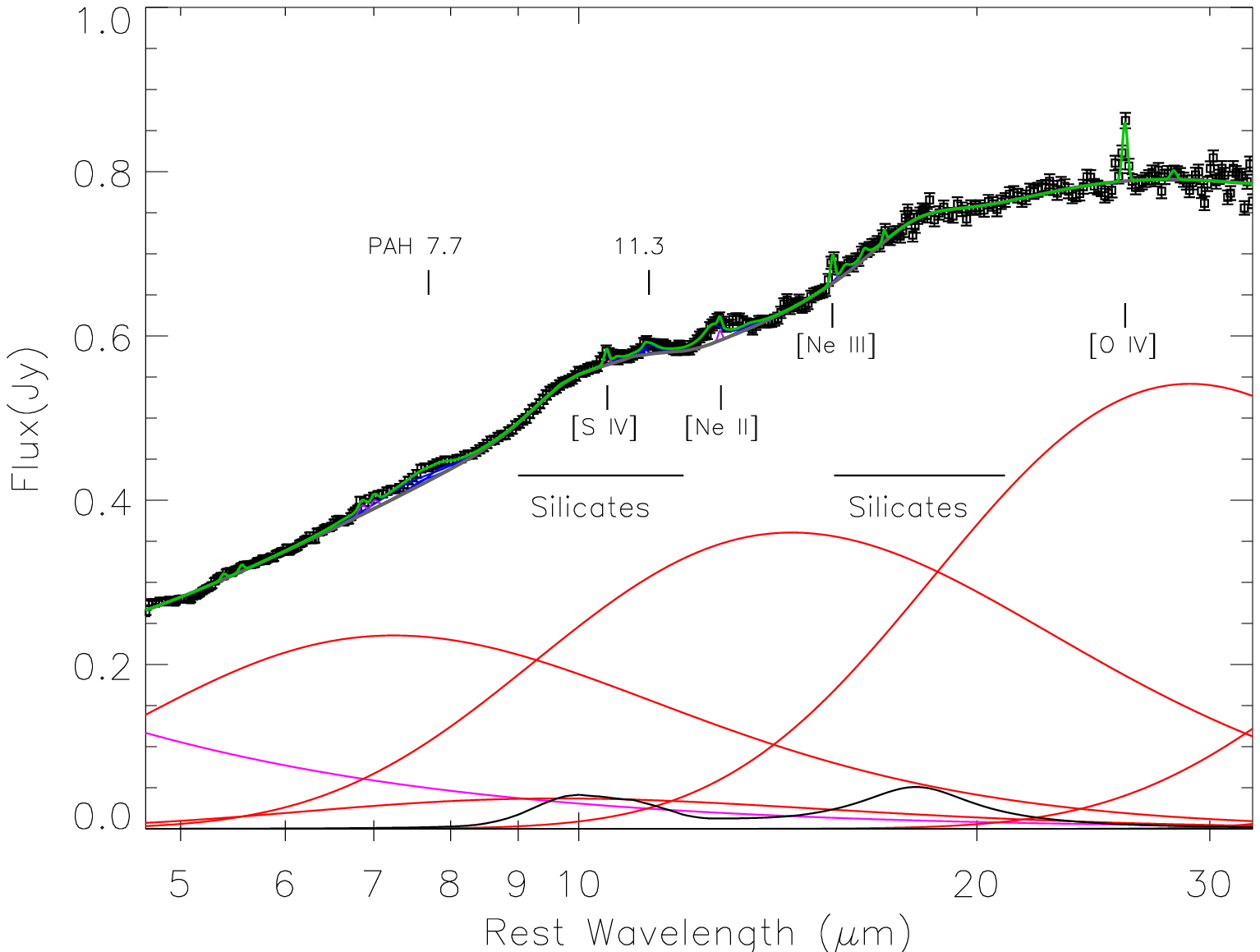}}
\caption{The best fit model (green) to the low resolution Spitzer IRS spectrum ($\times
1.077$) showing  all contributions (black
body continua, silicate, PAH, starlight, and narrow-emission lines).  
To contrast, grey curve shows contribution from  black bodies,
starlight, and silicate; additional contribution from PAH emission (blue),
and narrow-emission lines [\ion{S}{iv}], [\ion{Ne}{ii}], [\ion{Ne}{iii}] and 
[\ion{O}{iv}] (purple) are also shown.
Model components contributing to the aforementioned include 
silicate emission (bottom, black)
superimposed on a black body continuum (red) and hot dust or starlight (magenta; note that
it is not possible to distinguish the two without observations at
near-IR wavelengths).  The best
fit silicate emission temperature is $\sim 160$~K while the continuum
components have temperature range $50-400$~K. (The T=25K continuum component has
negligible contribution to the fit and hence not shown as part of the 
model here.)
\label{fig:IRS-lowres}
}
\end{figure}

As shown in Figure~\ref{fig:IRS-lowres}, the mid-IR continuum rises fairly
smoothly to a peak of 0.8 Jy at 30 $\mu$m.  Superimposed are weak, broad
spectral features from silicate emission at  10 $\mu$m and 18 $\mu$m. Weak
polycyclic aromatic hydrocarbon (PAH) resonances are also detected at 7.7
$\mu$m and 11.3 $\mu$m, indicating a weak starburst contribution to the mid-IR
emission \citep{schweitzer06, shi07}.

%%%%%%%%%%%%%%%%%%%%%%%%%%%%%%%%%%%%%%%%%%%%%%%%%%%%%%%%%%%%%%%%%%%%%%%%%%%%%%

%%%%%%%%%%%%%%%%%%%%%%%%%%%%%%%%%%%%%%%%%%%%%%%%%%%%%%%%%%%%%%%%%%%%%%%%%%%%%%

%%%%%%%%%%%%%%%%%%%%%%%%%%%%%%%%%%%%%%%%
%Table 6
\begin{table*}
\begin{center}
\caption{Mid-IR emission lines in IRAS13349+2438}
\label{mirlines}
\begin{tabular}{lccccc}
\hline\hline
{Feature}$^a$ & {$\lambda_{\rm rest}$$^b$} & {$\lambda_{\rm vac}$}$^c$ &
{$\mathring{W}$}$^d$ &
{\textit{FWHM}}$^e$ &
{Flux}$^f$ \\
 &
{\tiny (\mic)}    &
{\tiny (\mic)}   &
{\tiny ($\times 10^{-3}$~\mic)} &
{\tiny ($\times 10^{-2}$~\mic)} &
{\tiny ($\times 10^{-21}$~W~cm$^{-2}$)} \\
\hline
PAH(+\ion{H}{i}?)&7.678$\pm$0.007&7.700&4.91$\pm$0.99&15.27$\pm$0.10&11.87$\pm$2.40\\
$[$\ion{S}{iv}]&10.506$\pm$0.001&10.511&2.35$\pm$0.14&4.61$\pm$0.73&4.98$\pm$0.23\\
PAH(+\ion{H}{i}+\ion{He}{ii}+Mg$_2$SiO$_4$?)&11.243$\pm$0.001&11.250&5.14$\pm$0.95&16.14$\pm$0.09&7.97$\pm$1.41\\
PAH(+\ion{H}{i}?)&12.686$\pm$0.003&12.700&0.99$\pm$0.31&2.47$\pm$0.26&1.17$\pm$0.37\\
$[$\ion{Ne}{ii}$]$&12.819$\pm$0.001&12.813&3.15$\pm$0.40&3.10$\pm$0.14&3.51$\pm$0.48\\
PAH(+\ion{H}{i}?)&14.241$\pm$0.006&14.250&3.24$\pm$1.75&4.62$\pm$0.82&1.04$\pm$0.74\\
$[$\ion{Ne}{v}$]$&14.323$\pm$0.003&14.322&2.96$\pm$0.53&7.29$\pm$0.39&3.30$\pm$0.52\\
$[$\ion{Ne}{iii}$]$&15.560$\pm$0.002&15.555&3.39$\pm$0.61&1.96$\pm$0.13&2.48$\pm$0.54\\
$[$\ion{S}{iii}$]$&18.726$\pm$0.001&18.713&5.08$\pm$0.46&4.94$\pm$0.14&3.73$\pm$0.32\\
$[$\ion{Ne}{v}$]$&24.317$\pm$0.009&24.318&4.65$\pm$1.86&8.30$\pm$1.02&2.23$\pm$0.80\\
$[$\ion{O}{iv}$]$&25.906$\pm$0.002&25.890&28.35$\pm$3.10&5.92$\pm$0.23&9.43$\pm$1.16\\
\hline 
\end{tabular}
\end{center}
{$^a$}{Feature's name} \\
{$^b$}{Measured wavelengths (Rest frame)} \\
{$^c$}{Laboratory wavelengths (Vacuum)} \\
{$^d$}{Equivalent widths} \\
{$^e$}{Full-width at half-maximum} \\
{$^f$}{Flux}
\end{table*}

%%%%%%%%%%%%%%%%%%%%%%%%%%%%%%%%%%%%%%%%

\vspace{-0.2in}
\subsubsection{The Low Resolution Spectrum}
\label{sssec:LRS}

We fit the low-resolution spectrum using the IDL $\chi^2$-fitting routines
PAHFIT \citep{smith07} and MPFIT
\citep{2009Markwardt}\footnote{http://cow.physics.wisc.edu/~craigm/idl/idl.html},
modified to include a silicate emission component. The continuum model consists
of 6 fixed-temperature blackbody dust emission components
($T=25,50,100,200,300,400$ K) and a blackbody stellar component (5000 K). The
silicate emission is modeled as the product of a Galactic silicate opacity
curve \citep{smith07} and a single-temperature blackbody. This  assumes that
the silicate emission features come primarily from an optically thin region.
The flux ratio of the 10 $\mu$m--to--18 $\mu$m silicate features  depends
strongly on temperature, as does the shape of the 10 $\mu$m feature. The
best-fit temperature for the silicate emission is $\sim 200$~K.  Adding
additional temperature components did not improve the fit.

%%%%% DCH made significant changes here
%The silicate emission temperature is considerably lower than the sublimation
%temperature ($\sim 1400$ K). This is not surprising since the silicate emission
%features are emitted preferentially over a temperature range of $\sim 100-200$
%K.  The presence of dust at temperatures up to (and possibly exceeding) 400 K
%is indicated by the blackbody components of our continuum model. It is likely
%that even hotter dust temperatures are present and may be observed in the
%near-IR.  The peak of the SED (in $\nu F_\nu$) occurs at  $\sim 20 \mu$m,
%corresponding to a temperature of $\sim 200$ K. However, with a clumpy dust
%distribution, the bulk of this dust emission may still come from a region close
%to the sublimation radius.

The silicate emission temperature is considerably
lower than the sublimation temperature ($\sim 1400$ K). This is not
surprising since the silicate emission features are emitted
preferentially over a temperature range of $\sim 100-200$ K.  The
presence of dust at temperatures up to (and possibly exceeding) 400 K
is indicated by the blackbody components of our continuum model. Keck
optical interferometry at $K$-band suggests near-IR emission from
an inner radius of $0.92 \pm 0.06$~pc \citep{kishimoto:09}, reasonably
consistent with dust at temperatures near sublimation. The peak of the
SED (in $\nu F_\nu$) occurs at  $\sim 20 \mu$m, corresponding to a
temperature of $\sim 200$ K. However, with a clumpy dust distribution,
the bulk of this dust emission may still come from a region close to
the sublimation radius.  

%The dust sublimation radius is estimated to be
%$R_\mathrm{sub}=0.2(L_\mathrm{bol}/10^{46})^{1/2}$ pc \citep{laor98, laor03}.
%The mid-IR luminosity of IRAS 13349 is $\nu L_\nu (25 \mu \mathrm{m}) = 3.2
%\times 10^{45} = L_\mathrm{bol} f_\mathrm{c} f_\mathrm{i}$ erg s$^{-1}$, where
%$f_\mathrm{c} \sim 0.6$ is  a representative dust covering fraction and
%$f_\mathrm{i} \sim 0.5$ is a representative mid-IR anisotropy for quasars
%\citep{ogle06, shi05}.  Therefore, the dust sublimation radius is at
%$R_\mathrm{sub} \sim 0.2$ pc.  It is likely that the dust responsible for the
%mid-IR continuum is the same dust that absorbs the direct optical-UV spectrum,
%since it accounts for a large fraction of the absorbed and reprocessed
%bolometric luminosity of the quasar.

%However, quasars tend to have unusual dust emission spectra compared to
%Galactic ISM. The 10$\mu$m peak is shifted to the red, indicating either a
%depletion of small dust grains or more crystalline silicate dust (e.g.
%enstatite) rather than amorphous olivine.

%Given that the silicate features are in emission, just like in other normal
%quasars and Seyfert 1s, it is quite clear that we are not looking through a lot
%of dust, but rather are at a high-enough elevation angle above the torus to
%clear the densest gas, but perhaps just grazing the surface and looking through
%its atmosphere, which is what is causing the UV and X-ray absorption lines and
%the extinction of the UV continuum.

\subsubsection{The High Resolution Spectrum} \label{subsec:spitzerhires}

%%%%%%%%%%%%%%%%%%%%%%%%%%%%%%%%%%%%%%%%%%%%%%%%%%%%%%%%%%%%%%%%%%%%%%%%%%%%%%
\begin{figure}
\begin{tabular}{r}
\includegraphics[scale=0.5, width = 8.5 cm, trim = 10 0 0 0, clip]{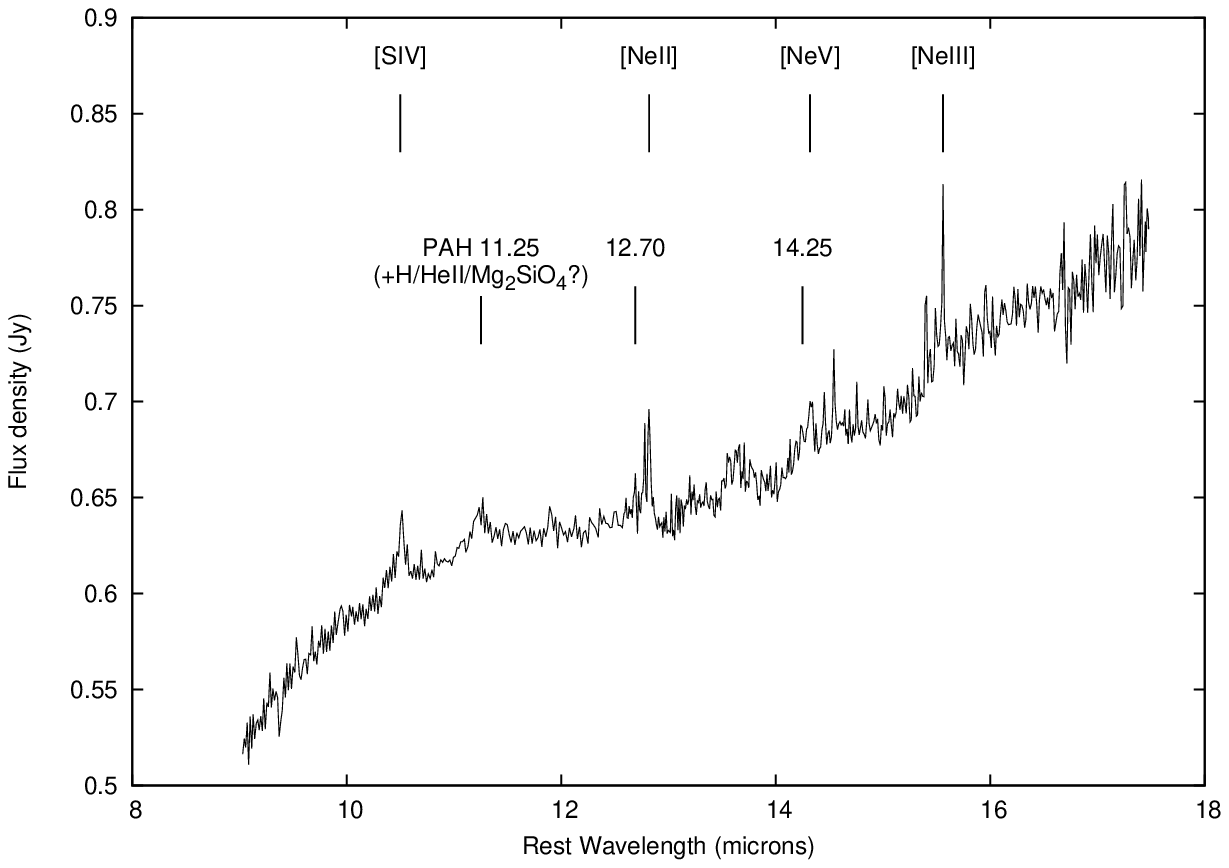}\\
\includegraphics[scale=0.5, width = 8.5 cm, trim = 10 0 0 0, clip]{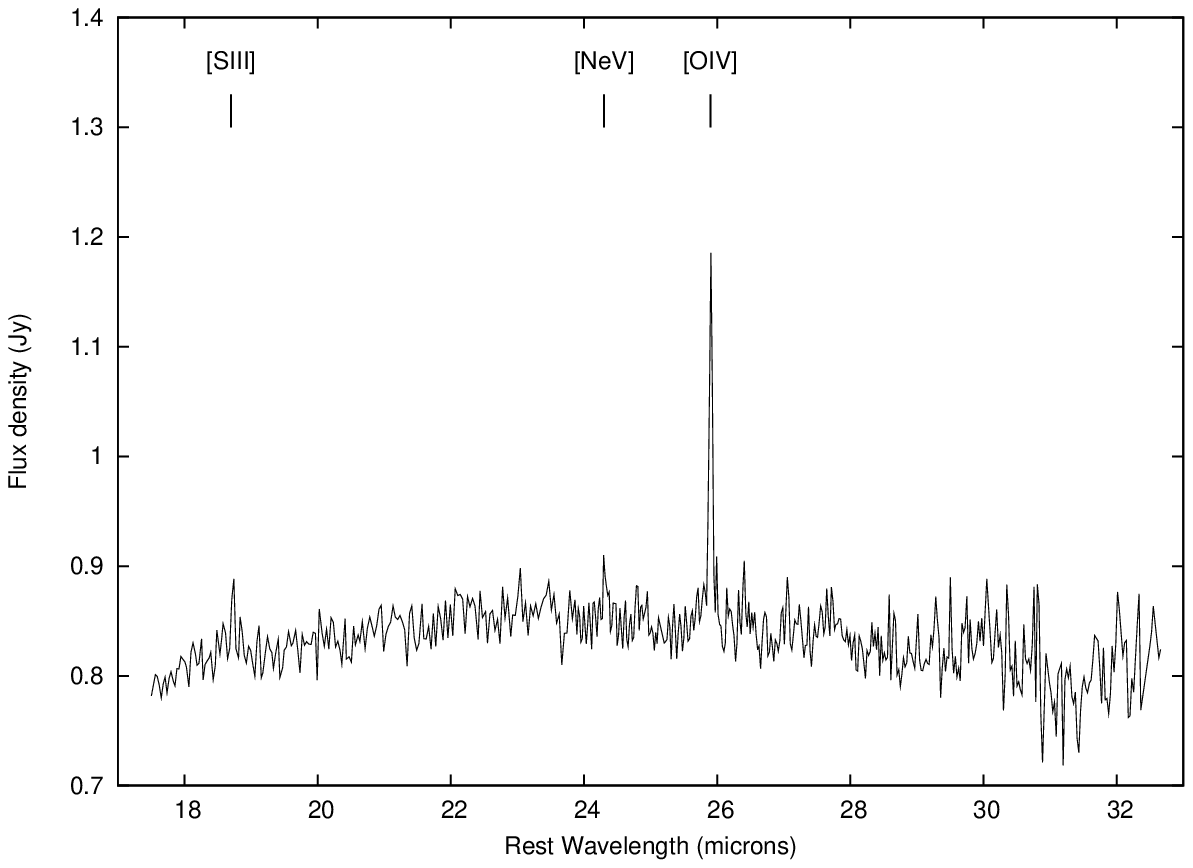}
\end{tabular}
\caption{{High resolution Spitzer IRS LH (bottom panel) and SH (top panel)
spectra of \ir13349. The spectrum shows dust features attributed to 10$\mu$m
(broad continuum bump from $9-13 \mu$m) and $18 \mu$m emission from dust in
silicates.  The $18 \mu$m is so broad that it is difficult to distinguish from
a blackbody peaking at 200~K.}
\label{fig:IRS-hires}
}
\end{figure}
%%%%%%%%%%%%%%%%%%%%%%%%%%%%%%%%%%%%%%%%%%%%%%%%%%%%%%%%%%%%%%%%%%%%%%%%%%%%%%

Dust signatures include a strong mid-IR excess and emission features
pointing to PAHs.  However, it should be noted that while we have
identified the 11.2\mic\, feature with PAH, additional contributions
from H and/or Mg-rich olivines (e.g. forsterite, Mg$_2$SiO$_4$; see
\citealt{kempler:QSOwinddust07}) are also a possibility.  Detailed
modeling to determine the relative contributions of  PAH to Mg-rich
olivines (if they exist in \ir13349) is beyond the scope of this
paper, however.

In addition to dust features, low-equivalent width, narrow forbidden emission
lines from a range of ionisation states of O, Ne, and S are detected (see
Table~\ref{mirlines} and Figure~\ref{fig:IRS-hires}).  In particular, the
$[${Ne}{V}$]$ and $[${O}{IV}$]$ lines originate from gas that is highly
photoionized by the UV continuum from the AGN. Moreover, the
$[$\ion{Ne}{iii}$]$, $[$\ion{Ne}{ii}$]$, and $[$\ion{S}{iii}$]$ lines are
likely dominated by AGN emission, although there may be contribution from
star-forming regions.

\cite{2010Tommasin} presented a comprehensive mid-IR high-resolution
spectroscopic survey of 91 Seyfert galaxies and derived useful
observational and semi-analytical diagnostics that we also use here 
to assess the degree to which \ir13349 is AGN- or starburst-dominated.
In particular, a good tracer of AGN contribution can be found in a
comparison of the $[$\ion{Ne}{v}$]$14.32\mic~:~$[$\ion{Ne}{ii}$]$12.81\mic\
ratio against $[$\ion{O}{iv}$]$24.32\mic~:~$[$\ion{Ne}{ii}$]$12.81\mic.~--~For
\ir13349, these values are respectively $0.800 \pm 0.223$ and $2.958
\pm 0.322$, pointing  to an 80--90\% AGN contribution to the \ir13349
IR emission, according to Fig.~8a of Tommasin et al.

%-------------------------------------------------------------------------------------

%-------------------------------------------------------------------------------------
%\section{The IRAS~13349+2438 Spectral Energy Distribution: Observations and Theory}
\section{The IRAS~13349+2438 Spectral Energy Distribution and its influences on the 
warm absorber: Observations and Theory}
\label{sec:SED:obsth}

In having considered the line detections separately in the different wavebands,
here we assess the \ir13349 warm absorber based on photoionisation modeling as
tied to this QSO's SED.  In particular, we investigate the impact of different
SED components (including the X-ray Compton power-law component, soft excess,
and big blue bump/accretion disk) spanning UV to X-rays on affecting warm
absorber thermodynamic stability conclusions.  We also investigate the methods
for assessing the \ir13349 black hole mass and accretion rate based on the
shape of the SED.

\subsection{The Observed Spectral Energy Distribution of IRAS 13349+2438}
\label{subsec:SedObs}

\begin{figure*}
\centerline{\includegraphics[width=0.7\textwidth,height=0.75\textheight,angle=270]{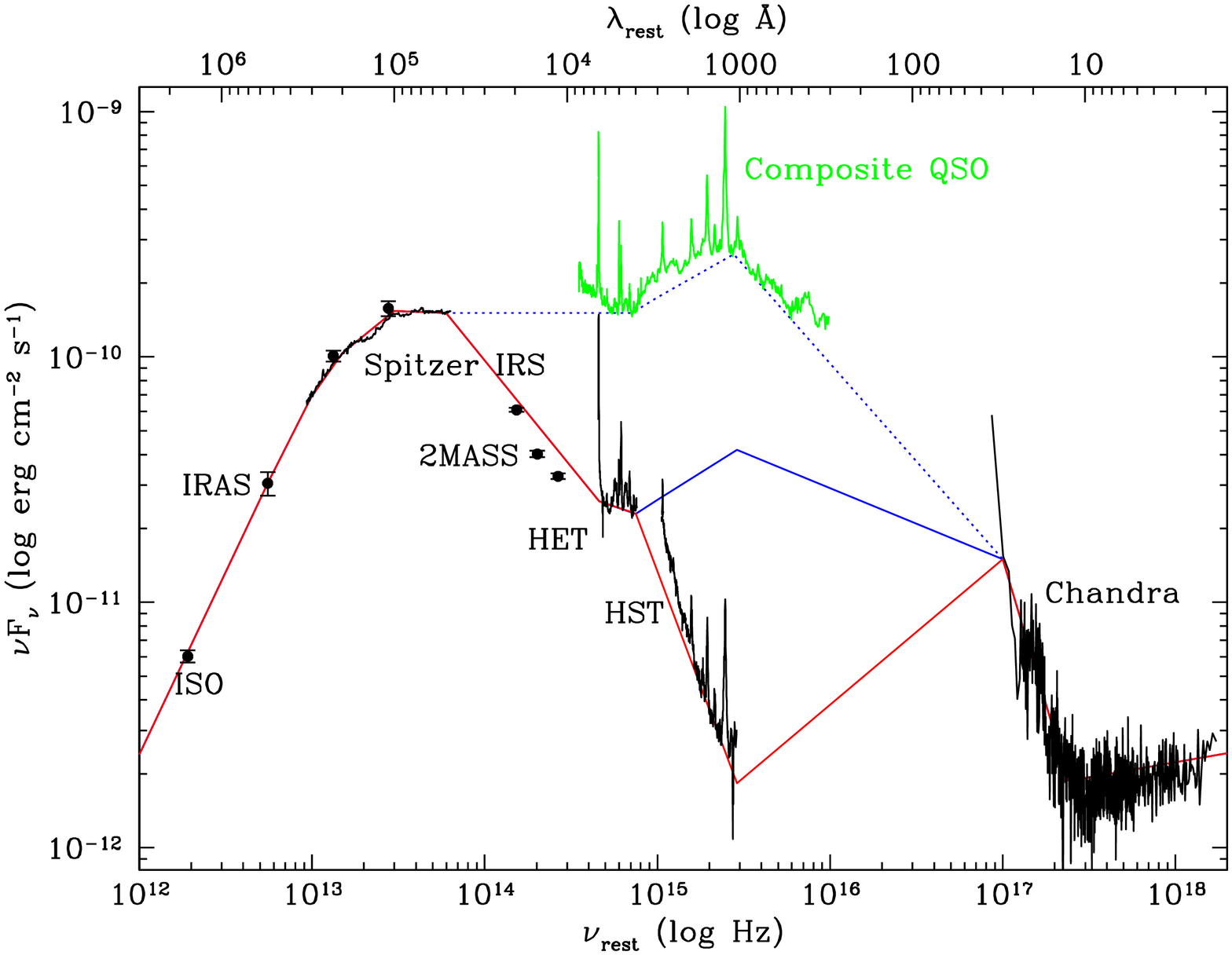}}
\caption{Observed and intrinsic Spectral Energy Distribution (SED) of \ir13349
based on our observations. The x axis (both $\nu$ and $\lambda$) is in the rest
frame. Solid points with error bars are archived data from  (1) ISO 170$\mu$m
data from \cite{12microAGNsample02}, as labeled.  (2) IRAS 12, 25, 60 $\mu$m
measurements of the IRAS Faint Source Catalog, Version
2.0~$-$\citealt{irasfaintsrccat}, and (3) 2MASS Large Galaxy Atlas of
\citealt{2MASS_lggal:03}.  Spectra are also shown (in solid black) of recent
observations with the Spitzer IRS (IR), the HET (optical), the HST STIS (UV),
and the Chandra HETGS (X-ray).  The ``Composite QSO" spectrum (solid green
line) has been renormalised upward by 0.85~dex so that its EUV peak lies
0.25~dex above the IR peak of the observed \ir13349\ spectrum, as is found in
the mean RQ QSO spectrum of \citet{elvis94QSOSED}.  {\bf Straight line
segments} between $10^{15}$~Hz and $10^{17}$~Hz show a range of simple
assumptions for what the \ir13349 SED can be in the EUV energy range where
there is no data.  At one extreme, we take our observed SED without any
corrections and decompose it into a set of power-law segments connecting the
observed HST spectrum to the lowest ($\sim$0.3~keV, rest) energy Chandra point
(solid red).  The solid blue line represents a ``corrected" UV spectrum where
the ``correction" is derived by assuming that the intrinsic spectrum is
represented by the Composite QSO spectrum normalised to the observed HET
spectrum of \ir13349 assuming a break at the peak of the Composite spectrum
(1000 \AA\ rest) that we again extrapolate down to $\sim$0.3~keV in the X-ray.
The dotted blue line follows the continuum established by the Composite
spectrum {\it renormalised} by the relative UV and IR peaks of the
\citet{elvis94QSOSED} mean RQ QSO SED, which we again extrapolate through the
EUV using a power law from the peak of the Composite spectrum at 1000~\AA\ to
the \chandra\ 0.3 keV soft X-ray point.
\label{fig:sed0}
}
\end{figure*}

To assess photoionization scenarios for the absorbing gas in \ir13349,
we first need to know the intrinsic spectral energy distribution (SED)
of the active nucleus that is illuminating the surrounding gas and dust.  To
start, we assemble an IR to X-ray SED based upon our observations and archival
data (see Figure~\ref{fig:sed0}).  In the IR, we start with 170~\micron  far IR
ISOPHOT measurements \citep{12microAGNsample02}, while mid-IR (12, 25 and
60~$\mu \rm m$) IRAS data are taken from the Faint Source Catalog of
\cite{irasfaintsrccat}, and near-IR photometric points are the 2MASS
1--2~\micron data from the Large Galaxy Atlas of \cite{2MASS_lggal:03}.  For
comparison, our Spitzer IRS spectrum is overlaid on these data after scaling to
match the 12~\micron and 25~\micron IRAS photometry (as noted in
\S\ref{subsec:spitzer}).  In the optical and UV bands, we correct our observed
HET and {\it HST} spectra for foreground Galactic extinction of $E(B-V) =
0.013$, which we obtain from NED, assuming a mean Galactic extinction curve
\citep{ccm89} with a ratio of total to selective extinction of $R_V = 3.1$.
Our \chandra\ spectrum in the X-ray is corrected for foreground Galactic
absorption as described in \S2.

For comparison, we generate a ``generic" composite optical-UV QSO spectrum
based on the Sloan Digital Sky Survey (SDSS) composite quasar spectrum of
\citet{sdssqso} (for $\lambda > 3200$~\AA\, rest) {\it and} the radio-quiet HST
composite quasar spectrum of \citet{telfer02} at shorter wavelengths.  
%We initially normalise this composite spectrum ({\color{blue} resulting in
%solid blue curve in Figure~\ref{fig:sed0}}) to match our extinction-corrected
%HET spectrum at 5700 \AA\ rest, where the emission is dominated by the
%continuum.  
If we normalise this composite spectrum to match our extinction-corrected HET
spectrum at 5700~\AA\ rest (resulting in the solid blue curve of
Figure~\ref{fig:sed0}; hereafter ``5700\AA\,-normalised-generic-composite")
what immediately stands out is the large deficit of UV and far UV flux in
\ir13349, which is likely reradiated by heated dust in the IR since the
observed SED peaks at 30~$\mu$m, rest. In the optical band, the HET spectrum of
\ir13349 is virtually identical to that of the SDSS composite portion of the
5700\AA\,-normalised-generic-composite, indicating that
%if we normalize the composite QSO spectrum to match our extinction-corrected
%HET spectrum at 5700 \AA\ rest (Fig.~\ref{fig:sed0}, solid blue; hereafter
%referred to as ``normalised composite"), the two are virtually identical,
%indicating that the HET spectrum of \ir13349\ is virtually identical to that
%of the normalised SDSS composite, indicating {\color{green}{that
the \ir13349 optical spectrum is only modestly reddened or extincted at these
wavelengths. (The analysis of \citealt{hopkins04} shows that the full SDSS
composite has an internal SMC-like extinction of $E(B-V) \sim 0.013$ at most.)
The Balmer decrement measured in our HET data is  $2.85 \pm 0.18$ for the ratio
of H$\alpha$--to--H{$\beta$}.  Correcting this for foreground extinction, it
becomes $2.817 \pm 0.18$.  To correct this to a nominal Case~B value of 2.76
implies $E(B-V) = 0.021$ for an SMC-like extinction law.  Comparing the
Galactic-extinction-corrected spectrum of \ir13349\ to the SDSS composite in
more detail, the ratio (in the continuum) is flat from 4000--6000~\AA; if
anything, \ir13349\ appears a bit bluer. So, internal extinction on the order
of 0.01 to 0.02 (with an SMC-like law) is consistent with $E(B-V) = 0.021$ as
derived from the Balmer decrement.~--~For this value, $A_{\rm V} = {\rm
E(B-V)}R_{\rm V} = 0.09 \pm 0.02$  taking $R_V = 4.05 \pm 0.8$ ala
\citet{2000Calzetti}.

%%%%%%%%%%%%%%%%%%%%%%%%%%%%%%%%%%%%%%%%%%%%%%%%%%%%%%%%%%%%%%%%%%%%%%%%%%%%%%
%%%%%%%%%%%%%%%%%%%%%%%%%%%%%%%%%%%%%%%%%%%%%%%%%%%%%%%%%%%%%%%%%%%%%%%%%%%%%%

Based on the 5700\AA\,-generic-normalised-composite spectrum 
%(solid blue curve in Figure~\ref{fig:sed0}), we find that the UV/optical
we find that the UV/optical peak (i.e. the peak of the solid blue SED) is 0.6
dex below the peak radiation in IR (at $\sim 10^5$~\AA).  According to the
median radio-quiet QSO SED in Figure 12 of \citet{elvis94QSOSED}, the UV peak
for a ``normal QSO'' should be 0.25 dex above the IR peak. Assuming \ir13349 to
be in this category (IR and optical spectra suggest this to be the case for
\ir13349) its UV/optical peak is then 0.85 dex below that of an unabsorbed
``normal QSO". 
%(We further note that if 68\% of the entire \citet{elvis94QSOSED} sample is
%considered, there is $\sim 0.4$ dex dispersion around $\Delta$ of the mean
%SED, where $\Delta$ is the difference in $log(\nu F_{\nu})$, of the UV/optical
%peak from the IR peak.) 
This suggests that either we are viewing the optical spectrum of \ir13349\
through a gray screen, or largely in scattered light, but from a scattering
region that again is largely colour neutral at wavelengths longward of 4000
\AA. At shorter wavelengths, the observed UV spectrum is likely some complex
mix of scattered and absorbed light from the central regions. The complex
geometry of this radiative transfer is difficult to unravel even using the
added benefit of polarisation information \citep[W92;][]{hines01}, so to
produce an SED that is fully corrected for these effects, we will assume that
scaling the composite QSO spectrum to the level implied by the median QSO SED
in \citet{elvis94QSOSED} is a good representation of the intrinsic SED of the
active nucleus of \ir13349.
%We show this corrected, intrinsic SED in Figure~\ref{fig:sed0}. Notice that
%the corrected SED 
The resultant SED (see Fig.~\ref{fig:sed0}, solid green and associated dotted
blue line connecting it to the IR and X-rays) looks like a qualitatively good
description for the intrinsic SED of \ir13349\ since the extreme ultraviolet
portion of the composite spectrum is declining smoothly to higher frequency in
a way that would provide a good match to the soft X-ray portion of the spectrum
observed with \chandra. Assuming this to be the intrinsic \ir13349 spectrum, at
its corrected UV luminosity of $L_{2500} = 8.1 \times 10^{29}~\rm
ergs~s^{-1}~Hz^{-1}$, we derived $\alpha_{ox} = -1.83$, a value that is $-0.25$
lower than the nominal $\alpha_{ox} = -1.58$ value expected from the
\citet{young10} relation.  This slightly lower $\alpha_{ox}$ we observe for
\ir13349 is consistent with the weak \c4\ absorption that we see in the HST
spectrum \citep[c.f.][]{gallagher06}.

%%%%%%%%%%%%%%%%%%%%%%%%%%%%%%%%%%%%%%%% 
% Table 7
\begin{table*}
\begin{center}
\caption{Luminostiy of \iras13349 in different energy bands}
\label{tab:sedlum}
\begin{tabular}{ccccccc}
\hline\hline
{SED}      &
$^{e}$ {$\rm L_{bol}$}      &
$^{f}$ {$\rm L_{IR}$} &
$^{g}$ {$\rm L_{Opt}$} &
$^{h}$ {$\rm L_{UV}$} &
$^{i}$ {$\rm L_{EUV}$} &
$^{j}$ {$\rm L_{X}$} 
\\
& 
{($\rm \times 10^{46} erg \, s^{-1}$)} & 
{($\rm \times 10^{46} erg \, s^{-1}$)} & 
{($\rm \times 10^{45} erg \, s^{-1}$)} & 
{($\rm \times 10^{45} erg \, s^{-1}$)} & 
{($\rm \times 10^{45} erg \, s^{-1}$)} & 
{($\rm \times 10^{44} erg \, s^{-1}$)} \\
\hline
$^{a}$ 1 & 1.42 & 1.19 & 0.72 & 0.36 & 0.68 & 5.55  \\
$^{b}$ 2 & 1.74 & 1.19 & 0.72 & 1.44 & 2.77 & 5.55 \\
$^{c}$ 3 & 3.90 & 1.51 & 3.72 & 9.73 & 9.89 & 5.55 \\
$^{d}$ 4 & 2.03 & - & 1.95 & 9.60 & 7.19 & 5.42 \\
\hline 
\end{tabular}
\end{center}
\noindent{{$^a$}{Observed SED - the solid red line in Figure \ref{fig:sed0}.} \\
{$^b$}{Corrected SED - the solid blue line in Figure \ref{fig:sed0} and in all panels of Figure \ref{fig:TheorySed}.} \\
{$^c$}{Corrected SED - the dotted blue line in Figure \ref{fig:sed0} and in all panels of Figure \ref{fig:TheorySed}.} \\
{$^d$}{Theoretical SED - the solid red line in bottom panel of Figure \ref{fig:TheorySed}.} \\
{$^e$}{Bolometric luminosity integrated from $\rm log \, \nu = 11.5-20$.} \\
{$^f$}{IR luminosity integrated from $\rm log \, \nu = 11.5-14.5$.} \\
{$^g$}{Optical (HET) luminosity integrated from $\rm log \, \nu = 14.5-14.87$.} \\
{$^h$}{Ultraviolet (HST) luminosity integrated from $\rm log \, \nu = 14.87-15.5$.} \\
{$^i$}{Extreme ultraviolet (unobserved) luminosity integrated from $\rm log \, \nu = 15.5-17.0$.} \\
{$^h$}{X-ray luminosity integrated from $\rm log \, \nu = 17.0-20$.}}
\end{table*}

%%%%%%%%%%%%%%%%%%%%%%%%%%%%%%%%%%%%%%%%

%See also \S\ref{sec:uvxsed} and Figure~\ref{fig:TheorySed} for a discussion of
%our theoretical efforts to recreate the intrinsic unextincted \ir13349 SED.
%We have used the radio-quiet SED from Elvis (1994) to normalise the
%``Composite QSO spectrum", so that the EUV peak lies 0.25~dex above the IR
%peak in $\nu f(\nu)$.  
We have also checked the Elvis et al. SED against more recent compilations of
QSO SEDs \citep{shang11QSOSED, richards06:QSOSED, hopkinsQSOSED:07} built from
various combinations of SDSS, Spitzer, FUSE, and HST data. The variation in the
UV-IR peak difference (in $\rm log(\nu F_{\nu})$) is less than a factor of 2
when comparing these different samples.  (Note that the Hopkins et al. SEDs use
the Richards et al. sample of  objects but adds an additional $\alpha_{\rm ox}$
correction based on the object's peak UV flux; also, the Richards sample of
QSO, in using primarily SDSS have a median redshift $z>0.5$). 

To assess luminosity contributions for the different wave band components of
these SEDs, we integrate the line-segment SEDs of Figure~\ref{fig:sed0} over
the energy bands noted in Table~\ref{tab:sedlum}; we assume
$H_0=73$~km~s$^{-1}$ Mpc$^{-1}$, $\Omega_{\rm M}=0.25$, and
$\Omega_{\Lambda}=0.75$.   These luminosities are then used as relevant for
calculations in subsequent sections, either in whole or in part. The bolometric
luminosity we derived from the  piecemeal integration is $L_{\rm bol} \sim
3.12 \times 10^{46}~\ergps$.

\subsection{Reconstruction of the  UV to X-ray SED and its components: theory
and observations} 
\label{sec:uvxsed}

The radiation from the central regions of AGN is likely to peak in the
extreme ultraviolet at $\sim 10-100 \ev$ ($\sim 124-1240$~\AA), an
energy range often suffering from extinction effects due to Galactic
and/or intrinsic dust.  As stated in \S\ref{subsec:SedObs}, such is
the case for \ir13349, which shows a large UV and far-UV deficit in
its observed SED, which we attribute to dust extinction, a conclusion
also supported by Spitzer IR observations (see \S\ref{sssec:LRS} for
details).  Likewise, the two most important spectral components
featuring in this {\it un}-observable energy range are the {\it disc
blackbody} from an accretion disc peaking between 10 - 100~eV (typical
of AGN), and the soft excess, typically manifesting at $<$~1~keV.
Here, using theoretical considerations in combination with our
multi-waveband observations of \ir13349, we derive what we believe to
be the most likely scenario for the intrinsic \ir13349 SED, sans
extinction.

The ionising continuum dictating the ionisation state of the absorbers
in AGN is primarily contained in the UV-to-X-ray range.  Therefore we
focus our discussion on this energy range, with special attention paid
to recreating the extincted UV part of the spectrum.

The general mathematical form we use in building the \ir13349
theoretical SED is:
%%%%%%%%%%%%%%%
\begin{eqnarray}
f(\nu) & \sim & \Big[ A_{pow}\,\nu^{-\alpha} + A_{se}\,\left\{\frac {2 \, \pi \,h} {c^2} \, \frac {\nu^3} {\exp(h\,\nu/k_B\,T_{se})\, - \,1} \right\} \nonumber \\
&& ~~~~~~~~~~~~~~~~~~~~~~~~ + A_{dbb}\,f_{dbb}(\nu, T_{in}) \Big] e^{-\frac{\nu}{\nu_{max}}} ~~~~~~~~~ \rm{(a)} \nonumber \\
f(\nu) & \sim & \Big[ A_{pow}\,\nu^{-\alpha} + A_{se}\,f_c(\nu, T_{in}, T_c, \Gamma_c) \nonumber \\
&& ~~~~~~~~~~~~~~~~~~~~~~~~ + A_{dbb}\,f_{dbb}(\nu, T_{in}) \Big] e^{-\frac{\nu}{\nu_{max}}}. ~~~~~~~~~ \rm{(b)} \nonumber \\
%\labequn{TheorySed}
\label{eqn:theorysed}
\end{eqnarray}
%%%%%%%%%%%%%%
where the first and third terms in both
Equations~\ref{eqn:theorysed}a~and~\ref{eqn:theorysed}b represent the
1-10 keV X-ray power-law and the big blue bump (modeled by ``disk-blackbody'') respectively. As can be
seen, the two equations differ only in the 2nd term, associated with
competing models describing the soft excess (see \S\ref{subsec:SE} for
details).  We proceed with a discussion of this component, to be
followed by more in-depth discussion of the  big blue bump: its
derivation, and the dependence of its shape on the central black hole
mass and accretion rate.

%%%%%%%%%%%%%%%%%%%%%%%%%%%%%%%%%%%%%%%%%%%%%%%%%%%%%%%%%%%%%%%%%%%%%%%%%%%%%%
\begin{figure*}
\centerline{\includegraphics[scale=0.7,height=0.99\textwidth, trim = 10 15 0 10, clip, angle=90]{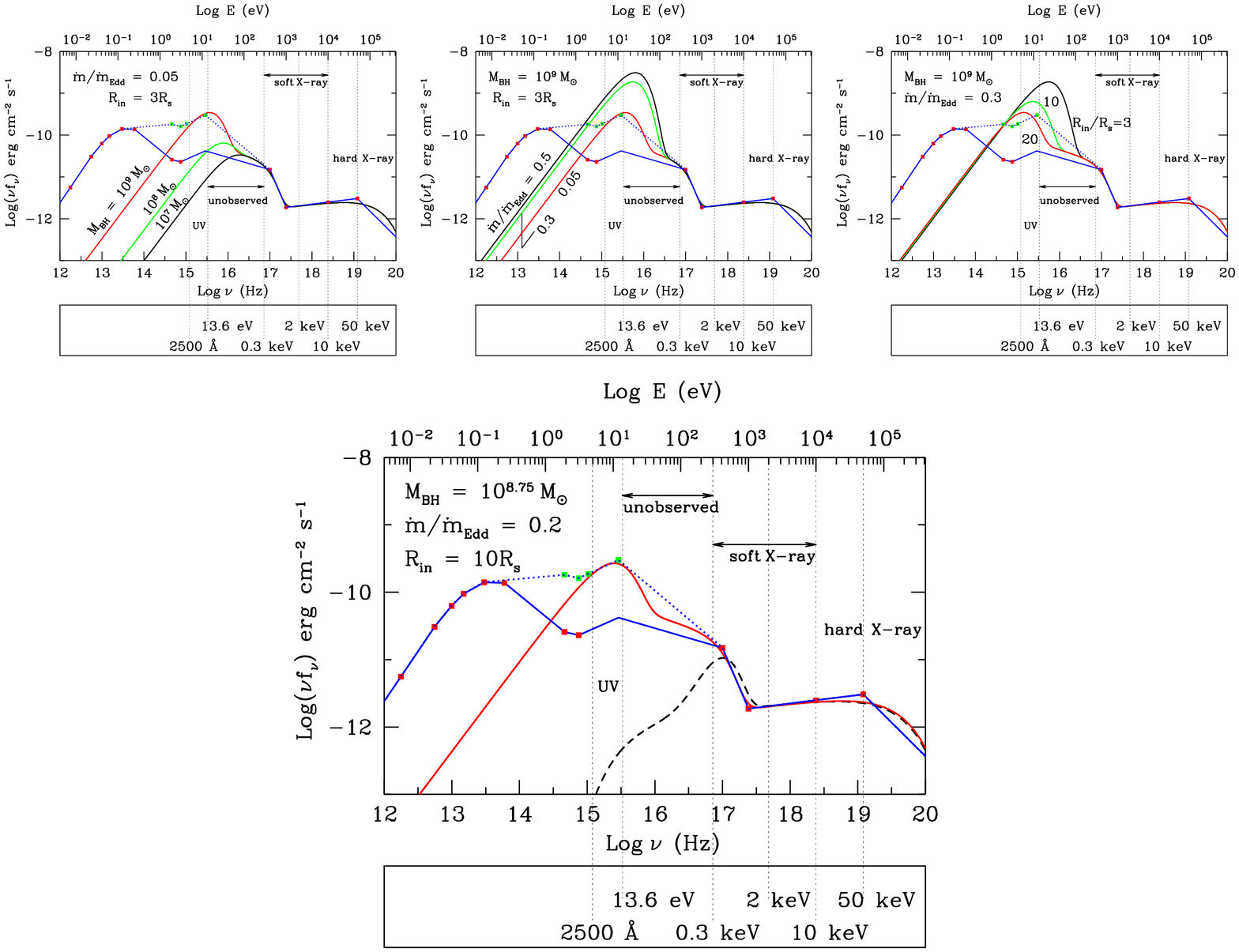}}
\caption{Theoretical SEDS calculated for \ir13349 based on
multi-wavelength data shown in Fig.~\ref{fig:sed0}.  The filled red
squares reflect the observed \ir13349 SED based on ISO, IRAS, Spitzer,
HET and Chandra.  Green squares are the renormalised observed
``generic-composite QSO" spectrum made up of a SDSS composite of $\sim 2000$
quasars \citep{sdssqso} longward of 3200 \AA\ and the HST RQ quasar
composite of \cite{telfer02}. The solid and dotted blue lines trace
out the same SED as in Fig.~\ref{fig:sed0}. {\bf (BOTTOM PANEL:)} The
theoretically computed SED (solid red) that  best matches the observed
\ir13349 SED corresponds to a black hole mass of $\rm{M_{BH}} \approx
10^{8.75} \,\rm{M_{\odot}}$, an accretion rate of
$\rm{\dot{m}/\dot{m_{Edd}}} \approx 0.2$ and $R_{in} = 10 R_s$. On the
other hand, if the accretion disc is completely ignored and one draws
a SED based only on the Chandra-HETG spectra, we get the dashed black
curve.  {\bf (TOP PANELS:)} Theoretical SEDs comparing spectral shape
differences for different combinations of $\rm{M_{BH}}$,
$\rm{\dot{m}/\dot{m_{Edd}}}$ and $R_{in}/R_s$ as discussed in
Section~\ref{sec:uvxsed}, for comparison and edification.
\label{fig:TheorySed}
}
\end{figure*}
%%%%%%%%%%%%%%%%%%%%%%%%%%%%%%%%%%%%%%%%%%%%%%%%%%%%%%%%%%%%%%%%%%%%%%%%%%%%%%

\subsubsection{The sub-keV soft excess}
\label{subsec:SE}

Observations of most AGN reveal that the SED between 2-10 keV is well
approximated by a power-law with spectral indices $\alpha \sim 0.9$
(photon indices $\Gamma \sim 1.9$). However, if this power-law is
extended to lower energies ($< 1 \kev$), for most type I AGN, there is
additional unaccounted radiation, which has come to be known as the
soft excess.  This  component is particularly important for shaping
the nature of the X-ray absorber \citep{chakravorty12}. However, the
physical origin of the soft excess is not understood, and presently 
the community largely treats it as a phenomenological spectral
component modeled as a blackbody (Equation~\ref{eqn:theorysed}a) with
temperature $T_{se} \sim 100 - 200 \ev$, which generally does a good
job of describing most observations.  Since this temperature is too
hot for the temperature, $T_{in}$, of the innermost stable circular
orbit of the accretion disc in AGN (see Equation~\ref{eqn:DbbTemp}),
the soft excess is often believed to be a separate spectral component
altogether, or a reprocessed (shortward in wavelength) extension of
the accretion disc component.  Our fit to the \chandra HETGS spectrum
of \ir13349 in \S3.1 gives a temperature for the soft excess of
$T_{se} = 99$~eV.

An alternative model $f_c(\nu)$ (Equation~\ref{eqn:theorysed}b) for
the {\it soft excess} is the thermal Comptonisation model {\it
nthcomp} \citep{lightman87,zdziarski96,zycki99} included in
\footnote{http://heasarc.gsfc.nasa.gov/docs/xanadu/xspec/}{\sc
xspec}~v.12.5 \citep{arnaud96}.  In this description, the seed photons
from the accretion disc are reprocessed by the thermal plasma to
generate sufficient photons at sub-keV energies to mimic the soft
excess. The high energy cut-off of this component is parametrised by
the electron temperature $T_c$, and the low energy rollover is
dependent on the effective temperature, $T_{in} = T(R_{in})$
(Equation~\ref{eqn:DbbTemp}) of the seed photons from the accretion
disc.  Between the low and high energy rollovers, the shape of the
spectrum is approximated by an asymptotic power-law with photon index
$\Gamma_c \approx \frac{9}{4} y^{-2/9}$, where $y$ is the Compton
y-parameter, which gives a measure of the extent of Compton
reprocessing; i.e., the larger the value of $y$, the greater the
fraction of photons reprocessed from the accretion disc.  For our
modeling efforts, we adopt $T_c = T_{se} = 100$~eV, consistent with
our Chandra soft X-ray data for \ir13349.  The associated
normalisation constant $A_{se}$ was determined from the ratio of the
$35 - 5$ \AA\, (0.35 - 2.5 keV) photon flux in the soft excess
component to that in the power-law, as seen in the Chandra HETGS data
for \ir13349.

%%%%%%%%%%%%%%%%%%%%%%%%%%%%%%%%%%%%%%%%%%%%%%%%%%%%%%%%%%%%%%%%%%%%%%%%%%%%%%
\begin{figure*}
\centerline{\includegraphics[scale=0.7,height=0.99\textwidth, angle=-90]{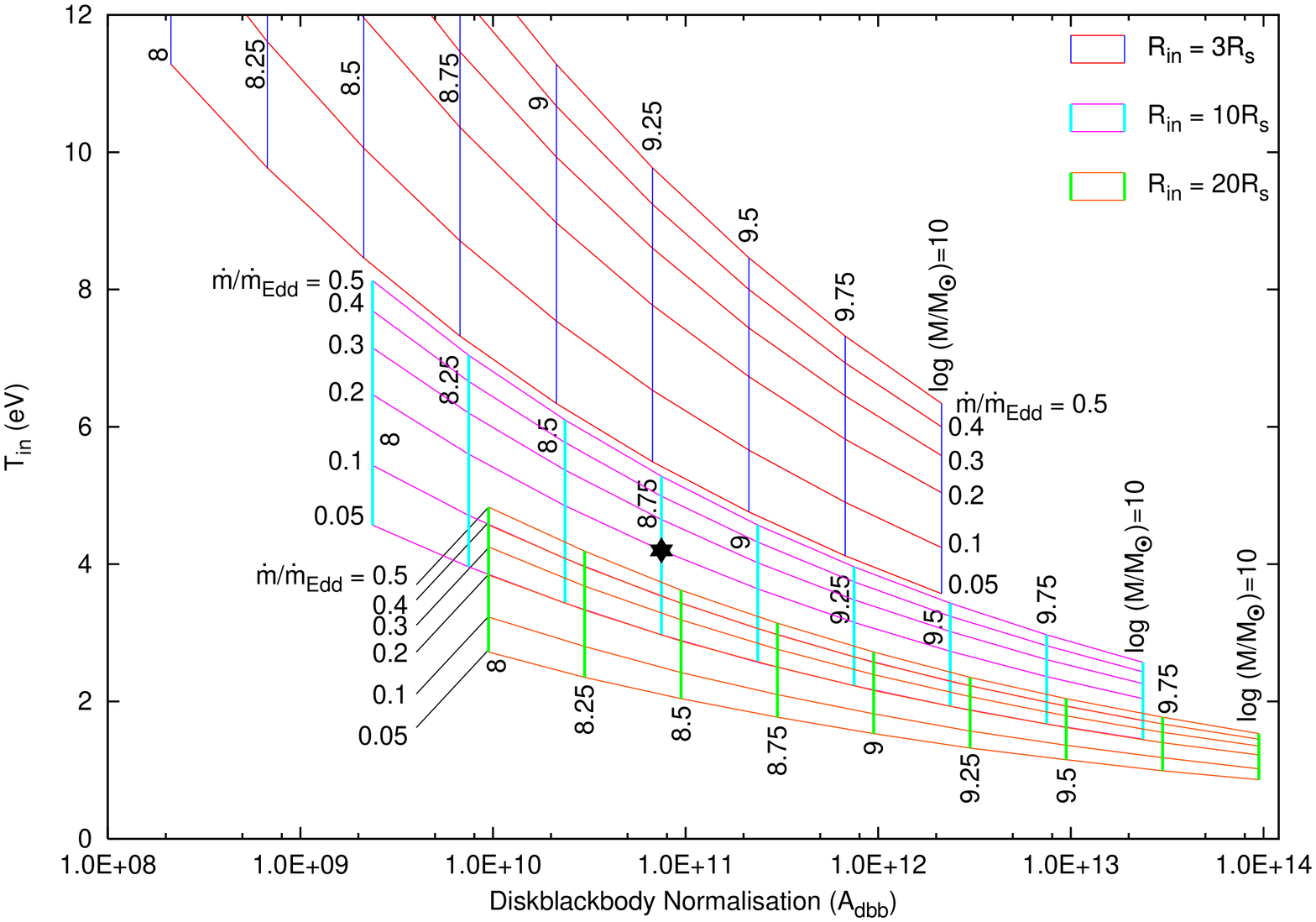}}
\caption{{{Based on grid runs using {\sc diskbb}, a theoretical
representation of all the SEDs that we have investigated to find the
black hole parameters that best match the composite QSO spectra
normalised to the flux of \ir13349. The big blue bump SED can be
mathematically described by $T_{in}$ and $A_{dbb}$ which we plot for
each combination of the parameters $\rm{M_{BH}}$,
$\rm{\dot{m}/\dot{m_{Edd}}}$ and $R_{in}$. The three different grids
(colour coded and labeled appropriately) of $\rm{M_{BH}}$ and
$\rm{\dot{m}/\dot{m_{Edd}}}$ correspond to three values of
$R_{in}/R_s$. The black star represents the best match SED (red curve,
lower panel of Figure~\ref{fig:TheorySed}) for $\rm{M_{BH}} \approx
10^{8.75} \,\rm{M_{\odot}}$, an accretion rate of
$\rm{\dot{m}/\dot{m_{Edd}}} \approx 0.2$ and $R_{in} = 10 R_s$. See
Section~\ref{subsec:ModelSed} for details.}}
\label{fig:TheoryParamGrid}
}
\end{figure*}
%%%%%%%%%%%%%%%%%%%%%%%%%%%%%%%%%%%%%%%%%%%%%%%%%%%%%%%%%%%%%%%%%%%%%%%%%%%%%%

The aforementioned models of the soft excess do not influence the UV
part of the spectrum and hence do not affect the predictions for the
shape of the big blue bump (see next section). The soft excess is an
important component influencing the ions which absorb the soft X-ray
radiation.  Our theoretical simulations show that there is a slight
difference in the predicted ion fraction of the various soft X-ray
ions, depending on choice of models for the soft excess. However,
while the Chandra HETGS data cover the energy range of the soft
excess, these are not sensitive enough to detect these relatively small
differences in ion abundances, and hence cannot be used to
differentiate between the \texttt{nthcomp} or \texttt{blackbody}
models.  Therefore, for the remainder of our analysis we adopt the
\texttt{nthcomp} model with $\Gamma_c =2.3$ and $T_c = 100 \ev$,
acknowledging that an alternative \texttt{blackbody} model with
$T_{se} = 100 \ev$ is likely to produce equally good results.

\subsubsection{The ``Big Blue Bump''}
\label{subsec:BBB}

Multi-wavelength observations suggest that AGN continua peak in the EUV
energy band and usually dominate the quasar luminosity (e.g.,
\citealt{elvis94QSOSED}).  This spectral component, often referred to
as the ``Big Blue Bump'', is considered to be the signature of the
presence of an accretion disc.  Yet, for many systems, it can only be
partially observed in the UV-EUV energy range. As such, an attempt to
reconstruct it requires careful theoretical considerations.

According to standard thin disc accretion theory \citep{ss73},
radiation from the accretion disc may be modeled as a sum of local
blackbodies emitted from the different annuli of the disc at different
radii, the temperature of the annulus at radius $R$ being
\begin{equation} 
T(R) = 6.3 \times 10^5 \left(\frac{\dot{m}} {\dot{m}_{\rm Edd}} \right)^{\frac{1}{4}} \left(\frac {M} {10^8M_{\odot}} \right)^{-\frac{1}{4}} \left(\frac{R}{R_s}\right)^{-\frac{3}{4}}\rm{K} 
\label{eqn:DbbTemp}
\end{equation}
\citep{peterson97bk,accretionpower:frank02} where $\dot{m}$ is the accretion
rate of the central black hole of mass $M_{BH}$, $\dot{m}_{\rm Edd}$ is its
Eddington accretion rate and $R_s$ is the Schwarzschild radius.  The
normalisation constant $A_{dbb}$ for this spectral component is given by
\begin{equation}
A_{dbb} = \left\{ \frac {R_{in}/\rm{km}} {D/(10\,\,\rm{kpc})} \right\}^2 \cos\theta
\labequn{DbbNorm}
\end{equation}
for an observer at a distance $D$ whose line-of-sight makes an angle $i$ to
the normal to the disc plane. $R_{in}$ is the radial distance of the innermost
stable annulus of the accretion disc from the black hole. Thus, the radiation
from the accretion disc has direct dependence on the mass of the black hole and
its accretion rate. As such, the shape of the spectral energy distribution can
provide important diagnostic power for assessing the fundamental parameters of
the black hole.

We acknowledge that more rigorous models exist for modeling the big blue bump,
that involve real radiative transfer in the accretion disc (see e.g.
\citealt{blaes01, hubney00, hubney01, hui05}) and/or the black hole spin
effects (see e.g. \citealt{davis05, davis06}).  For the same black hole mass
and accretion rate, relative to a simple ``disk-blackbody'', these
more-involved models change the location of the peak and the shape of the EUV
spectrum, as UV and EUV flux gets absorbed and re-emitted at longer wavelengths
in these models.  For models which include the black hole spin, the peak of the
accretion disk spectrum is pushed to higher energies for increasing black hole
spin.  Thus, qualitatively we can expect that using these models would result
in higher black hole mass as compared to the best-fit results obtained by using
``disk-blackbody'', for the same accretion rate.  However, since the UV
spectrum of \ir13349 is heavily extincted, we cannot make observationally based
distinctions between the different accretion disc models.  Given this, in the
next section we opt for the less-computationally expensive ``disk-blackbody''
model \citep{xspecdiskbb1, xspecdiskbb2} to assess the mass and accretion rate
of the black hole.  While we present results based on fits to {\sc diskbb}  to
allow for easier comparison to work on this and other AGN by other authors,  we
acknowledge that the model {\sc ezdiskbb} \citep{ezdiskbbref} may be the
theoretically more sound model, and therefore also discuss our subsequent SED
results based on a comparison of the two, where relevant.  In brief, the major
difference in the two models is most notable at $R < 10 R_{\rm in}$, whereby
the {\sc diskbb} predicted temperature continues to increase in contrast to
{\sc ezdiskbb};  this difference stems primarily from the {\sc ezdiskbb}
imposed boundary condition that the viscous torque be zero at the  inner edge
of the disc.

\subsubsection{Determining the mass and accretion rate of the \ir13349 black
hole}
\label{subsec:ModelSed}

%%%%%%%%%%%%%%%%%%%%%%%%%%%%%%%%%%%%%%%%%%%%%%%%%%%%%%%%%%%%%%%%%%%%%%%%%%%%%%
\begin{figure}
\centerline{\includegraphics[scale=1.0,width=12 cm, trim = 50 290 50 35, clip, angle=0]{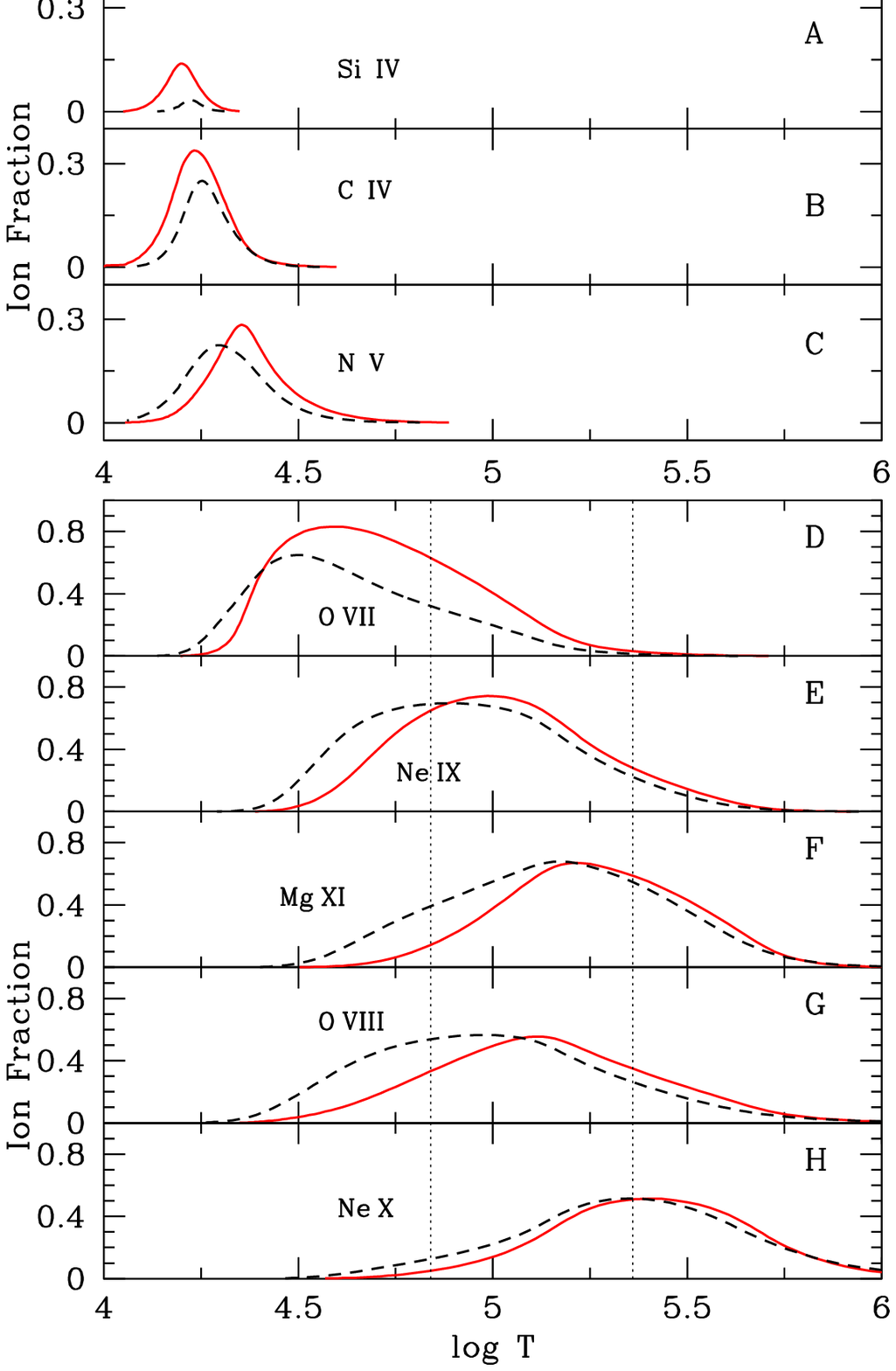}}
\caption{{Ion fractions versus gas temperature as a function of the two
relevant ionising SEDs. The line styles follow the same scheme used in
Figure~\ref{fig:TheorySed} : the solid red curves are obtained using the full
UV to X-ray SED whereas the dashed black curves are a result of an X-ray only
SED based on only the Chandra-HETG continuum. The panels A-C are for the UV
ions and D-H are for the X-ray ions as labeled. The dotted vertical lines
running across panels D-H mark the range $2.0 \leq \log \xi \leq 3.0$
corresponding to the full UV to X-ray ionising SED. This is the range of $\xi$
within which the ion fractions for \ion{Ne}{ix}, \ion{Mg}{xi} and \ion{O}{viii} peak. The importance
of this range of the ionisation parameter is discussed in Sections
\ref{subsubsec:WaModel} and \ref{subsec:wathermo}.} 
\label{fig:TheoryIF}
}
\end{figure}
%%%%%%%%%%%%%%%%%%%%%%%%%%%%%%%%%%%%%%%%%%%%%%%%%%%%%%%%%%%%%%%%%%%%%%%%%%%%%%

Using the observed \ir13349 SED (see \S\ref{subsec:SedObs}
discussion), in combination with Equation~\ref{eqn:theorysed}, and the
\citet{ss73} model (applied to Equation~\ref{eqn:theorysed}, 3rd
term), we generate a series of SEDs for different values of $\dot{m}$
and $M_{\rm BH}$ to obtain the best match to the \ir13349 SED
(Figure~\ref{fig:sed0};  \S\ref{subsec:SedObs}  for details). We carry
out an iterative, step-by-step approach to reconstruct the UV-EUV SED,
with permutations of the relevant parameters, as described below.

We begin by running models using a coarse grid of various combinations
of the parameter values $\rm{\dot{m}/\dot{m}_{\rm Edd}} = (0.05, 0.1,
0.5, 1.0)$ and $\rm{M_{BH}} = (10^7, 10^8, 10^9 \,\rm{M_{\odot}}$ for
each $\rm{\dot{m}/\dot{m}_{\rm Edd}}$,) with a fixed $R_{\rm
in}=3R_{\rm S}$.  For this initial run, we find that
$\rm{\dot{m}/\dot{m}_{\rm Edd}} = 0.05$ and $\rm{M_{BH}} = 9.0$ gives
a reasonable match to the observed \ir13349 SED (see top left panel of
Figure~\ref{fig:TheorySed} comparing theoretical models with UV
composite QSO spectra, green data points).  It should be noted that we
can expect a range of acceptable matches between theory and
observations for different permutations of key parameters, primarily
due to the fact that Equation~\ref{eqn:theorysed}, which we use for
modeling the big blue bump, is degenerate between mass and accretion
rate.  The top panels of Figure~\ref{fig:TheorySed} demonstrate the
effects on the shape of the theoretical SED when various combinations
of $R_{\rm in}$, $\rm M_{BH}$, and $\rm{\dot{m}/\dot{m}_{\rm Edd}}$
are fixed/varied.

Therefore, to reduce uncertainties, we use a black hole mass derived from the
{\sc H}$\beta$ line width ($\sigma_{\rm H\beta} = 1948~ \pm 3.7 \kmps$;
see Section~\ref{sec:opticallines}), and the optical continuum
luminosity $\lambda L_{5100} = 6.28 \times 10^{44}~\rm erg~s^{-1}$,
based on the renormalised, green Composite QSO spectrum in
Figure~\ref{fig:sed0}.  Based on the \citet{McGill08} relation,
\begin{equation}\rm
log  \,M_{BH} = 7.68 + 2\, log\, \frac{\sigma_{\rm H\beta}}{1000}~\kmps + 0.518 \,log\, \frac{\lambda L_{5100}}{10^{44}}~erg~s^{-1}
\end{equation}
we derive for \ir13349, $\rm log~M_{BH} = 9.10~M_\odot$; the McGill relation
has an rms scatter in $\rm log~M_{\rm BH} = 0.09$. Barring better constraints,
we use $\rm M_{BH} \sim 10^{9}~\rm M_\odot$ for input into
Equation~\ref{eqn:theorysed} to determine the accretion disc component of our
theoretical intrinsic SED.  Given that this $M_{\rm BH}$ estimate is close to
our ``best fit" value from the initial coarse grid of runs based on a fixed
$R_{\rm in}=3R_{\rm S}$, we next run a finer grid of models allowing all the
aforementioned three parameters free.  Figure~\ref{fig:TheorySed} (bottom
panel) shows the best theoretical SED match (red curve) to the UV-EUV flux
distributions based on our hypothesised extrapolation of observed data in the
UV-EUV spectral domain (green points). The best match SED has optimal
parameters $\rm M_{BH}=10^{8.75} M_\odot$, $\rm{\dot{m}/\dot{m}_{\rm
Edd}}=0.2$, and  $R_{\rm in}=10 R_{\rm S}$.  Figure~\ref{fig:TheoryParamGrid}
shows $T_{in}$ versus $A_{dbb}$ for all theoretical values of $M_{\rm BH}$,
$\rm{\dot{m}/\dot{m}_{\rm Edd}}$ and $R_{\rm in}$ that we have investigated,
including where in these curves our best fit lies. For thoroughness, we also
perform the same exercise using {\sc ezdiskbb}, and find that while  the ``best
fit" SED shape remains the same, the same exercise with {\sc ezdiskbb} gives a
higher accretion rate for  fixed values of $R_{\rm in}$ and $M_{\rm BH}$.  From
our trials at finding the best match based on {\sc ezdiskbb}, the relevant SED
parameters bracket a similar best range found using {\sc diskbb}, i.e.
$10^{8.50} < \rm M_{BH} < 10^{9.25} M_\odot$, $0.2 \leq
\rm{\dot{m}/\dot{m}_{\rm Edd}} \leq 0.7$, and  $3 < R_{\rm in} \leq 10 R_{\rm
S}$.

The shape obtained for the big blue bump is then complemented with X-ray
power-law ($\S\ref{sec:uvxsed}$), and soft excess (discussed in
\S\ref{subsec:SE}) components to derive the full UV-XRay SED
(Figure~\ref{fig:TheorySed} bottom panel - red) needed for input to
photoionization modeling.--See luminosity information associated with SED~\#4
of Table~\ref{tab:sedlum}. (Note that $L_{\rm IR}$, which is primarily
attributed to reradiation by dust that has absorbed part of the optical-UV
radiation is excluded here, since it has no bearings on the actual disc
spectrum.)   Note that if the accretion disc component is completely ignored
and a SED is constructed on the basis of only the Chandra HETGS X-ray data for
\ir13349, then one would obtain the black dashed curve shown in the bottom
panel of Figure~\ref{fig:TheorySed}.  Such an SED was used by \citet{holczer07}
to study the warm absorber in \ir13349. In Sections \ref{subsec:UV2XraySED} and
\ref{subsec:wathermo}, we discuss differences in, and effects on, derived
absorber properties based on an ionising SED that accounts for only X-ray
contributions versus one which also includes disc contributions.

%%%%%%%%%%%%%%%%%%%%%%%%%%%%%%%%
\subsubsection{Advantages to using the full UV to X-ray SED for assessing the
location(s) of the  absorbing gas}
\label{subsec:UV2XraySED}

Approximately half of all low-z AGN show high-ionisation UV absorption lines
and signatures of even higher ionisation warm absorption in the X-ray data.  In
assessing the  role of the disc ionising radiation for affecting our
conclusions on UV and X-ray warm absorber properties (e.g.  common versus
distinct origins; continuous vs. clumpy clouds),  we use {\sc xstar} to
generate ion fractions (Figure~\ref{fig:TheoryIF}) as a function of the
temperature of the absorbing gas for the UV (panels A-C) and the X-ray (panels
D-H) ions, using an SED that includes the big blue bump, the sub-keV soft
excess and the X-ray power-law (Figure~\ref{fig:TheorySed} lower panel - red
curve) versus an SED that only accounts for the X-ray spectral components,
namely the power-law and the black body soft excess, as determined from
continuum fits to the X-ray data (Fig.~\ref{fig:TheorySed} black dashed curve). 

One of the most important ions in the X-ray warm absorber is He-like
\ion{O}{vii}.  As can be seen in Figure~\ref{fig:TheoryIF}, its ion fraction is
under-predicted by $\sim 25\%$ if the big blue bump is excluded from the
ionising continuum (panel D). The relative abundance of \ion{O}{vii} and the
H-like \ion{O}{viii}, another of the important ions in the X-ray warm absorber,
are sensitively inter-related. Hence under-prediction of the \ion{O}{vii} ion
would be associated with a complementary over prediction of the \ion{O}{viii}
(panel G). \ion{Mg}{xi} is also significantly over-predicted by a SED which
excludes the big blue bump. Furthermore, it can be seen that the inclusion of
the big blue bump predicts a larger ion fraction for the UV ions (panels A-C),
and hence a higher likelihood that they are produced in the same gas as that
responsible for \ion{O}{vii} (panel-D) seen in X-ray absorbers. This is
contrary to the conclusions one would arrive at, if considering an X-ray-only
SED.

\subsection{The ionised dusty absorbers in \ir13349}
\label{subsubsec:WaModel}

%----------------------------------------------

%%%%%%%%%%%%%%%%%%%%%%%%%%%%%%%%%%%%%%%%%%%%%%%%%%%%%%%%%%%%%%%%%%%%%%%%%%%%%%
\begin{figure*}
%\centerline{\includegraphics[scale=1, width = 0.8\textwidth, angle=0]{FIGS/chandra/IRAS-WAFIT.ps}}
%\centerline{\includegraphics[scale=1, width = 0.9\textwidth, angle=270]{FIGS/chandra/iras13_highres_withlabel.ps}}
\centerline{\includegraphics[scale=1, width = 0.9\textwidth, angle=0]{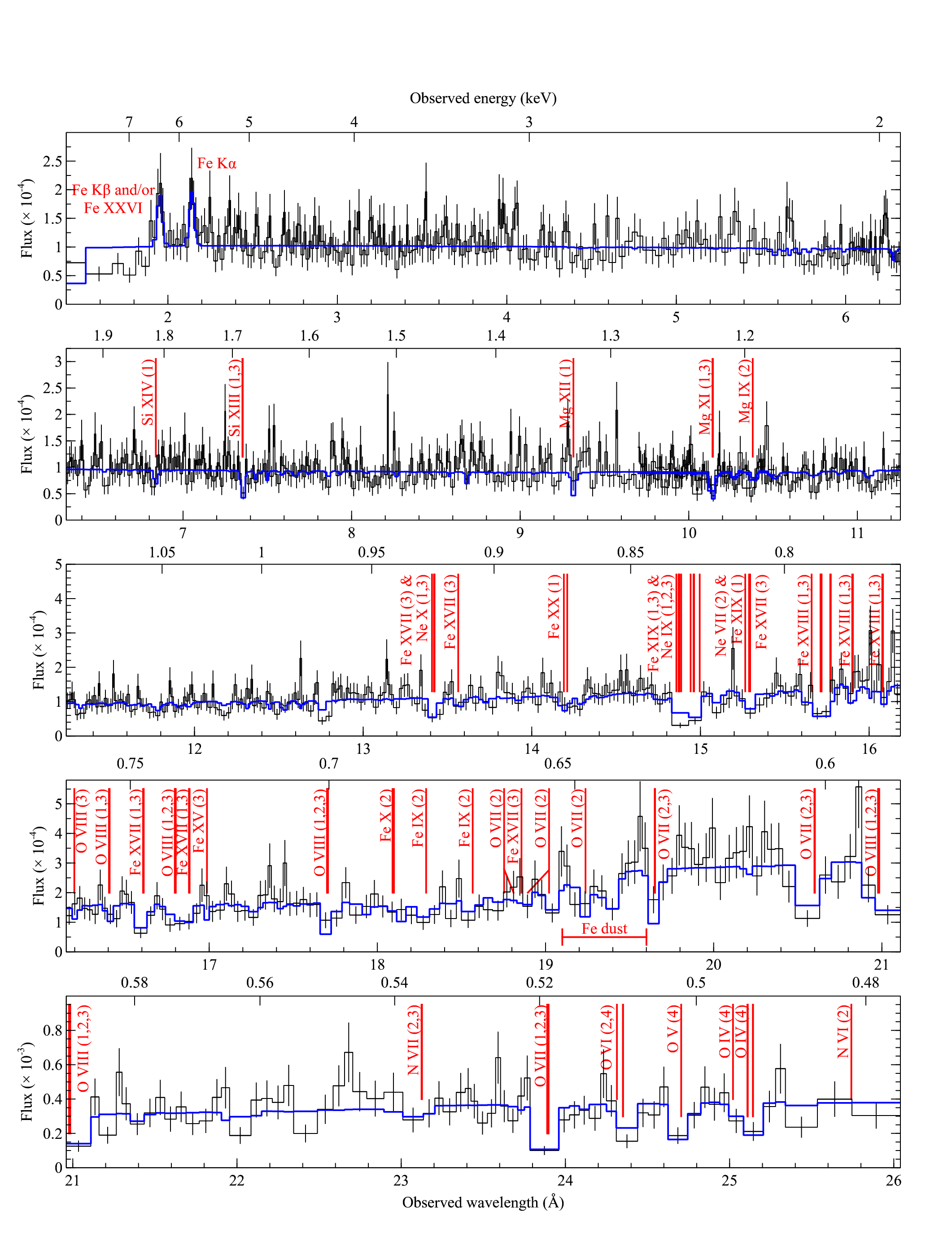}}
%\centerline{\includegraphics[scale=1, width = 1.0\textwidth, angle=0]{FIGS/chandra/post-ref-fig12.ps}}
\caption{The best fit to the Chandra HETGS spectrum of \ir13349 (in the
observed frame). Numbers associated with the ion
labels correspond to contributions from one or multiple absorbers as
described in Section~\ref{subsubsec:WaModel} and Table~\ref{tab:xstar_mod}.
Note that while this figure is shown with adaptive binning ($15\, \rm cts \, bin^{-1}$) 
to facilitate easier viewing,
our fitting results are based on {\it constant binning} at the HETGS resolution. 
Only predicted lines with equivalent width $W_\lambda > 0.005$~m\AA\, are labeled.
\label{fig:hetgxstarfit}
}
\end{figure*}
%%%%%%%%%%%%%%%%%%%%%%%%%%%%%%%%%%%%%%%%%%%%%%%%%%%%%%%%%%%%%%%%%%%%%%%%%%%%%%

Using {\sc xstar 2.2.1}, we generate ``warm absorber" models with ion
populations that reflect the ionising spectrum of our preferred \ir13349 SED
(see Figure \ref{fig:TheorySed}, bottom panel, red SED; \S\ref{subsec:ModelSed}
for details). We employ the analytic version of {\sc xstar} which enable line
broadening and optical depth calculations in real time as relevant to the data.
In addition, the granularity associated with analytic models are only in $\xi$,
which we mitigate by creating population files sampled with $\Delta(\log\xi) <
0.1$ . (In contrast, one can expect table models to have granularity in all
free parameters.~--~Private Communication: Tim Kallman.) 

%%%%%%%%%%%%%%%%%%%%%%%%%%%%%%%%%%%%%%%
% Table 8
%\begin{table}
\begin{table*}
\begin{center}
\caption{Best fit parameters based on fits to the Chandra HETGS data.}
\label{tab:xstar_mod}
\begin{tabular}{ccccc}
\hline\hline
{$^a$} BB norm ($10^{-4}$) & {{$^b$} BB kT} & {$^c$} PL norm ($10^{-3}$) & {{$^d$} PL $\Gamma$} \\
$\rm{ph\,cm^{-2}\,s^{-1}}$ & (eV) & $\rm{ph\,cm^{-2}\,s^{-1}}$ & {} \\
\hline
\multicolumn{4}{c}{The continuum} \\  
$1.87_{-0.07}^{+0.11}$ & $107_{-0.1}^{+0.1}$ & $1.30_{-0.02}^{+0.01}$ & $1.98_{-0.02}^{+0.02}$ \\ 
\hline
\hline 
& {$^e$} Line Flux ($\times 10^{-6}$) & {$^f$} Centre & {$^g$} Width & Width \\
& $\rm{ph\,cm^{-2}\,s^{-1}}$ & (keV) & (keV) & (km~s$^{-1}$) \\ 
\hline 
\multicolumn{5}{c}{The Iron Emission Lines} \\
\ion{Fe}{i} K$\alpha$ & $4.2_{-1.85}^{+0.98}$ & $6.42_{-0.03}^{+0.05}$ & $0.04_{-0.02}^{+0.06}$ & $1868_{-936}^{+2801}$ \\
\ion{Fe}{i} K$\beta$ and/or \ion{Fe}{xxvi} (?) & $4.0_{-1.86}^{+2.19}$ & $7.04_{-0.04}^{+0.04}$ & $0.05_{-0.05}^{+0.06}$ & $2129_{-2129}^{+2555}$ \\ 
\hline
\hline 
& {$^h$} $\rm log \, \xi$ & {$^i$} $\rm v_{blue}$ & {$^j$} $\rm b$ & {$^k$} $\rm N_{H}$ \\ 
& & (km~s$^{-1}$) & (km~s$^{-1}$) & ($10^{21} \rm{cm^{-2}}$) \\
\hline 
\multicolumn{5}{c}{4 components of the warm absorber} \\
WA-1 & $1.67_{-0.12}^{+0.15}$ & $-798_{-78}^{+18}$ & $ 395_{-62}^{+59}$ & $1.01_{-0.07}^{+0.13}$ \\
WA-2 & $3.38_{-0.07}^{+0.10}$ & $-678_{-115}^{+68}$ & $492_{-85}^{+95}$ & $3.26_{-0.55}^{+0.57}$ \\ 
(WA-3) & $3.00_{-0.07}^{+0.00}$ & $-758_{-48}^{+71}$ & $95_{-38}^{+76}$ & $2.39_{-0.43}^{+0.54}$ \\
(WA-4) & $-0.94_{-0.06}^{+0.22}$ & $-1298_{-197}^{+127}$ & $455_{-146}^{+315}$ & $0.14_{-0.04}^{+0.05}$ \\
\hline
\hline 
& {$^l$} $\rm N_{Dust}$ & {$^m$} $\rm v_{blue}$ & 
$\frac{\rm N_{Dust}}{\rm N_{Gas} + N_{Dust}}$  \\
& ($10^{17} \rm{cm^{-2}}$) & (km~s$^{-1}$) & \\
\hline 
&\multicolumn{3}{c}{Iron based Dust} &\\ 
& $1.23_{-0.33}^{+0.21}$ & $-104_{-275}^{+407}$ & 98.27\% & \\
\hline 
\end{tabular}
\end{center}
$^{\rm a}${Blackbody normalisation}. $^{\rm b}${Blackbody temperature}. $^{\rm c}${Power-law normalisation}. $^{\rm d}${Power-law photon index}. \\
$^{\rm e}${Gaussian emission line normalisation}. $^{\rm f}${Line central energy}. $^{\rm g}${Line width}. \\
$^{\rm h}${Log of the ionization parameter (See Section~\ref{subsec:wathermo} for detailed definition)}. \\
$^{\rm i}${Outflowing velocity (negative indicates blushift)}. $^{\rm j}${Turbulent/thermal velocity}. \\
$^{\rm k}${Column density}. $^{\rm l}${Column density of iron in condensed matter (i.e. dust/molecule) form}. $^{\rm m}${Dust outflow velocity (negative indicates blushift)}.
\end{table*}
%\end{table}

%%%%%%%%%%%%%%%%%%%%%%%%%%%%%%%%%%%%%%%%

Using codes we developed for the {\sc isis} \citep{isisref} fitting/analysis package
in combination with the aforementioned {\sc xstar}-generated models,
we initially fit the data (binned to the HETGS resolution) with 9 absorber components spanning the
$-4 < \log \xi < 4$ range, and find that it is the high $\xi$ ($\log \xi > 1$)
gas which drive the fit.  
%The respective statistics were -- With Dust $\Delta C = 1.134 (2822.105/2540),
%\,\, \chi^2 = 1.101 (2740.434/2540)$.  Without Dust $\Delta C = 1.131
%(2819.127/2540), \,\, \chi^2 = 1.106 (2756.748/2540)$.
In also investigating dustless (Section~\ref{subsec:uta}) and dusty
(Section~\ref{subsec:xfedust}) absorption, we find that a statistically good
fit describing the HETGS data (0.5--1.3 keV MEG and 1.2--8.9 keV HEG chosen to
maximise both spectral resolution and throughput) is a power-law plus blackbody
continuum absorbed by two ionised absorbers [hereafter \warm1 ($\log \xi  =
1.68_{-0.16}^{+0.14}$, $\log \rm{N_H} = 21.03_{-0.09}^{+0.05}$) and \wa2 ($\log
\xi = 3.26_{-0.08}^{+0.05}$, $\log \rm{N_H} = 21.66_{-0.06}^{+0.05}$)] at
similar ($\sim$ 700--800\,\kmps) velocities, and iron dust (see
Section~\ref{subsec:xfedust} for details), intrinsic to the source.  The
reduced Cash statistic $\Delta C$ for this model is 1.147 (2887.127/2540) and
the corresponding reduced $\chi^2$ = 1.102 (2774.783/2540). While the fit
naturally tend towards the aforementioned two-absorption+dust model,  we
consider additional absorbers, driven by our own theory considerations based on
thermodynamic stability arguments (see Section~\ref{subsec:wathermo}; also
Section~\ref{subsec:wa3}) and the detection of the Unresolved Transition Array
in the previous \xmm \citep{sako13349:01} study of \ir13349
(Section~\ref{subsec:uta}). While these additional absorbers aid in fitting a
few additional weak lines, they do not add significantly to the overall
improvement in fit statistics ($\Delta C$ = 2842.34/2540 = 1.133, $\chi^2$ =
2741.884/2540 = 1.093 for the addition of WA-3 and WA-4).  -- See
Table~\ref{tab:xstar_mod} and Figure~\ref{fig:hetgxstarfit} for details on the four-absorber+dust model.  We also
test for the presence of additional absorbing material from within the Milky
Way and find no strong evidence for its presence along the \ir13349
line-of-sight.

\begin{figure*}
\centerline{\includegraphics[scale=1.0, height = 19 cm, trim = 45 0 35 0, clip, angle=90]{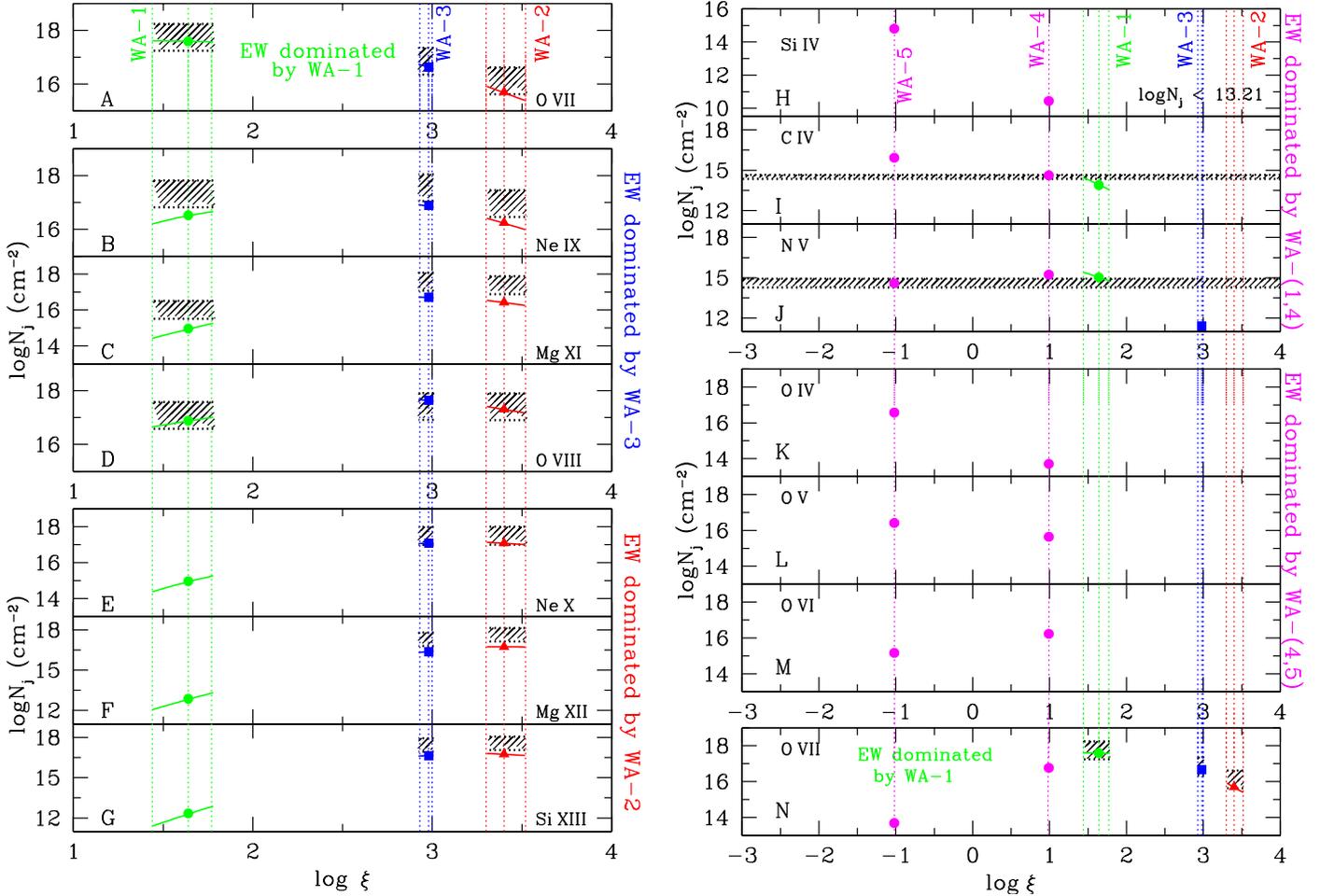}}
\caption{The column densities of the various X-ray (panels A-G on the left and
N on the right) and UV (panels H-J on the right) found in the high resolution
spectra of \ir13349. Panels K-M, on the right, give the column densities for
some of the intermediate theory predicted ions, barely detected in the X-rays.  The
horizontal hashed regions in the left panels A to G, correspond to derived \nj
based on fits to the line series (see Table~\ref{tab:xseries} and
Section~\ref{subsec:xplasmastudy}), but weighted  according to the amount that
the X-ray warm absorbers 1~to~3 contributes to it.  Respectively, from left to
right, green, blue and red points correspond to the best fit \nj as determined
by {\sc xstar} fits for absorbers  \warm1, \w3, \wa2.  It can be seen that OVII
(panel A) is associated with \warm1, whereas ions in B--D and E--G panels
correspond to \w3 and \wa2 respectively.  In the right panels I and J, the
horizontal shaded strips represent the ionic column densities \nj derived from
the UV observations, assuming a Doppler parameter of $b = 100$~\kmps (Table
\ref{tab:UvALines}). The right panels H--N have a different $\log \xi$ range,
in order to accommodate WA-4 and WA-5 (marked with vertical dotted magenta
lines).  
\label{fig:TheoryNh}
}
\end{figure*}

On the intrinsic \ir13349 absorption, Figure \ref{fig:TheoryNh} shows the
observed and predicted column densities (\nj) of the various ions we have found
in the UV and X-ray spectra of IRAS 13349+2438, based on both the plasma
diagnostic approach of Sections~\ref{sec:analysis}, and the {\sc xstar}
photoionization fitting here. As  demonstrated in the figure, the series
fitting and {\sc xstar} modeling agree with each other on the predicted column
densities for many of the detected ions. For instance, it can be seen that for
\ion{O}{vii} (Figure~\ref{fig:TheoryNh}, left panel A),  the column
density predicted from fitting the line series (horizontal shaded regions)
matches with the {\sc xstar} predicted \nj.  In addition, the figure is
instructive in revealing the extent to which each absorber is responsible for
the different ionic columns. Taking the example of \ion{O}{vii}  again, the
dominant optical depth contribution comes from \warm1 and the \nj due to \wa2 
falls short by 1.91~dex. On the other hand, for H-like \ion{Ne}{x},
\ion{Mg}{xii} and \ion{Si}{xiii} (panels E--G) the higher ionisation phase of
WA-2 satisfies the observed optical depth due to the respective ion.  In the
following, we consider the impact of adding additional ionized absorbers.

\subsubsection{A third warm absorber? - \w3}
\label{subsec:wa3}

It is interesting to note that the \nj for He-like \ion{Ne}{ix} and
\ion{Mg}{xi} do not fall in the range predicted by either \warm1 or \wa2.  This
makes sense according to panels E and F of Figure~\ref{fig:TheoryIF} which
indicate that these ions peak at a temperature consistent with an ionisation
$2.0 < \log\xi < 3.0$.  Indeed, theory considerations based on thermodynamic
stability arguments (see \S\ref{subsec:wathermo} for details) point to more
than two discrete zones of absorbers$--$when a 3rd absorber (\w3: $\log\xi  =
2.98^{+0.02}_{-0.05}$, $\rm{N_H} = 2.30^{+0.52}_{-0.47} \times 10^{21}$)
confined within the ionisation range dictated by our theory considerations (\S\ref{subsec:wathermo}) is
added to the fits, the aforementioned ions are better fit, although changes in
fit statistics with the addition of this absorber is minimal ($\Delta C$ =
1.143 \,(2872.129/2540) and reduced $\chi^2 = 1.098 \, (2759.126/2540)$).~--~See
Figure~\ref{fig:hetgxstarfit} for ion contributions from \w3.
We will return to WA-3 in Section~\ref{subsec:wathermo} in the context of
theoretical considerations based on the thermodynamic stability of the
absorbers in \ir13349. In short, the present X-ray data are such that the S/N
is not large enough for it to drive the fitting results toward greater than a
two-zone absorber, although when more ionisation zones are included, the
observed column densities for H- and He-like Ne and Mg are better matched to
that derived based on our plasma diagnostic approach of Section~\ref{subsec:xplasmastudy}.

%%%%%%%%%%%%%%%%%%%%%%%%%%%%%%%%%%%%%%%%%%%%%%%%%%%%%%%%%%%%%%%%%%%%%%%%%%%%%%
%%%%%%%%%%%%%%%%%%%%%%%%%%%%%%%%%%%%%%%%%%%%%%%%%%%%%%%%%%%%%%%%%%%%%%%%%%%%%%

\subsubsection{UTA Contributions}
\label{subsec:uta}

As stated, our best model fit to the \ir13349 required a dust component
for the X-ray fitting. For thoroughness, we investigate whether
additional (gas phase) absorber components can take its place. In particular,
L-~to~M-shell photoexcitations in lower charge states of \ion{Fe}{i} to
\ion{Fe}{xvi} give rise to absorption at rest wavelengths 14 -- 17.5 \AA\,
($\sim 0.7-0.89$~keV), $\sim$~the same spectral region where the dust absorption
is required.  These features, dubbed the {\it Unresolved Transition
Array} or UTA (see \citealt{feuta_calc:01} and \citealt{gu07} for theory), was
initially detected by \cite{sako13349:01} based on an \xmm RGS study of
\ir13349.  To test for the prevalence and strength of these lines based on our
method of analysis (i.e. one which tie line strengths to the observed UV-Xray
ionising SED), we explore fits which include additional absorber components
that may account for them. (As a reminder, note that we began with an initial
9-absorber fit, covering the range $\rm -4 \le log \xi < 4$, that would have
accounted for all UTA lines, if significant.)

\begin{figure}
\centerline{\includegraphics[scale=1.0,width=12 cm, trim = 50 80 50 35, clip, angle=0]{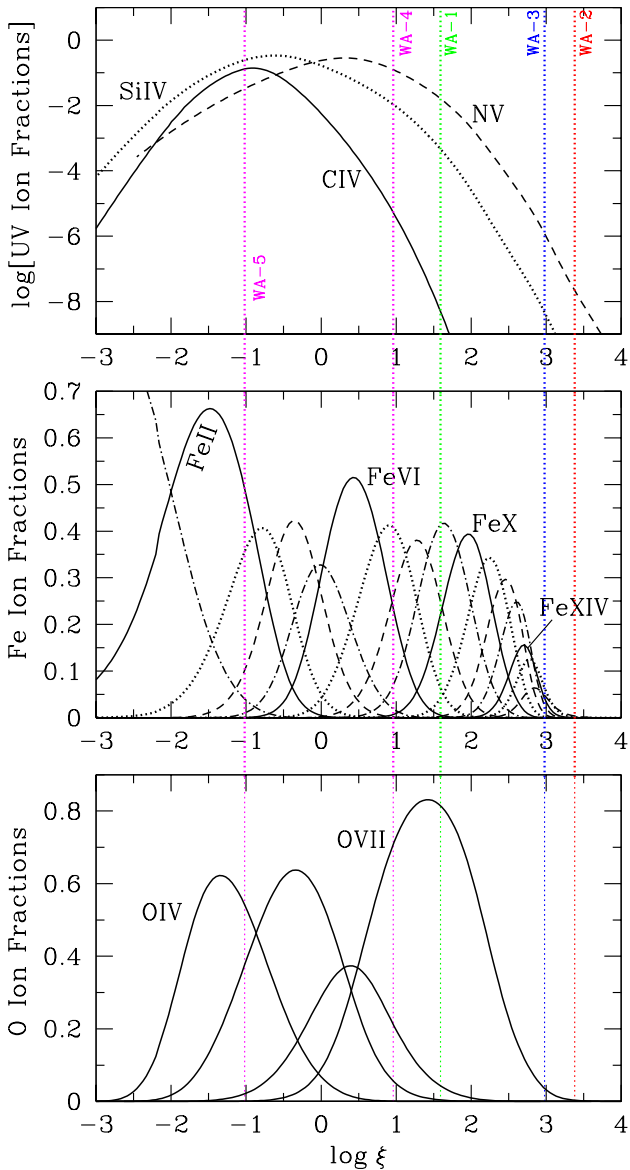}}
\caption{
The distribution of ion fractions for the \ion{Fe}{i}--\ion{Fe}{xvi} ions
(middle panel), responsible for the UTA, as a function of
the ionization parameter. For comparison, we also plot the ion fraction
distributions of some of the relevant, prominent UV (top panel) and X-ray ions
(O IV--VII, bottom panel). Vertical dotted lines mark the best fit 
$\xi$ values of the various warm absorber components 
(see Section~\ref{subsec:uta} for details). 
\label{fig:UtaIF}
}
\end{figure}

We begin by checking the range of ionization parameters where the ion fractions
of \ion{Fe}{I} to \ion{Fe}{xvi} peak (middle panel of Figure \ref{fig:UtaIF}).
As can be seen in the figure, {\sc wa-(1-3)} primarily account for the range of
$\log \xi > 1.49$, which while fitting for UTA ions \ion{Fe}{vii} to
\ion{Fe}{xvi}, do not fully account for the \ion{Fe}{ii} to \ion{Fe}{vi}
transitions that populate the spectral region between
17--17.5~\AA\,~(0.71--0.73 keV rest; observed~$\sim$~18.8-19.4~\AA=0.64--0.66
keV). Figure \ref{fig:UtaIF} shows that the ``missing" low-ionization UTA lines
should fall between $-3.0 \le \log \xi \le 1.0$.  As such, we include two
additional warm absorber components: {\sc wa-4} (forced within $-1.0 \le \log
\xi \le 1.0$) and {\sc wa-5} (forced within $-3.0 \le \log \xi \le -1.0$) to
account for all ``missing" UTA producing ions.  The resultant fits ({\sc wa-4}:
$\rm log\, \xi \sim 0.99 \; {with} \; \log N_H \sim -1.78$, and {\sc wa-5}:
$\rm log \xi \sim -1.02 ; {with} \; \log N_H \sim -1.86$) primarily affect
\ion{O}{iv} to \ion{O}{vi} with the strongest contribution to \ion{O}{v} (24.8
\AA\, observed) and \ion{O}{iv} (25.2 \AA\, observed) absorption, while the fit
statistics ($\Delta C = 1.138;\, \chi_\nu^2 = 1.099)$ are still poorer than our
best fit 4WA+dust fit, although marginally.  (See Section~\ref{subsec:xfedust}
for details on fitting for dust in X-ray spectra.) Indeed, as can be seen in
Figures~\ref{fig:TheoryNh}~and~\ref{fig:UtaIF}, {\sc wa-4} contributes noticeably to the column density
for these lower ionization oxygen (K--M) and UV ions (H--J), while {\sc wa-5}
give \nj values which are higher than those constrained by our HST measurements
of \ion{Si}{iv} and \ion{C}{iv}.  Nevertheless, this argues strongly for the UV
and low-$\xi$ X-ray warm absorbers 3-5  having similar origins, as also borne
out based on kinematics (see Section~\ref{subsec:wakinematics}).  (For
subsequent fits, we exclude {\sc wa-5} because it does not obviously contribute
to the fits in a positive way.) It can be seen from Figure~\ref{fig:DustOrUta}
(blue) that the UTA alone cannot account for all the absorption in this
spectral region.

%%%%%%%%%%%%%%%%%%%%%%%
\subsubsection{Direct X-ray detections of iron dust}
\label{subsec:xfedust}
\begin{figure}
\centerline{\includegraphics[scale=1.0,height= 8 cm, angle=-90]{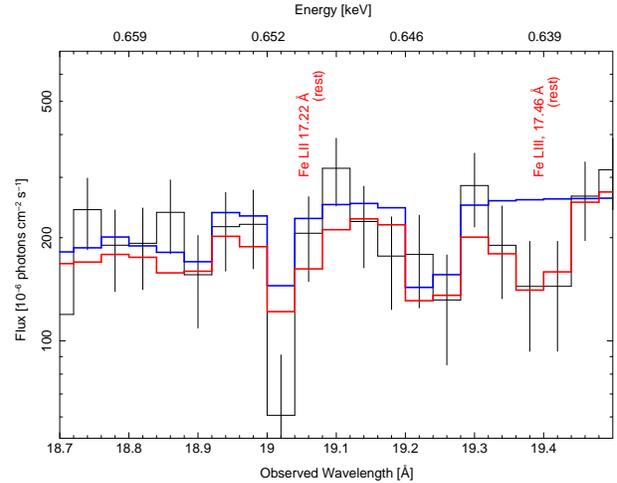}}
\caption{The Chandra MEG observed spectrum zoomed into the UTA and iron dust spectral
region, at the redshift of \ir13349 (black).  It is clear that ionised gas
alone (5-warm absorbers including the UTA) is unable to explain all the
absorption in this region (blue; see Section~\ref{subsec:uta} for detailed
discussion on UTA), and additional absorption with origins in iron-dust in
\ir13349 is also needed (red; Section~\ref{subsec:xfedust}).
\label{fig:DustOrUta}
}
\end{figure}

As just described, the UTA, while present, is unable to account for
all absorption in the 17--17.5~\AA\,~(0.71--0.73 keV rest;
observed~$\sim$~18.8-19.4~\AA\,=\,0.64--0.66) spectral region.  
%As can be seen in Figure~\ref{fig:DustOrUta} (blue), this is particularly
%noticable between 19.3-19.5~\AA\, (observed; $\sim 0.63-0.64$~keV).
It was initially proposed by \cite{jcl_mcg6wa1:01}, based on a \chandra HETGS 
study of the Seyfert galaxy \mcg6, that dust can be directly detected in high resolution 
spectra of X-ray bright objects.  For iron-based dust, these features appear
in the form of Fe~L~{\sc (iii, ii, i)} photoelectric edges between (lab) $\sim 14.7-17.5$~\AA\,
(0.7-0.84~keV; overlapping with the UTA spectral region). (Fe-based dust in 
astrophysical environments can also be measured at $\sim$~7~keV~Fe~K~--~see~\citealt{leeravel:xafs05},
but this is not relevant to the \chandra HETGS capabilities.)
As can be seen in Figure~\ref{fig:DustOrUta} (blue), the UTA is unable to
account for much of the absorption between 19.3-19.5~\AA\, (observed; $\sim 0.63-0.64$~keV),
consistent with the Fe~L~{\sc iii} edge (redshifted to $z_{\small{IR13349}}$)~-~in the context of 
\ir13349, this argues strongly for direct detections of iron-based  dust 
in the AGN environment.

%leeravel:xafs05

%%%%%%%%%%%%%%%%%%%%%%%%%%%%%%%%%%%%%%%%%%%%%%%%%%%%%%%%%%%%%%%%%%%%%%%%%%%%%%
%%%%%%%%%%%%%%%%%%%%%%%%%%%%%%%%%%%%%%%%%%%%%%%%%%%%%%%%%%%%%%%%%%%%%%%%%%%%%%

%As mentioned, fits to the X-ray data reveal direct spectral signatures
%of dust at the redshift of \ir13349 consistent with the 700~eV (lab)
%photoelectric edge of Fe~L.  
To investigate the direct detection of iron-based dust in  \ir13349, we
incorporate cross sections associated with condensed matter forms of iron into
our 4-absorber fit to determine an ionic column for iron in grains to be
$N_{\rm Fedust} \sim (1.2 \pm 0.21) \times 10^{17} \, \psqcm$.~(See
\citealt{leexafs:09} for details on X-ray methodology for determining the
quantity and composition of interstellar dust.)  Using the ISM abundance values
of \cite{ism_ref:00}, this translates to an equivalent Hydrogen column $N_{\rm
H} \sim (4.5 \pm 1.30) \times 10^{21}\, \rm cm^{-2}$.  While the X-ray data are
not of sufficient S/N to distinguish the exact iron composition (e.g. pure Fe
vs. FeO vs. $\rm Fe_2 O_3$ vs.  $\rm Fe_2 O_4$), a conservative estimate of an
observable transmission signal, requires that the iron-based grains have
thickness $\rm t \sim 0.1-0.8~\mu m$, based on the formalism for transmission
$T = e^{-\mu \rho t}$, where $\rho$ is the density of the specific compound as
defined from ``The Handbook of Chemistry and Physics"; the values for the
attenuation length were obtained from the
\footnote{http://www-cxro.lbl.gov/optical\_constants/}CXRO at LBL, and
\footnote{http://physics.nist.gov/PhysRefData/FFast/html/form.html}NIST.
Furthermore, the fits strongly suggest that at least 90\% of the Fe is locked
up in grains, {with a minor contribution from gas phase iron,
$N_{\rm Fe-gas} \sim 2.1 \times 10^{15} \, \psqcm$. }

% Susmita New
%{\bf
%We acknowledge that \citet{holczer07} had reported the strong neutral Fe
%feature at observed wavelength $\sim 19.3 \AA$ (rest $\sim 17.5 \AA$) and
%attribute it to either neutral gas or dust. In our fits, in addition to the WA components and the Fe dust absorber, we have also used neutral Fe gas as an absorption component at the redshift of the source. The column density due to the neutral gas was fit to be $N_{\rm Fegas} = 2.1 \times 10^{15} \, \psqcm$, two orders of magnitude lower than that of Fe dust. Hence we rule out neutral Fe gas as a significant contributor to the strong absorption feature at $\sim 19.3 \AA$ (observed) and attribute it to Fe dust.
%}

%%%%%%%%%%%%%%%%%%%%%%%

%%%%%%%%%%%%%%%%%%%%%%%%%%%%%%%%%%%%%%%%%%%%%%%%%%%%%%%%%%%%%%%%%%%%%%%%%%%%%%
%%%%%%%%%%%%%%%%%%%%%%%%%%%%%%%%%%%%%%%%%%%%%%%%%%%%%%%%%%%%%%%%%%%%%%%%%%%%%%

\subsubsection{The absorber kinematics }
\label{subsec:wakinematics}

\begin{figure}
\centerline{\includegraphics[scale=1.0,width = 8 cm, angle=0]{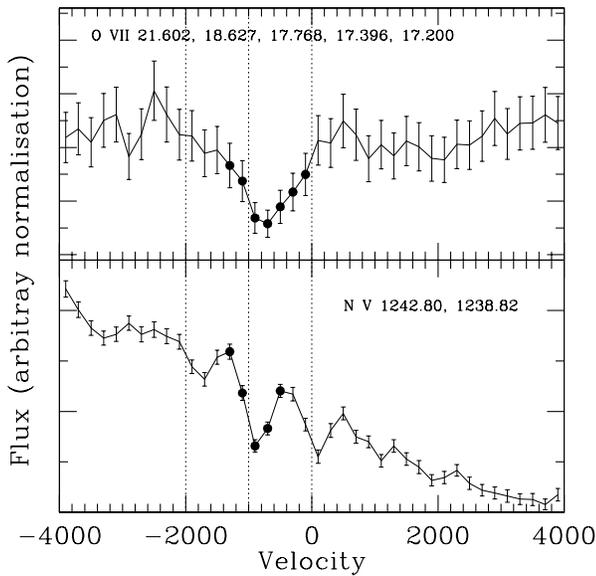}}
\caption{A velocity profile comparison of the X-ray and UV absorbers, based on
representative ions show that the UV and low-$\xi$ X-ray warm absorbers likely
have a common origin.  {\emph (TOP):} velocity spectrum combined from the five
strongest (X-ray) resonance lines of \ion{O}{vii} based on \chandra MEG X-ray
spectra, and {\emph (BOTTOM)} velocity spectrum combined from two strong
\ion{N}{v} absorption lines seen in HST STIS spectra.  Details for how these
are generated can be found in Section~\ref{subsec:xplasmastudy}.
\label{fig:compare_velprofile}
}
\end{figure}

Of interest, the outflow velocities for the X-ray absorbers we detect are best
matched to the higher velocity ($\sim 950$\kmps) UV outflowing component.~--See
e.g. Figure~\ref{fig:compare_velprofile} comparison of the HETGS view of
\ion{O}{vii} versus STIS view of \ion{N}{v}.
%{\bf The velocity match is demonstrated in Figure
%\ref{fig:compare_velprofile}.  Both OVIII (from Chandra HETGS data) and NV
%(from HST STIS data) show velocity components $\sim 800 - 900$ km/s.}
Figure~\ref{fig:TheoryNh} shows that the observed column densities for the UV
ions \ion{C}{iv} and \ion{N}{v} (panels I and J) are consistent, within errors,
with the {\sc xstar} predicted column densities for the low $\xi$ phase of the
warm absorber, determined from fitting to the Chandra spectra (OVII, panel N).
Our analysis shows that the low-$\xi$ (\warm1 and WA-4) can have linked UV and
X-ray absorption from \ion{C}{iv} to \ion{O}{vii} and some \ion{O}{viii},
although the latter is primarily associated with the higher-$\xi$ \wa2 and \w3.

%%%%%%%%%%%%%%%%%%%%%%%%%%%%%%%%%%%%%%%%%%%%%%%%%%%%%%%%%%%%%%%%%%%%%%%%%%%%%%
%%%%%%%%%%%%%%%%%%%%%%%%%%%%%%%%%%%%%%%%%%%%%%%%%%%%%%%%%%%%%%%%%%%%%%%%%%%%%%

Furthermore, it would appear that the lower ionisation phases are
associated with a higher outflowing velocity.  
 Our lower limits on the 
derived values for $v_{\rm blue}$ are roughly consistent
with the upper limit findings by \nocite{sako13349:01} 
Sako et al., (2001; $v_{\rm blue} = 420_{-180}^{+190}$~\kmps) and 
\nocite{holczer07} Holczer et al. (2007; $v_{\rm blue} = 300 \pm 50$~\kmps) 
{\it if} we account for the difference in choice of redshift value,
i.e. $z=0.10764$ \citep{ir13349-z} used by previous authors versus the $z = 0.108530$
value derived in this paper based on higher resolution HET data.
We do not find any evidence for the $\sim +20$~\kmps outflow reported
by Sako et al. however, and our fitted turbulent velocity widths are slightly
lower than the $v_{\rm turb} \sim 640$~\kmps values reported by both
Sako et al., and Holczer et al.

%We note that our
%derived values for $v_{\rm blue}$ and the turbulent width $b$ are {\it
%in}consistent with previous findings by \citet{sako13349:01}
%($420_{-180}^{+190} \,\, \rm{and} \,\, -20_{-330}^{+200}$\kmps; $b
%\sim 1500$ \kmps FWHM) based on XMM-Newton RGS data, and
%\citet{holczer07} ($300 \pm 50$\kmps; $b\sim 640$\kmps) based on this
%\chandra data set.  The discrepancy with \citet{sako13349:01} can
%likely be attributed to differences in spectral resolution between the
%RGS and HETG spectrographs; we do not understand the discrepancy with
%the \citet{holczer07} analysis, and can only attribute it to a difference
%in fitting method that we cannot assess.

%%%%%%%%%%%%%%%%%%%%%%%
%-------------------------------------------------------------------------------------

\section{Discussion}
\label{sec:discussion}

\subsection{Thermodynamics and structure of absorber}
\label{subsec:wathermo}

A topic of active debate is whether
warm absorbers in AGN are in a continuous medium or are an ensemble of
discrete clumpy media which are in different thermodynamic phases in
near pressure equilibrium with each other.  The answer to this
question has interesting consequences for the geometry of the AGN
environment, as a whole. Here, we use stability curves to demonstrate
that the choice of ionising SED (e.g. X-ray versus the full source
SED) for \ir13349 is very relevant in the conclusion for one scenario
over the other.

%%%%%%%%%%%%%%%%%%%%%%%%%%%%%%%%%%%%%%%%%%%%%%%%%%%%%%%%%%%%%%%%%%%%%%%%%%%%%%
\begin{figure}
\centerline{\includegraphics[scale=0.5,width=8 cm, angle=0]{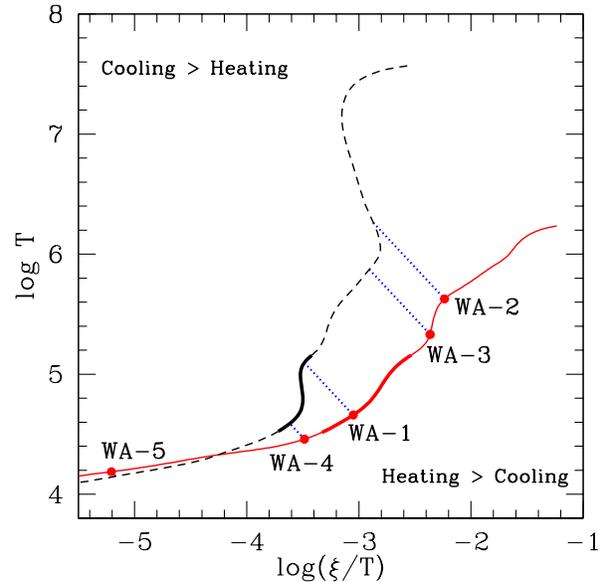}}
\caption{Stability curves illustrating differences in the thermodynamic
behaviour of the absorbing gas as illuminated by the full UV--X-ray SED  (solid
red curve corresponding to the SED in solid red shown in  Figure
\ref{fig:TheorySed}, bottom panel), versus an X-ray only ionising SED (black
dashed curve corresponding to same representation of ionising SED in
Figure~\ref{fig:TheorySed}, bottom panel). The unstable regions identified by
the \cite{holczer07} analysis of this Chandra X-ray data based on an X-ray only
ionising SED is highlighted in bold black.   Also in bold on the red curve is
the aforementioned unstable region translated over the same $\Delta T$ for the
stability curve drawn based on a UV (i.e. accretion disc) inclusive SED.  The
filled circles on the red curve are the points corresponding to \warm1, \wa2, \w3,
{\sc WA-4} and {\sc WA-5} as predicted by the {\sc xstar} fits to the Chandra HETGS
data and the HST STIS data and/or theoretical considerations, as discussed in
Section \ref{subsubsec:WaModel}. As can be seen, no unstable regions are
identified by our analysis. The dotted blue lines connect the
points with the same $\xi$ values in the two curves, thus showing that the
temperature for all the $\xi$ values are higher for the gas ionised by the
X-ray only SED, except for WA-5, where the temperatures are equal. 
\label{fig:Scurve}
}
\end{figure}
%%%%%%%%%%%%%%%%%%%%%%%%%%%%%%%%%%%%%%%%%%%%%%%%%%%%%%%%%%%%%%%%%%%%%%%%%%%%%%

Stability curves are thermodynamic phase diagrams of temperature
($\log T$) versus pressure ($\log(\xi/T)$, where
$\xi\equiv$~ionisation parameter. It is an effective theoretical tool
often used to discuss the thermodynamics of photoionized gas
associated with the X-ray and UV absorbers \citep[e.g.][and the
references therein]{kmt81gastheory, krolikkriss01:wawinds,
reynoldsfabianWa95, chakravorty09}.  By definition, an isobaric
perturbation of a system in equilibrium is represented by a small
vertical displacement from the stability curve; such perturbations
leave $\xi/T \sim L/pR^2$ constant, which for constant $L/R^2$ leave
the pressure unchanged. Any `system' located on the part of the curve
with positive slope is a stable thermodynamic system, because a
perturbation corresponding to an increase in temperature leads to
cooling, while a decrease in temperature leads to heating of the
gas. If the stability curve is characterised with discrete allowed
values of temperature at the same pressure, it points to a cloud-like
absorber, whereas a wind-like scenario will have a continuous
distribution of allowed temperature and pressure.  Before proceeding,
it is informative to briefly discuss  the definition of the ionisation
parameter in the context of stability curves.  In the definition $\xi
\sim L/nR^2$, it is conventional to use the luminosity $L$ in the
energy range $13.6 \ev - 13.6 \kev$ (i.e $1 - 10^3$ Rydberg). The warm
absorber properties are, however, determined by the photon
distribution in the soft X-rays ($E \gtrsim 100 \ev$) and not by
photons with energy $E << 100 \ev$.  Hence, in the making of warm
absorber stability curves, authors often use modifications in the
definition of the ionisation parameter. For example, \citet[and
references therein]{chelouche05} use an ionisation parameter $U_x$
which considers the ionising flux only between $540 \ev \,\,\rm{to}
\,\, 10 \kev$.  \citet{chakravorty12}, on the other hand, simply
normalise the stability curves by using the ratio of the total
luminosity to the luminosity in the X-ray power-law component in the
SED. However, it is to be noted that these modifications simply shift
the entire stability curve along the $\log(\xi/T)$ axis and does not
change the nature of the kinks or curves in the stability curve. For
example, the range ($\Delta[\log(\xi/T)]$ or $\Delta T$) of the stable
or unstable regions of the stability curves would remain unchanged
with such modifications of the ionisation parameters. As such, all the
qualitative properties of the stability curves and comparisons between
the stability curves of different SEDs, discussed subsequently, would
remain the same even if better definitions of the ionisation parameter
are used.

In the context of AGN warm absorbers, several authors
\citep[e.g.][]{kmt81gastheory, krolikkriss01:wawinds, gehrels93} have
commented that the gas in the $4.5 < \log T < 5$ region is thermally
unstable, since the cooling function $\Lambda(T)$ has a negative
gradient in this temperature range. As such, the ions
detected on the stable regions of the stability curve bracketing this
temperature range would be attributed to cloud-like absorbers where
the different phases are in pressure equilibrium with each other.
Based on the findings for just such an ionisation gap in \ir13349 by
\citet{holczer07} using data from the \chandra HETGS (the same data
we are re-analysing in this paper) and \citet{sako13349:01} with the
\XMM RGS, cloud-like clumpy absorbers would appear to be the
conclusion one would draw for this quasar. (We note however that
\citealt{sako13349:01} acknowledge that their stability curves do not
show thermodynamically unstable regions for \ir13349.)
In a comparison with the \citet{holczer07} results, the fact that their
absorption measure distribution (AMD) methodology, despite assuming a continuous
distribution of ions at the onset, concludes for thermodynamically
unstable regions, further strenghtens the case that independent of fitting technique (AMD vs.
photoionisation modelling as that employed here e.g.),
{\it it is the ionising SED that is the critical factor.}
%And while the \citet{holczer07} results, based on absorption measure distribution (AMD) 
%methodology which assumes a continuous distribution of ions at the oneset,
%%(rather than the type of photoionization modelling employed here), 
%in showing that the fits to \ir13349 shows a depletion of absorption in the 
%range $4.5 < \log T < 5$ leading to a conclusion for thermal instability
%further strenghtens the case that independent of fitting technique (AMD vs.
%photoionisation modelling as that employed here e.g.),
%it is the ionising SED that is the critical factor. 

%%%%%%%%%%%%%%%%%%%%%%%%%%%%%%%%%%%%%%%%%%%%%%%%%%%%%%%%%%%%%%%%%%%%%%%%%%%%%%
\begin{figure*}
\centerline{\includegraphics[scale = 1, width = 18 cm, trim = 20 420 20 125,
clip, angle=0]{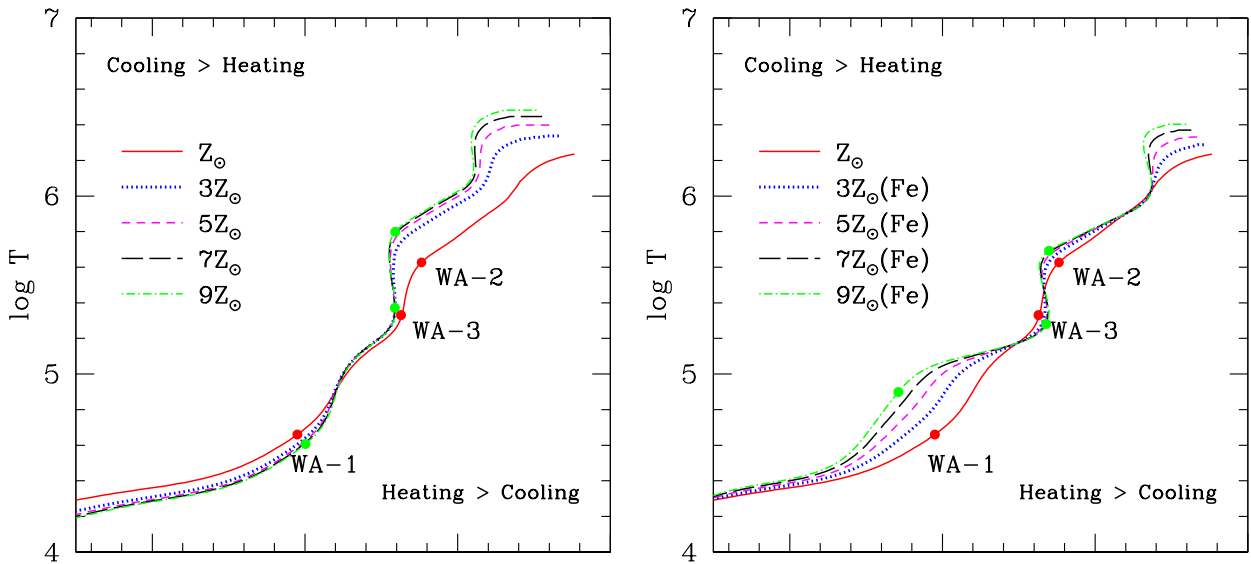}} 
\caption{{Stability curves corresponding to super-Solar gas of $3\zsol (\rm blue),
5\zsol (pink), 7 \zsol (black), {\rm and} \,\, 9 \zsol ({\rm green})$, where
(LEFT) $\zsol$ refers to {\it all} Z-elements, and (RIGHT) 
$\zsol$ refers only to iron, i.e. only iron abundance are altered here.
For comparison, in both panels, solar abundance stability curves are shown in
solid red, and plotted points correspond to {\sc xstar} predicted ionisation phases
for the warm absorber.
}
\label{fig:superSolScurve}}
\end{figure*}
%%%%%%%%%%%%%%%%%%%%%%%%%%%%%%%%%%%%%%%%%%%%%%%%%%%%%%%%%%%%%%%%%%%%%%%%%%%%%%

\citet{reynoldsfabianWa95} and \citet{chakravorty12} discuss and demonstrate
that a moderate to strong soft excess component makes the stability curve more
stable at $10^5$ K \citep[see also][]{kmt81gastheory}. \citet{chakravorty12}
further demonstrate in general for AGN that the inclusion of the accretion disc
spectral component in the EUV range in addition to a significantly strong soft
excess may completely remove the apparent thermal instability in the
aforementioned temperature range.  In this paper, we revisit the conclusions
for the thermodynamic stability conditions for the \ir13349 warm absorber gas
in the context of different SEDs: one that {\it ex}cludes the disc components
(aforementioned X-ray studies of \ir13349), and one which includes them (this
paper).  For our simulations of the photoionized gas thermodynamic properties,
we used {\sc xstar} to model a spherical shell of solar metallicity gas having
$N_{\rm{H}} = 5 \times 10^{21} \, \rm{cm^{-2}}$ and particle density
$n_{\rm{H}} = 10^9~\rm{cm^{-3}}$, illuminated by the aforementioned SEDs. It
should be noted that the stability curves are insensitive to the choice of
$10^3 \le n \le 10^{12} \rm cm^{-3}$ for the \ir13349 ionising spectra.  It can
be seen from Figure~\ref{fig:Scurve} that the choice of ionising spectrum
strongly influences the conclusions for gas thermodynamic stability. For the
two curves shown, we use thicker lines to highlight the temperature range
conventionally established to be unstable (negative slope) for photoionized gas
in AGN environments. It can be seen that an X-ray only (disc-free) ionising SED
leads to the prediction (black dashed curve) for a narrow range of $\xi$
coincident with the aforementioned temperature range over which the warm
absorber is mildly unstable. Indeed, it is exactly the finding for ions on
either side of this unstable phase seen in the black dashed curve that led
\citet{holczer07} to their conclusions for \ir13349. In contrast, when we
include the UV as part of the ionising input, no such instability is seen,
thereby pointing to a flow that is more wind-like and continuous in \ir13349.
The effect of the excess UV photons due to the addition of the accretion disk
component, facilitates the formation of OVII and other ions with similar
ionization potential (see panel D of Figure~\ref{fig:TheoryIF}).  OVII acts as
a cooling agent and hence lowers the temperature of the gas, which results in
thermodynamic stability in the region of \warm1. Moreover, the accretion disk
component also helps to lower the Compton temperature of the gas. Hence the
temperature of the curve in the region of WA-3 is lowered, rendering this part
of the curve stable, as well. Thus, while our analysis reveals similar
ionisation states found from previous studies, the additional consideration of
the UV-influence drives us to a conclusion that the ionised absorber in
\ir13349 should be more continuous and wind-like than discrete clouds.
 
Taking the thermodynamic instability-free {\it red} curve in
Figure~\ref{fig:Scurve}, it is interesting to further consider the conditions
driving the fit results.  Specifically, as discussed in
Section~\ref{subsubsec:WaModel}, the components \warm1 and \wa2 are
statistically `sufficient' to fit the Chandra HETGS data. However there are no
thermodynamically unstable phases between these two points to separate them ---
as seen in the figure (red curve). In other words, there is a range of $\xi$
which is theoretically allowed by thermodynamic calculations and yet is not
required while fitting the data.  While one explanation for this may be the
insufficient S/N of the available data, there may be additional more physically
motivated answers. Specifically, (1) there may be an absence of absorbing
material of sufficient column density at certain intermediate ionisations
(which may be related to intermediate distances from the black hole) to imprint
significantly on our spectrum.  Indeed, for \ir13349, when we `force fit' a
third component \w3 to the data, its physical conditions ($\xi$, $\rm{N_H}$ and
$\rm{v_{blue}}$) come out to be intermediate between that of \warm1 and \wa2.
Further, (2) when we include additional theoretically motivated component WA-4
and WA-5 to account for the entire UTA, we indeed populate the entire stability
curve suggesting presence of material in a continuous distribution of
ionisation states
%(2) The warm absorber may not be seeing the EUV radiation from the accretion
%disc, in its entirety. If the warm absorber is shielded from the central
%source, by some line of sight material, in such a way that it sees the X-ray
%only SED, it could have unstable phases as indicated by the black dashed curve
%in Figure~\ref{fig:Scurve}. Probing this interesting possibility for \ir13349
%further is beyond the scope of this paper.  
The stability conditions of a super-solar warm absorber would be distinctly
different from that of a gas with Solar abundances \citep[see][]{chakravorty09,
chakravorty12}.  Since the outflowing gas from the nuclear regions of AGN may
indeed have super-solar metallicities (e.g., \citealt{arav07:xuvagn}),  we
explore this idea further for \ir13349 in the next section.

\subsubsection{Super-solar metallicities}
\label{sec:supersolarmetals}

\citet{chakravorty09, chakravorty12} show that the stability curves for a
super-solar absorber is more likely to show discretely separated multi-phase
structure. In Figure~\ref{fig:superSolScurve} we show the stability curves for
different super-solar metallicites of the absorber.  Independent of whether
only iron, or all metals are varied, it can be seen that the higher metallicity
gas starts exhibiting thermodynamic instability at $\log T > 5.2$, and the high
ionisation phase \wa2 ($\rm log\, \xi \sim 3.4$; top-most, green circle) falls in a part of the curve
after it recovers from this instability. However, there is no distinct unstable
region between the low \warm1 ($\rm log\, \xi \sim 1.7$) and intermediate \w3 ($\rm log \,\xi \sim 3$) 
ionisation phases which would
mean a continuous distribution of phases in this range of ionisation
parameters. WA-3 falls exactly at the turnover where the high metallicity
curves become unstable at $\log T
> 5.2$, suggesting that while \wa2 and \w3 are thermodynamically
separated, they are nevertheless in pressure equilibrium.  It is of further
interest to note that the \wa2 and \w3 values derived from {\sc xstar} fits to
the \chandra data fall on opposite ends of the unstable regions in
Figure~\ref{fig:superSolScurve}.--This may not be coincidental,  since gas in
the unstable regions would preferentially migrate towards stable region, by
cooling or heating. Ultimately, if there are discrete (rather than continuous)
absorbers in IRAS~13349,  at least some components of the gas must have
super-solar metallicities.

We would like to caution that our discussion of super-solar
metallicities is not quantitatively absolute and should only be
considered as a qualitative impression of our expectations. A more
rigorous study would involve fitting the data with super-solar
population files in {\sc xstar} and drawing conclusions based on the
best-fit metallicity of the absorber. Given the S/N of our data,
however, such detailed fits are not warranted in this paper.

%\input{ir13349dust3.tex}
%\input{ir13349dust3_finaldraft1.tex}
%%%%%%%%%%%%%%%%%%%%%%%%%%%%%%%%%%%%%%%%%%%%%%%%%%%%%%%%%%%%%%%%%%%%%%%%%%%%%%
\begin{figure}
\centerline{\includegraphics[scale=0.5,width=8.5 cm, trim = 70 0 0 50, clip, angle=0]{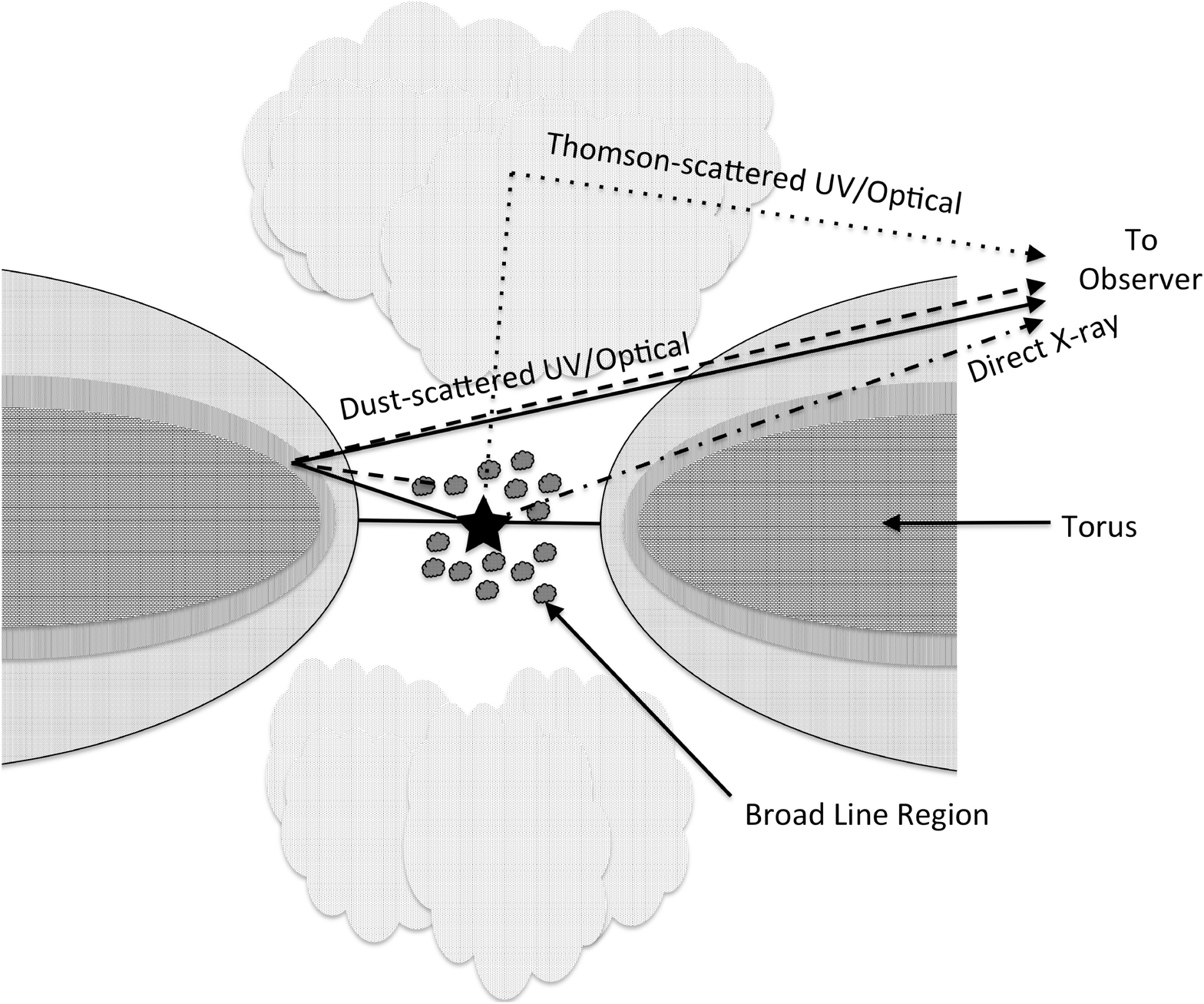}}
\caption{{\bf Inferred geometry of \ir13349.
See \S\ref{subsec-dwa} for details.}
\label{fig:geometry}
}
\end{figure}
%%%%%%%%%%%%%%%%%%%%%%%%%%%%%%%%%%%%%%%%%%%%%%%%%%%%%%%%%%%%%%%%%%%%%%%%%%%%%%

\subsection{The \ir13349 dusty warm absorber}
\label{subsec-dwa}
Given the \cite{reynolds97:dustywa} finding back in the \asca era that
$\sim 20$\% of the warm absorber Seyfert~1s have significant optical
reddening, it is not surprising  that there is now a remarkable
convergence of evidence  for dusty ionised winds from observations
with more powerful satellites.  That \ir13349 harbours dust has long
been established with IR  (\citealt{ir13349:discovery,barvainis87}),
UV/optical \citep{wills92,hines01} and X-ray \citep{brandt96, brandt97} studies.
Using the superb multi-wavelength data set presented here, in
combination with knowledge based on prior work, we present a
geometrical model for the \ir13349 line-of-sight (LOS), and location
of dust, based on dust modelling of the absorbed  optical-UV spectra
(Section~\ref{subsec:bestdustfit}) , {\it and} direct IR and X-ray
measurements (Section~\ref{subsec:xirdust}) of dust properties.

\subsubsection{A model for the \ir13349 dust} \label{subsec:irdustmodel}
As we showed in \S\ref{subsec:SedObs}, the direct UV and optical flux
from \ir13349\ is substantially lower than would be expected for a
typical QSO.  Additionally, the high polarisation rising to shorter
wavelengths \citep{hines01} is indicative of scattering by dust.
\citet{hines01} attempted to model the UV/optical flux and
polarisation of \ir13349\ as a combination of a direct view of the
nucleus reddened by an intervening screen, plus dust-scattered light
from biconical regions filling the poles above the obscuring
torus. Using dust with SMC-like properties, they were able to obtain a
reasonable match to the shape of the polarised flux spectrum, but the
predicted total flux spectrum did not match the observations---the
optical flux was much too high, and the reddening caused the predicted
UV flux to drop far below the observed intensity.  When one also
considers the shorter-wavelength data from STIS presented in this
paper, the fact that the UV flux is much higher than predicted by
their model suggests that there could be additional scattering that is
colour neutral, e.g., like the Thomson mirror of hot electrons in NGC
1068 \citep{am85-obs}.  This idea is reminiscent of the original
suggestion by \citet{wills92} that the polarised light in \ir13349\
was due to Thomson scattering, although that idea proved inconsistent
with the wavelength dependence into the near-UV observed by
\citet{hines01}.  For our proposed geometry here, the
Thomson-scattered component would play a minor role in the
polarisation, and only account for the flux at the shortest
wavelengths.

To test this idea, we developed a simple heuristic model as
illustrated in Fig.  \ref{fig:geometry}.  As shown in the figure, our
direct line of sight (the dash-dot line) to the active nucleus passes
through the upper atmosphere of the obscuring torus.  The atmosphere
consists of a mixture of highly ionised gas and dust that has not yet
been evaporated by the incident radiation.  This line of sight is
heavily reddened in the UV/optical band, or even completely blocked;
in the X-ray, since the gas is highly ionised, there is little
absorption by lighter elements, and the active nucleus is visible \citep{brandt97}.  A
second line of sight (the solid line and the dashed line) passes
through less dense portions of the atmosphere that are more
transparent at UV/optical wavelengths, but this line of sight only
gives a view of the active nucleus as reflected by dust from the far
wall of the obscuring torus.  Finally, a third line of sight passes
above the atmosphere of the torus, and it has a view of the nucleus
that is Thomson scattered from hot electrons in the thermal wind
escaping along the polar cavities of the torus.

Assuming the active nucleus plus BLR has a spectrum identical to the
composite QSO spectrum (defined as $f_{\rm c}$),  this model can be
analytically expressed as:

\begin{equation} 
F_{\rm optUV}(\lambda)  = \left [ F_{\rm direct}(\lambda) + F_{\rm scat}(\lambda) + F_{\rm Th}(\lambda)
\right ]\times 10^{-0.4A_{\rm V}g(\lambda)} 
\end{equation} 
where $A_{\rm V}$ is the optical absorption within our Galaxy and
$g(\lambda)\,=\,A_{\lambda}/A_{\rm V}$ is the extinction law in the
Milky Way, as defined in \citet{1999Fitzpatrick}.   The other
components within the \ir13349 galaxy are: \\

\noindent $\bullet$ {\bf Absorbed direct flux in the torus, torus
  atmosphere, and galaxy:} 
We define the LOS flux passing directly through the torus by:
\begin{eqnarray}
\lefteqn{
F_{\rm direct} (\lambda) = \Gamma_{\rm d} \, f_{\rm c} (\lambda)}\\
&  & \times 10^{-0.4\ (A_{{\rm V}_{\rm {tor}}}\, g_{\rm {tor}}(\lambda)+A_{{\rm V}_{\rm {atm}}}\, g_{\rm {atm}}(\lambda)+A_{{\rm V}_{\rm {gal}}}\, g_{\rm {gal}}(\lambda))} \nonumber
\end{eqnarray}
where $\Gamma_{\rm d} $ is the absolute normalisation of the composite
spectrum, $A_{{\rm V}_{\rm {tor}}}$ and $g_{\rm tor}(\lambda)\,=\,
A_{\lambda_{\rm{ tor}}}/A_{\rm V_{\rm tor}}$ are the optical
absorption and the wavelength-dependent extinction law within the
torus, $A_{\rm V_{\rm atm }}$ and $g_{\rm atm}(\lambda)\,=\,
A_{\lambda_{\rm atm}}/A_{\rm V_{\rm atm}}$ are the optical absorption
and the wavelength-dependent extinction law within the torus
atmosphere, and $A_{\rm V_{\rm{gal}}}$ and $g_{\rm gal}(\lambda)\,=\,
A_{\lambda_{\rm gal}}/A_{\rm V_{\rm gal}}$ are the optical absorption
and the wavelength-dependent extinction law within the galaxy,
respectively.\\

\noindent $\bullet$ {\bf Absorbed dust-scattered UV/optical flux in the
  torus atmosphere and diffuse medium:}
\begin{eqnarray}
\lefteqn{
F_{\rm scat} (\lambda)= \gamma_{\rm scat} \, \Gamma_{\rm d} \, f_{\rm c} (\lambda) \left (
  1 + \frac{\lambda}{\lambda_{\rm c}}\right )^{-\alpha} } \\
& & \times 10^{-0.4\ (A_{\rm V_{\rm{atm}}}\, g_{\rm{atm}} (\lambda)+A_{{\rm V}_{\rm{gal}}}\, g_{\rm{gal}}(\lambda))} \nonumber
\end{eqnarray}
where $\left (1 + \frac{\lambda}{\lambda_{\rm c}}\right
)^{-\alpha}$ stems from the wavelength dependence of the scattering
($\alpha=4$ is the Rayleigh limit), $\lambda_{\rm c}$ is the critical
wavelength of the scattering, characteristics of the typical dust
grain size, and $\gamma_{\rm scat}$ is the fraction of direct light which is dust-scattered.\\ 

\noindent $\bullet$ {\bf Absorbed Thomson-scattered UV/optical flux in the
  diffuse medium:}
\begin{equation}
F_{\rm Th} = \gamma_{\rm {Th}}\, \Gamma_{\rm d}\, f_{\rm c}(\lambda)\times 10^{-0.4\ A_{\rm V_{\rm {gal}}}\, g_{\rm {gal}}(\lambda)}
\end{equation}
where $\gamma_{\rm {Th}}$ is the fraction of direct light which is
Thomson scattered.

\subsubsection{The wavelength-dependent extinction within \ir13349}
\label{subsec:bestdustfit}

With this model in mind, we consider, through $\chi^2$ fitting, which
combination of dust absorption and  scattering  best-describes the
\ir13349 {\it observed} optical-UV spectrum.  We used the
Levenberg-Marquardt minimisation to determine  goodness of fit, based
on the IDL code {\sc mpfit} \citep{2009Markwardt}.  However, we do not
have any information about the extinction within \ir13349, neither in
the galaxy nor in  the torus and torus atmosphere. We therefore
considered several permutations of existing laws previously published
for the MW \citep{1999Fitzpatrick}, the average SMC and LMC
\citep{2003Gordon}, and starburst galaxies \citep{2000Calzetti}.  The
worst fits were obtained with the MW and/or LMC extinction, due to the
presence of the $2175$~\AA\ bump in these laws. This points towards
\ir13349\ lacking the unknown dust population responsible for this
$2175$~\AA\ absorption feature in the MW and the LMC, in good
agreement with several studies showing that the dust content in AGNs
is SMC-like \citep[see e.g.][and references therein]{2001Crenshaw, hopkins04}. As
expected, the use of the SMC and/or  the Calzetti laws gives better
results, but the fits are still poor, with reduced $\chi^2\,\ge\,3$.

We therefore build our own SMC-like extinction laws for the \ir13349\
torus, torus atmosphere, and galaxy as:
\begin{equation}
g_{\rm tor}(\lambda)\,=\,\left
  (\frac{5448~\textrm{\AA}}{\lambda}\right)^{\beta_{\rm tor}}
\end{equation}
in the torus,
\begin{equation}
g_{\rm atm}(\lambda)\,=\,\left
  (\frac{5448~\textrm{\AA}}{\lambda}\right)^{\beta_{\rm atm}}
\end{equation}
 in the torus atmosphere,
\begin{equation}
g_{\rm gal}(\lambda) \,=\,\left(\frac{5448~\textrm{\AA}}{\lambda}\right)^{\beta_{\rm gal}}
\end{equation}
in the galaxy. These laws are built so that they are equal to 1 in the
$V$ band, and we assume the dust population is different in the torus,
the torus atmosphere and the rest of the galaxy. In our fit, the
spectral indices $\beta_{\rm tor}$, $\beta_{\rm atm}$, and $\beta_{\rm
gal}$ are free parameters along with the optical extinction within the
torus $A_{\rm V_{\rm tor}}$, the atmosphere $A_{\rm V_{\rm atm}}$, and
the galaxy $A_{\rm V_{\rm gal}}$, the fraction of dust- and
Thomson-scattered light $\gamma_{\rm scat}$, $\gamma_{\rm Th}$, and
the critical dust-scattering wavelength $\lambda_{\rm c}$. The
absolute normalisation of the composite spectrum $\Gamma_{\rm d}$ is
fixed to the level discussed in Section~\ref{subsec:SedObs}, the
dust-scattering spectral index $\alpha$ is fixed to 4 (Rayleigh
limit), and the optical extinction within the MW is fixed to $A_{\rm
V_{Gal}} = 0.04$ ala the \cite{wills92}  derived value for
\ir13349. \\

%%%%%%%%%%%%%%%%%%%%%%%%%%%%%%%%%%%%%%%%%%%%%%%%%%%%%%%%%%%%%%%%%%%%%%%%%%%%%%
\begin{figure}
\centerline{\includegraphics[width=8cm]{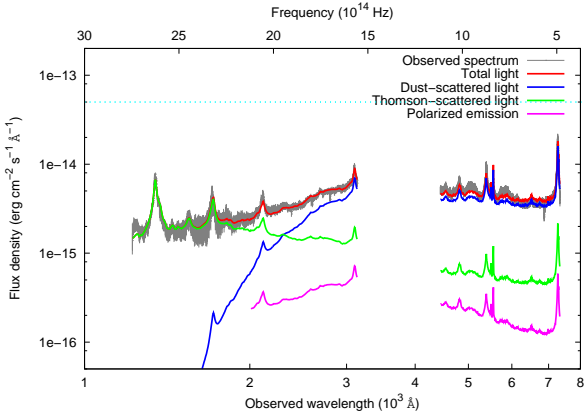}}
\caption{Best-fit model for the UV/Optical spectrum of \ir13349
(observed spectrum in grey).  The full modelled spectrum is the red
line. The Thomson-scattered component is the green
line, and the dust-scattered component is the blue line. We also
include the polarised light (magenta) as measured in Hines et
al. (2001). Our fit results point to the direct light having marginal
contribution, pointing to almost complete obscuration by the torus in
the UV and optical.
\label{fig:bestfitirdust}
}
\end{figure}
%%%%%%%%%%%%%%%%%%%%%%%%%%%%%%%%%%%%%%%%%%%%%%%%%%%%%%%%%%%%%%%%%%%%%%%%%%%%%%

\noindent$\bullet$\,{\bf Best-fit dust model:}\hspace{0.1in} In our
fitting, we are unable to constrain the fraction of dust-scattered
light $\gamma_{\rm scat}$ as it systematically pegs to 1. We can
interpret this as due to (1) an overestimation, by the composite
spectrum, of the nucleus continuum in the optical or, (2) the presence
of another contributing component in the optical domain.  We therefore
fix $\gamma_{\rm  scat}=0.2$, a conservative value chosen to reflect
the absolute normalisation of the absorbed UV spectrum needed to bring
\ir13349 in line with unabsorbed QSO behaviour (see
Section~\ref{subsec:SedObs} for details).  Likewise, it appears
difficult to constrain any torus parameters,  since all fits tends
towards removing the direct light. This strongly suggests that the
direct light is completely blocked by the torus, and so we fit using
only the dust- and Thomson-scattered light.  Our best fit
(Figure~\ref{fig:bestfitirdust}  and Table~\ref{bestfit}) points
towards a \textit{steep} extinction law within the \textit{torus
atmosphere} ($\beta_{\rm {atm}}\sim3.01$), while the extinction within
the galaxy is flatter, with $\beta_{\rm {gal}}\sim$2.11, but steeper
than that of the SMC. The typical grain size derived from the
dust-scattering function  $a\approx\lambda_{\rm
c}/2\pi\sim1.21$~$\mu$m, is consistent with large grains  within the
torus atmosphere absorbing the dust-scattered light in the far-UV,
where the emission is dominated by Thomson-scattering, with
$\gamma_{\rm Th}\sim2.4$\%.  We have checked that the lower
normalisation for the UV peak ($\sim$factor-of-two compared with the
\citealt{elvis94QSOSED} value we use; see Section~\ref{subsec:SedObs}
for details) predicted by the \cite{hopkinsQSOSED:07} SED give roughly
similar dust fit results.

We attempted to, but ultimately could not include polarisation in any
reasonable way in these fits to garner believable measurements,
although the conclusions based on our polarisation-free fits  are
consistent with those of \cite{hines01} based on polarisation studies.
Specifically, Hines et al. showed that polarisation is
wavelength-dependent,  therefore pointing to dust-scattering
dominating over wavelength-{\it in}dependent
Thomson-scattering. Similarly, our fits point strongly to
dust-scattered light dominating the 2000 and 8000~\AA\, spectral
region.

\subsubsection{Direct X-ray and IR measurements of the IRAS~13349 dust}
\label{subsec:xirdust}
Using additional knowledge gained from {\it direct} X-ray and IR
measurements of the \ir13349 dust properties as enabled by the superb
spectral resolution of the Chandra HETGS and Spitzer IRS, we further
consider the location of the different composition grains we detect. \\

\noindent $\bullet$~{\bf Chandra X-ray spectroscopic detection of iron grains:}\hspace{0.1in}

As described in Section~\ref{subsec:xfedust}, we {\it directly} detect
iron-based dust in the X-ray corresponding to $N_{\rm H_{FeGrains}} =
(4.5 \pm 1.30) \times  10^{21}\, \rm cm^{-2}$ and grain size $\sim
0.1-0.8$~\micron.   According to the radiative transfer calculations
of \cite{kama09},  iron sublimates between $\sim$~1000--2700K as a
strong function of densities (i.e. pressures).  The protoplanetary
disk conditions that they consider has typical vapour densities of
$\sim 10^{-12}$--$10^{-10} \rm g\, cm^{-3}$ which translates to
particle density $\sim 10^{10}$--$10^{12} \rm atoms \, cm^{-3}$,
assuming 55~\,$\rm grams\;moleFe^{-1}$, not dissimilar to the particle
densities  that one would expect for the inner AGN clouds. More
importantly, at these densities, iron sublimates at 1400K, a higher
temperature than silicates at 1100-1300K.  As such, in the context of
our geometrical model for the \ir13349 viewing geometry
(Figure~\ref{fig:geometry}), we can place  the iron in the upper, less
optically thick layers of the torus atmosphere,  with the silicates
more towards the interior.   Furthermore, in adopting the formalism
\begin{equation}
R_d \simeq 0.4 \, \left(\frac{L}{10^{45} \rm erg \,s^{-1}}\right)^{1/2} \left(\frac{1500K}{T_{\rm sub}}\right)^{2.6} \rm pc
\end{equation}
derived by \cite{nenkovatorus:08}, we find $\sim 2.7-5.2$ pc for the
inner radius of cloud distributions in a clumpy torus for sublimation
temperatures T$\sim 1100-1400$K and $L_{\rm bol} \sim 3 \times 10^{46}
\rm \ergps$ (see Table 7).
%%%%% DCH
The Keck Interferometer measurements of \cite{kishimoto:09} point to dust at $\sim 0.92$~pc, which 
suggests either dust T $\sim$ 2000K or larger dust
grains (e.g. 1.2\micron\, as derived based on spectral fitting in Section~\ref{subsec:irdustmodel}) 
-- see also \citep[][e.g.]{sitko:93}.

%{\color{blue} DEAN: In our telecon, you made some comment about B. Wills wanting
%Fe-filings in the interior of the torus.  Do you want to add some text to this
%effect, and other? I actually think the iron is pure Fe metal (personal bias), 
%but can't tell with the present X-ray data.  ALSO, the dust-to-gas ratio for
%seems to point to $>$70\% of the fe locked up in grains, but again, the data is
%too ratty to say for sure, but we could still make the point, if you have
%any thing to say to that effect.}

%Calculating a Schwarzschild radius $R_{\rm S} \sim 3 \times 10^{14} \rm cm$ 
%($9.4 \times 10^{-5}$pc) for the $10^9\,M_{\odot}$ IRAS~13349 black hole,

\vspace{0.1in}
\noindent $\bullet$~{\bf Spitzer IRS spectroscopic detection of
  silicates and PAHs}

As discussed in Section~\ref{subsec:iranalysis}, the Spitzer IRS
spectra show weak 7.7~\micron and 11.3~\micron PAH, as well as broad
10~\micron and 18~\micron silicate emission.  The PAHs are associated
with the  outer starbursting part of the galaxy and therefore play no
role in the dust absorption and scattering effect considered
previously.  Even if the obscuration is a supernova heated starburst
disk  (see e.g. theory in
\citealt{fabian98:starbursttorus,wada02:starbursttorus,thompson05:starbursttorus,ballantyne08:starbursttorus}),
in the IRAS~13349 host galaxy, any PAHs will have to be far enough
away from the strong radiation to have little effect.  Else, for the
detected silicates, given that they are in emission, just like in
other normal quasars and Seyfert 1s, it is quite clear that we are not
looking through a lot of dust, but rather are at a high-enough
elevation angle above the torus to clear the densest gas, but perhaps
just grazing the surface and looking through its atmosphere, which is
what is causing the UV and X-ray absorption lines and the extinction
of the UV continuum.

Furthermore, the 10$\mu$m peak is shifted to the red, indicating
either a depletion of small dust grains or more crystalline silicate
dust (e.g.  enstatite) rather than amorphous olivine.  The  opacity
at the 1.844~keV (6.724~\AA, rest) Si~K edge is too low for us to
verify the type of silicates with X-ray data.  However, the fact that
the dust we do detect in the X-rays is primarily locked up in iron,
argues strongly for the silicates to be Mg-rich (e.g. forsterite).
\cite{kempler:QSOwinddust07} argue for forsterite  in the
PG~2112+059 BAL QSO wind based on Spitzer studies; one might imagine a
similar dusty outflow for \ir13349, but as stated in
Section~\ref{subsec:spitzerhires},  assessing the level forsterite may
contribute to the 11.2~\micron\ feature is beyond the scope of this
paper.
%{\color{red} But if the dust is
%  Mg-rich and crystaline, then we should see the 11.2 micron
%  Forsterite feature... I don't think we do. - DCH} \cite{kempler:QSOwinddust07}

In further estimating where the silicate dust responsible for the
broad IR emission may reside, we derive $R_\mathrm{sub} \sim 1.2$~pc,
using the relation $R_\mathrm{sub}=0.2(L_\mathrm{bol}/10^{46})^{1/2}$
pc \citep{laor98, laor03}.  To arrive at this, we take the mid-IR
luminosity of IRAS 13349 $\nu L_\nu (25 \mu \mathrm{m}) = 3.2 \times
10^{45} = L_\mathrm{bol} f_\mathrm{c} f_\mathrm{i}$ erg s$^{-1}$,
where $f_\mathrm{c} \sim 0.6$ is  a representative dust covering
fraction and $f_\mathrm{i} \sim 0.5$ is a representative mid-IR
anisotropy for quasars \citep{ogle06, shi05}.  Based on this, it is
likely then that the dust responsible for the mid-IR continuum is the
same dust that absorbs the direct optical-UV spectrum, since it
accounts for a large fraction of the absorbed and reprocessed
bolometric luminosity of the quasar.

\subsection{The dusty wind in IRAS~13349: torus or accretion disk origin?}
\noindent Given the excellent qualitative agreement between our
proposed model  for the dust contributions
(Section~\ref{subsec:bestdustfit}) and the observed spectrum, our
picture in Figure~\ref{fig:geometry} must be close to reality. One
additional piece of evidence in support of our model can be gleaned
from the existing polarisation data.  \citet{hines01} note a
difference in polarisation angle for the broad emission lines compared
to the continuum.  The line centres have their angles rotated by ~5
deg, but the polarisation stays fairly constant. One can interpret
this rotation in terms of the differing angle between the scattering
region and the sources of the incident radiation, meaning that the BLR
is displaced from the optical/UV continuum source by several degrees
as viewed from the scatterer. Also, since the emission lines are not
depolarised, the angles from across the full BLR to the scatterer must
be within a fairly narrow range so that their polarisations do not
cancel out. So, the size of the BLR as seen from the scatterer must be
small.  If we take the difference in polarisation angle as the offset
of the BLR from the continuum source, then, given a distance for the
BLR from the nucleus, we can deduce the distance of the scatterer from
the continuum source.  Scaling up reverberation mapping results for
BLR sizes \citep{kaspi_revmap:05} to our extinction-corrected UV
luminosity for \ir13349\ ($\rm L_{1350} = 7.95 \times
10^{45}~\ergps$), the BLR would be at a radius of $\rm r = 7 \times
10^{17}$ cm.  The scatterer would then be at $8 \times 10^{18}$ cm, or
2.6 pc, For comparison, the dust sublimation radius $R_\mathrm{sub}
\sim 1.2$~pc we calculate in Section~\ref{subsec:xirdust} (same value
from \citealt{barvainis87}), and the inner edge of the torus is
expected to be at $\sim 1~\rm pc (L/10^{45})^{0.5} = 2.8$ pc
\citep{krolik88}.  This is remarkably consistent with the optical
light being scattered light from the far wall of the torus.  In
adopting the classic picture presented  whereby an obscuring torus is
always present  (\citealt{a93-review}; see also
\citealt{lawrence-torus:82}) our view of \ir13349\ then is at high
inclination,  barely skimming the surface of the torus.

This picture presents a problem then for scenarios of accretion-disk
winds (e.g., \citealt{elvis00}), where the wind streamlines favour the
equatorial plane, parallel to the surface of the torus. Nevertheless,
it is of interest to consider further an alternative to the torus  for
the obscuration, that would allow the wind to have an accretion disk
origin. Recently, \cite{czerny11:dustwind} proposed a model for the
creation of dusty winds at $\approxgt 1000 R_{\rm S}$ by drawing
parallels between clouds on quasar disk surfaces to  similar
temperature and pressure conditions that cause dust formation in AGB
stars. In the context of the Elvis quasar model, the material in this
wind would then be the cause of the obscuration, in lieu of the torus.
But can this satisfy our condition for a high inclination viewing
angle for IRAS~13349?   Based on the wind flow lines shown in Fig. 3
of  \cite{risalittiwind:10}, the scale height at which our
line-of-sight intersects the dusty wind is $z = 200R_{\rm S} =
0.09$~pc at  1000$RR_{\rm S}^{-1}$ for a $10^{9} M_\odot$, quite close
to the disk plane.  Alternatively, a warped disk  as the source of
obscuration \citep{lawrence-ad:08},  although  for IRAS~13349, it
would have to be contorted in such a way as to satisfy the known
effects caused by dust scattering, absorption, and polarising.  Else,
if we again consider the putative torus, one might argue that our line
of sight on the near side of the torus passes under the streamlines of
the accretion disk wind, but then the  line of sight from the nucleus
to the far wall of the torus and back again must pass directly through
the wind, and at multiple angles.

While the accretion disk wind idea is a desirable one, it is
ultimately more difficult to reconcile, compared to a torus wind, for
IRAS~143349.  The line widths in our STIS spectrum are similar to the
narrow, few hundred \kmps\ absorption lines common in other nearby AGN
\citep{crenshaw03, kriss02, dunn07}, as expected for the thermal winds
of \citet{krolik95, krolikkriss01:wawinds}.   Also, the velocity
shifts we detect in both UV and X-ray are slow compared to the
ultra-fast outflows expected from the disk
(\citealt{tombesi10-RQ-ufo1}; see also
\citealt{chartas_apm08279_qsowind:02,pounds_pg1211_qsowind:03} for the first claims of such
outflows in Seyferts).

If there is no disk wind in this object, then perhaps the conditions
that drive it are just not right.  The disk winds of \citet{murray95a,
murray95b} require a high ionisation parameter and and a high column
density at the inner edge of the wind, and \ir13349\ does not have a
high column density of either X-ray or UV-absorbing gas.  The
manifestation of a disk wind in an AGN may not be a universal property
that depends only on orientation, but also on other intrinsic
parameters such as the accretion rate or the Eddington ratio.

\subsection{Similarities to non-black hole astrophysical systems}
\label{subsec:simgeometries}
It is interesting that many apparent parallels can be drawn between
\ir13349,  a massive energetic black hole systems  fuelling a whole
galaxy, and other less powerful/smaller systems.  For example, the
relatively large grains found in the torus of \ir13349 are also
encountered in the dusty winds of luminous blue variable \citep[LBVs,
see e.g.][]{1998Robberto, 2003Hara}. In these extreme massive stars,
the maximum grain size is set proportionally  to the mass-loss rate of
the high-opacity wind, and dust is likely created during stellar
eruptions, when this mass-loss reaches very high levels
\citep[0.001--0.01~$M_\odot$/yr,][]{2011Kochanek}.  In \ir13349, the
high-opacity component (characterised by the high optical extinction
suffered by the direct light), and the dust appear to be co-located,
suggesting perhaps also dust formation in the \ir13349 outflows, be it
from the torus or disk.   In addition, based on sublimation
temperature arguments, our placement for the iron- grains (detected in
the Chandra X-ray spectra)  in the rim and upper layers of the torus
with the silicates (detected  in Spitzer IR data) in the interior,
gives a picture qualitatively similar to that arrived at by
\cite{chiang:dustystars:01} for the dust in the surface layers  of
T~Tauri and Herbig Ae disks.

\begin{table}
\begin{center}
\begin{tabular}{ccl}
\hline
\hline
Parameters&Best-fit value& Definition\\
\hline
$\gamma_{\rm scat}$&0.2 (fixed)&Fraction of dust-scattered light\\
$\lambda_{\rm c}$ ($\mu$m)&$7.60\pm0.62$&Dust-scattering critical wavelength\\
$\alpha$&4 (fixed)&Dust-scattering index\\
$A_{\rm V_{\rm atm}}$&$0.14\pm0.02$&Optical extinction in atmosphere\\
$\beta_{\rm atm}$&$3.01\pm0.08$&Spectral index of extinction in atmosphere\\
$\gamma_{\rm Th}$&$0.024\pm0.003$&Fraction of Thomson-scattered light\\
$A_{\rm V_{\rm gal}}$&$0.05\pm0.01$&Optical extinction in galaxy\\
$\beta_{\rm gal}$&$2.11\pm0.11$&Spectral index of extinction in
galaxy\\
$\chi_{\rm r}^2$ (d.o.f)&1.66 (862)\\
\hline
\end{tabular}
\end{center}
\caption{Best-fit parameters. Uncertainties are given at the 2$\sigma$
  confidence.}
\label{bestfit}
\end{table}

%\bibliographystyle{apj3}
%\bibliography{jcl.bib}
%\end{document}

\section{Summary}
Although \ir13349 shows many signs of being an obscured QSO, the
Spitzer spectra show it as a typical QSO.  The continuum peaks at 30
$\mu$m; the silicate features at 10 and 18 $\mu$m are in emission, and
PAH features are weak.  The emission lines are dominated by highly
ionised species such as $[$\ion{Ne}{v}$]$ and $[$\ion{O}{iv}$]$. Furthermore, even though \ir13349 exhibits
extreme Eigenvector-1 characteristics, HET optical spectra argue 
against its classification as a narrow-line-Seyfert~1. 
The high
signal-to-noise and $R \sim 1300$  HET data has also allowed us to improve redshift
measurements for \ir13349, locating it at z=0.10853.

In the ultraviolet, our low-resolution STIS spectra cover the observed
wavelength range from 1150--3180 \AA. 
These spectra show for the
first time blue-shifted absorption from Ly$\alpha$, \ion{N}{v} and
{\ion{C}{iv}, with components at systemic velocities of $-950~\kmps$ and
$-75~\kmps$.  The lines are unresolved, and they have intrinsic widths
with Doppler parameters that are $< 200~\rm \kmps$.  As seen before
(Hines et al. 2001), the UV spectrum is heavily reddened, but at the
shortest wavelengths in the UV, below an observed wavelength of 1600 \AA,  
the spectrum begins to
recover in flux and appears less heavily reddened.

Our HETGS spectrum of \ir13349 covers X-ray wavelengths from 2--38 \AA
(observed).  The continuum is well described by a power law with photon index
$\Gamma = 1.9$.  At low energies we observe a soft X-ray excess that
can be modelled as a black body with $kT \sim 100$~eV.  \ir13349 is
well known for the ionised absorption that is prominent in its X-ray
spectrum (Brandt et al. 1996, 1997; Sako et al.  2001; Longinotti et
al. 2003; Holczer et al. 2007).  Our spectrum shows absorption lines
attributable to at least two warm absorbers at velocities comparable
to the higher velocity UV absorber.  The lowest ionisation \warm1 is
co-located with the UV absorber.

Using simultaneous \chandra HETGS and HST/STIS spectra obtained in
February 2004, contemporaneous ground-based optical spectra from the
HET, and archival {\it Spitzer} IRS spectra from 2005 as well as IRAS,
ISO, and 2MASS photometry data, we have constructed a broad-band view
of the SED of \ir13349.  If we compare the observed SED of \ir13349 to
the median QSO spectral energy distribution (SED) of Elvis et
al. (1994), we see that the IR and X-ray bands are a good match, but
the UV and optical are severely suppressed.  This suggests that these
portions of the spectrum are either heavily obscured and/or only
viewed in scattered light.  By normalising the peak of the infrared
continuum and the absorption-corrected X-ray continuum to the Elvis et
al.  (1994) SED, we determine that the unreddened portion of the
optical continuum represents a fraction of only 20\% of the intrinsic
light from \ir13349.

If we correct the observed SED of \ir13349 for scattering and
extinction, it peaks at a rest wavelength of $\sim$1000 \AA.  A simple
sum-of-thermal black body thermal disc spectra provides a good fit to
the SED.  We arrive at a black hole mass $M_{\rm BH} \sim10^{9}~\Msun$
for \ir13349 based on theoretical fits to the UV-Xray SED, {\it and}
the H$\beta$ line width, independently.  Theoretical considerations
comparing different ionising SEDs reveal that including the UV
(i.e. disc) as part of the ionising continuum has strong implications
for  the conclusions one would draw about the thermodynamic stability
of the warm absorber.  Specific to \ir13349, we find that an Xray-UV
ionising SED leads to the conclusion for a continuous distribution of
ionisation states in e.g. a smooth flow (this paper), versus discrete
clouds in pressure equilibrium (previous work,
e.g. \citealt{holczer07}).

To explain the shape of the SED, we developed a geometrical model in
which we view the nuclear regions of \ir13349 along a line of sight
that passes through the upper atmosphere of an obscuring torus.  This
sight-line is largely transparent in X-rays since the gas is ionised,
but, as previously discussed, it is completely obscured by dust that blocks a
direct view of the UV/optical emission region.  20\% of the intrinsic
UV/optical continuum is scattered into our sight line` by the far wall of an
obscuring torus.  An additional 2.4\% of the direct light, which likely dominates the UV emission,
is Thomson-scattered into our line-of-sight by another off-plane component of highly ionized gas.
%%%%% DCH
Our model suggests that the direct line-of-sight is
probably completely obscured by dust. In the standard Unified Scheme,
such a configuration should produce a spectrum dominated by
narrow-emission lines, and the object would be classified as a Type 2
QSO.  However, our model also suggests that nearly all of the
UV/optical light we see is scattered from the central engine.  Similar
inferences have been made to explain the observed properties of other
infrared-selected QSOs --- 2MASSI J130005.3163214
\citep{schmidt_qso:02} and 2MASX J10494334+5837501
\citep{schmidt_qso:07} --- and the BALQSOs Mrk~231
\citep{gallagher_bal:05}. In each case, the view of the central
engine, and the scattered light, are significantly obscured and the
broad-line emission is revealed primarily in scattered light.  For
Mrk~231, imaging polarimetry constrains the scattering region to be
very compact. For \ir13349 we also argue that the scattering material
is fairly close to the BLR. It is likely that viewed from a higher
inclination, these compact scattering regions would also be obscured
and the object would be classified as a Type 2 QSO.  In that case, the
BLR could be revealed by light scattered much farther from the central
engine, in an ionisation/scattering (bi)cone in direct analogy with
other highly polarised Type 2 QSOs (e.g. \citealt{hines:ir13349:93,
hines95, hines_qso:99, tran:00, zakamska:05}).

Future modeling of \ir13349 can take our picture presented above as a
starting point to calculate more detailed radiative transfer models of
the scattering processes that include polarisation calculations. These
can provide more quantitative constraints on the viewing angles and
sizes of the scattering regions in our complex view of this intriguing
AGN.  We note further that few grating-resolution X-ray spectra exist
for quasar-luminosity objects, and it is important to acquire as many
as possible so that their properties can be compared reliably with
those of the well-studied local Seyfert galaxies.  Similarly, only a
few grating-resolution spectra exist for strong \ion{Fe}{ii}, weak
$[$\ion{O}{iii}$]$ galaxies with soft X-ray spectra that are at the
extremes of Eigenvector-1, as is \ir13349.

\section*{acknowledgements}
We are grateful to Dave Huenemoerder, John Houck, Tim Kallman, and
Aneta Siemiginowska  for helpful dialogue and/or computing-related
assistance, and the referee for a thorough reading of the paper.  
We acknowledge the generous support of Chandra Guest
Observer grant GO4-5110C for support of this work.  Funding was also
provided by NASA through grants for HST program number 10088 from the
Space Telescope Science Institute, which is operated by the
Association of Universities for Research in Astronomy, Incorporated,
under NASA contract NAS5-26555. WNB acknowledges the support from NASA ADP
grant NNX11AJ59G.  The HET is a joint project of the University of Texas at
Austin, the Pennsylvania State University, Stanford University,
Ludwig-Maximillians-Universitat Munchen, and Georg-August-Universitat
Gottingen. The HET is named in honor of its principal benefactors, William P.
Hobby and Robert E. Eberly.  This research has made use of NASA's Astrophysics
Data System Bibliographic Services.

\bibliographystyle{apj3}
\bibliography{jcl.bib}
%\bibliography{Lee_Iras13349_Resubmit01_Vrsn03.bib}

\end{document}